\documentclass[12pt,prd,aps,showpacs,tightenlines,preprint,nofootinbib,floatfix]{revtex4-1} 

\usepackage{epsfig,color,graphicx}
\usepackage{multirow}
\usepackage[dvipsnames]{xcolor}
\usepackage{hyperref}
\usepackage{amsmath,amssymb}
\usepackage{rotating}
\usepackage[caption=false]{subfig}
\usepackage{bm}
\usepackage{york}

\begin{document}
\preprint{CERN-TH-2018-227}


\title{Evidence for charm-bottom tetraquarks and the mass dependence of heavy-light tetraquark states from lattice QCD}

\newcommand\york{Department of Physics and Astronomy, York University, Toronto, Ontario, M3J 1P3, Canada}

\newcommand\cern{Theoretical Physics Department, CERN, CH-1211 Geneva 23, Switzerland}

\author{Anthony~Francis$^1$}
\email{anthony.francis@cern.ch}
\author{Renwick~J.~Hudspith$^2$}
\email{renwick.james.hudspith@googlemail.com}
\author{Randy~Lewis$^2$}
\email{randy.lewis@yorku.ca}
\author{Kim~Maltman$^2$}
\email{kmaltman@yorku.ca}

\affiliation{$^1$\cern,$^2$\york}

\date{\today}

\begin{abstract}
We continue our study of heavy-light four-quark states and find evidence from lattice QCD for the existence of a 
strong-interaction-stable $I(J^P)=0(1^+)$ \udcb tetraquark 
with mass in the range of $15$ to $61$ MeV below $\bar{D}B^*$ threshold. Since this range includes the electromagnetic $\bar{D}B\gamma$ decay threshold, current uncertainties do 
not allow us to determine whether such a state would decay 
electromagnetically, or only weakly. We also perform a study 
at fixed pion mass, with NRQCD for the heavy quarks, simulating 
$qq^\prime \bar{b}^\prime \bar{b}$ and 
$q q^\prime \bar{b}^\prime\bar{b}^\prime$ 
tetraquarks with $q,\, q^\prime =ud$ or $\ell s$ and variable, 
unphysical $m_{b^\prime}$ in order to investigate 
the heavy mass-dependence of such tetraquark states. We 
find that the dependence of the binding energy follows a 
phenomenologically-expected form and that, though
NRQCD breaks down before $m_{b^\prime}=m_c$ is reached, the
results at higher $m_{b^\prime}$ clearly identify the $ud\bar{b}^\prime \bar{b}$ channel as the most likely 
to support a strong-interaction-stable tetraquark state 
at $m_{b^\prime}=m_c$. This observation serves to motivate the 
direct \udcb simulation. Throughout we use dynamical 
$n_f=2+1$ ensembles with pion masses $m_\pi=415$, $299$, 
and $164$ MeV reaching down almost to the physical point, a 
relativistic heavy quark prescription for the charm quark, and 
NRQCD for the bottom quark(s).
\end{abstract}


\maketitle

\section{Introduction}

Many theoretical efforts since the formulation of QCD  
have hypothesized the existence of exotic states containing
four or more quarks and/or antiquarks (for a recent review see \cite{Ali:2017jda} and references therein). It is only
in the last decade that unambiguously exotic states, 
including the
hidden-charm pentaquark states recently
discovered at LHCb \cite{Aaij:2015tga} and at least some
of the XYZ states \cite{Ali:2017jda}, which fail to fit
into the standard quark model picture, have begun
to be observed experimentally.
These experimental results have shown that complicated 
many-quark structures do exist in nature and the goal for theorists is to investigate why some such multi-quark structures are preferred, and to elucidate the mechanisms
underlying their existence. The mechanisms behind the binding of such configurations should help to provide 
greater insight into the complex phenomena of QCD in
the non-perturbative realm.

In this work we use lattice QCD to investigate
configurations of two
light quarks and two heavy antiquarks in channels expected
to be favorable to the formation of bound, exotic tetraquark
states. In general, such four-quark bound states have not yet been definitely proven to exist experimentally, but there are indications, both from 
models and from lattice QCD \cite{Zouzou:1986qh,Lipkin:1986dw,SilvestreBrac:1993ss,
Semay:1994ht,Pepin:1996id,Brink:1998as,Barnes:1999hs,Gelman:2002wf,
Vijande:2003ki,Janc:2004qn,Ebert:2007rn,Vijande:2007rf,Zhang:2007mu,Lee:2009rt,Vijande:2009kj,Yang:2009zzp,Valcarce:2010zs,Carames:2011zz,Ohkoda:2012hv,Hyodo:2012pm,Silbar:2013dda,Karliner:2017qjm,Eichten:2017ffp,Du:2012wp,Chen:2013aba,Bicudo:2015vta,Francis:2016hui}, that they should exist.
A benefit of employing the lattice approach is that 
unphysical quark masses can be used as input simulation parameters, allowing for an extended investigation of the
underlying binding mechanisms.

In a prior work \cite{Francis:2016hui} we predicted the existence of \udbb and \lsbb tetraquarks with quantum numbers $I(J^P)=0(1^+)$
and ${\frac{1}{2}}(1^+)$, respectively, using lattice QCD. 
The focus on these channels was motivated 
by features of the splittings in the heavy baryon spectrum.
The key observation is that a pair of heavy antiquarks in 
a color $3_c$ configuration will serve as the source of a nearly static color $3_c$ field, analogous to that produced
by the heavy quark in a singly-heavy baryon. From the pattern of
splittings in the heavy baryon sector, it is clear that a
strong spin-dependent attraction exists for light quark $ud$ 
or $\ell s$ pairs in Jaffe's \cite{Jaffe:2004ph} ``good-diquark'' configuration (color $\bar{3}_c$ and $I=J=0$ or $I=1/2, J=0$, for $ud$ or $\ell s$, respectively). 
The strength of this attraction, moreover, increases the lighter the light 
quark mass. In a doubly-heavy $I(J^P)=0(1^+)$ or 
${\frac{1}{2}}(1^+)$ $qq^\prime \bar{Q}\bar{Q}^\prime$ tetraquark channel, this good-diquark attraction is available to a localized
four-quark state, but not to the lowest-lying asymptotic
two-meson state in the same channel, where the spin-dependent
interactions of the light quarks with their heavy antiquark partners from the same meson are suppressed by the heavy 
quark mass. The attractive $\bar{Q}\bar{Q}^\prime$ $3_c$
color-Coulomb interaction provides a further contribution
to binding in the localized tetraquark configuration not
available to two separated mesons. This picture leads to
the expectation that bound $ud\bar{b}\bar{b}$ and 
$\ell s\bar{b}\bar{b}$ tetraquark states should exist in
the $I(J^P)=0(1^+)$ and ${\frac{1}{2}}(1^+)$ channels, with 
binding of order $150-200$ MeV for the former and 
a reduced binding for the latter. These semi-quantitative
expectations were confirmed by explicit lattice
simulations, in which we found states bound by
$\left|\Delta E_{ud\bar{b}\bar{b}}\right|=189(10)(3)$ and 
$\left|\Delta E_{\ell s\bar{b}\bar{b}}\right|=98(7)(3) \text{ MeV}$ 
relative to the corresponding two-meson thresholds, 
$BB^*$ and $B_sB^*$ respectively, at the physical mass point. With such binding
energies, these states are not only strong-interaction 
stable, but can, in fact, decay only weakly. Other lattice studies using static prescriptions of the bottom quarks and heavier than physical sea quarks \cite{Richards:1990xf,Mihaly:1996ue,Stewart:1998hk,Michael:1999nq,Pennanen:1999xi,Cook:2002am,Detmold:2007wk,Bali:2011gq,Wagner:2011ev,Cheung:2017tnt} analyzing similar quantities have observed attractive potentials in this channel and some \cite{Brown:2012tm,Bicudo:2012qt,Bicudo:2015vta,Bicudo:2016jwl} have indicated binding as well.

Since the discovery of the doubly-charmed $\Xi_{cc}$ baryon at LHCb \cite{Aaij:2017ueg}, sum rule calculations and phenomenological models, see e.g. \cite{Eichten:2017ffp,Czarnecki:2017vco,Mehen:2017nrh,Karliner:2017qjm}, have also led to the identification
of these channels as favorable to doubly heavy tetraquark
binding. In the model calculations, an important role is also 
played by the attractive natures of the light-quark
spin-dependent interactions and short distance 
color-Coulomb potential for heavy antiquark pairs in a 
$3_c$ color configuration \cite{Ali:2017jda}. 

Assuming the above picture correctly captures the basic
physics involved in the binding observed in the 
\udbb and \lsbb systems,
one should see binding which grows as the heavy
quark mass is increased since the short-distance,
$3_c$ color-Coulomb attraction should scale as the 
reduced mass of the heavy anti-diquark system. One 
should also see contributions to the binding which are, to a
first approximation, independent of the heavy quark mass
corresponding to the good-light-diquark attraction
in the static heavy quark limit. Corrections to this
limit should produce binding corrections proportional to the inverse of the heavy quark mass. The increase with decreasing heavy-quark 
mass of the net residual light-heavy spin-dependent 
attraction in the two-meson (vector-pseudoscalar) threshold
will also reduce
the tetraquark binding relative to this threshold
and produce binding corrections proportional 
to the inverse of the heavy quark mass. 

These qualitative expectations can be tested by extending
our previous study to include unphysical values of the
masses of one or both of the two heavy antiquarks. The
only limitation is the NRQCD approach to treating the
$\bar{b}$, which breaks down before the charm quark
mass is reached. As we will see, the results of this
variable-heavy-mass study confirm the picture outlined above. 
Although the NRQCD-based approach does not allow us to
push this study down to the charm mass, the pattern of bindings obtained 
as the variable heavy mass (or masses) is (are) decreased below $m_b$ can, nonetheless, be used to identify
which channel (or channels) involving one or two charmed
antiquarks is (are) most likely to support bound tetraquark states. These considerations lead us to focus our attention,
and direct simulation efforts, on the most favorable of 
these channels, which turns out to be the $ud\bar{c}\bar{b}$ channel.

In this paper we will first detail the variable heavy mass
study outlined above, and then discuss the results of our
direct simulations of the $ud\bar{c}\bar{b}$ channel.
Throughout, we will use the same three dynamical, fixed lattice spacing, $n_f=2+1$ PACS-CS ensembles 
employed in our previous \udbb study.
These ensembles have pion masses $m_\pi=164,\, 299$ and 
$415 \text{ MeV}$.
A relativistic prescription will be used for the charm quark
and, as before, NRQCD for the bottom quark. Evidence is 
presented for the existence 
of an $I(J^P)=0(1^+)$ \udcb
tetraquark bound with respect to the lowest free two-meson
threshold, $\bar{D}B^*$, in this channel.

For the variable-heavy-mass study, we will focus our attention on 
the ensemble with $m_\pi=299\text{ MeV}$, and study the heavy anti-diquark mass dependence for unphysical tetraquark 
candidates \qqpQQp, with $q=u,q^\prime=d,\, s$ and either 
$\bar Q^\prime \neq \bar Q$ or $\bar Q^\prime = \bar Q$.

\section{Phenomenology of heavy-light tetraquarks}\label{sec:pheno}

We focus on heavy-light tetraquark candidates
\qqpQQp that can be pictured as a combination of a ``good''
light-diquark $qq^\prime$ and heavy anti-diquark $\bar{Q}\bar{Q}^\prime$
with $q q^\prime =ud$, or $\ell s$ and $Q$, $Q^\prime = b$
or $c$. The color $\bar{3}_c$, $J=0$, flavor-antisymmetric
good-light-diquark configuration
is accessible only when the heavy anti-diquark is in a color
$3_c$. Assuming no spatial excitation between the heavy
antiquarks, the heavy anti-diquark spin is necessarily
$J_h=1$ when $Q=Q^\prime$. The favored tetraquark 
configuration is then $J^P=1^+$.

In the limit of infinitely heavy $Q,Q^\prime$, 
$m_{Q,Q'}\rightarrow\infty$, the attractive nature of the color-Coulomb potential guarantees a bound ground state of the 
\qqpQQp type \cite{Heller:1985cb,Carlson:1987hh,Manohar:1992nd}. 
Whether a bound state is realized away from this limit, in particular, when $Q, Q^\prime$
are charm or bottom quarks, depends on the details of non-perturbative effects in such systems.

The phenomenological arguments outlined in the previous
section, based on observed splitting patterns in the
heavy baryon system, were shown in Ref.~\cite{Francis:2016hui} to 
suggest the likelihood of the existence of tetraquark bound states 
of the $qq^\prime\bar{b}\bar{b}$-type with tetraquark binding increasing with
decreasing light quark mass(es). 

The lattice results of Ref.~\cite{Francis:2016hui} not 
only confirmed the existence of these bound states, but produced
binding energies for physical light quark masses in line with those expected based on the heavy baryon spectrum arguments.
Assuming the picture underlying this successful prediction
is correct, and the good-diquark contribution to binding indeed increases with decreasing light quark mass, this implies
it is imperative to have access to light quark masses
as close to physical as possible in lattice simulations
of such tetraquark channels. This is a firm prediction of this binding mechanism, 
one that may seem counterintuitive given
the common experience with lattice calculations in other channels, for example, 
in the study of multi-baryon states, where a decrease in constituent quark masses 
also decreases the binding energy (for a collection of recent results and presentation 
of the issues faced see \cite{Iritani:2017rlk} and references therein).

It is possible to further test the qualitative physical picture underlying this understanding of the $ud\bar{b}\bar{b}$ and
$\ell s\bar{b}\bar{b}$ tetraquark binding observed in
Ref.~\cite{Francis:2016hui} by studying related systems
with variable heavy antiquark mass(es). This study is carried out using the same NRQCD action used previously, in 
Ref.~\cite{Francis:2016nmj}, for the physical bottom quark case.
A brief outline of the NRQCD framework, together with details of the 
implementation of the variable $b^\prime$ mass, are provided in
Appendix A. The NRQCD heavy-mass parameter, $m_Q$, was tuned by measuring 
the dispersion relation of the spin-averaged $\Upsilon$ and $\eta_b$. We also 
computed static propagators, allowing the 
extrapolation of $m_{b^\prime}$ to $\infty$ to be carried out 
for the $Q^\prime\neq Q$ case. For the variable heavy
mass study, we focus, to be specific, on the 
intermediate ensemble, $E_M$, with $m_\pi =299$ MeV
and $m_\pi L=4.4$, and 
consider unphysical bottom quark masses $\simeq 6.29, 4.40, 1.93, 1.46, 0.85, 0.68, 0.64$ and $0.60$ 
times the physical bottom quark mass. Lower values are not accessible in this approach 
due to the breakdown of the NRQCD approximation. Denoting 
such unphysical bottom quarks by $b^\prime$, we investigate 
$qq^\prime \bar{b}\bar{b}^\prime$ and 
$qq^\prime\bar{b}^\prime \bar{b}^\prime$ tetraquark channels. 

Given the qualitative physical picture outlined above, we expect
there to be a contribution to tetraquark binding from the color-Coulomb attraction between the two heavy antiquarks in the color
$3_c$ configuration which scales linearly with the reduced mass
of the heavy anti-diquark system. There should also be a contribution to the binding 
which, for a given light-diquark channel, should be independent of the heavy quark 
masses, reflecting the attractive nature of the good-light-diquark configuration in the 
infinite heavy quark mass limit. 
Finally, there should be contributions to the binding
resulting from the presence of residual heavy-light 
interactions, scaling as the inverse of the heavy quark mass(es), in both the tetraquark 
and two-meson threshold states.
Contributions of the former type should scale as 
${\frac{1}{m_{h1}}}+{\frac{1}{m_{h2}}}$ for tetraquark
states with heavy antiquark masses $m_{h1}$ and $m_{h2}$.
Residual interactions between a heavy quark and light-diquark are also present in the heavy baryon systems.
Comparing the $\Sigma_h-\Lambda_h$ and $\Xi^\prime_h-\Xi_h$ splittings 
for $h=b,c$, one finds heavy baryon
residual interactions depending on both the inverse of
the heavy quark mass and the type ($ud$ or $\ell s$) of
light-diquark. We thus expect the coefficient of 
${\frac{1}{m_{h1}}}+{\frac{1}{m_{h2}}}$ for the residual
heavy-light tetraquark interactions to be different
for tetraquarks containing $ud$ and $\ell s$ good diquarks.
With the ratio of observed charm and bottom vector-pseudoscalar
splittings in good agreement with expectations based on
the assumption that these scale as the inverse of the
relevant heavy quark mass, the contributions to tetraquark
binding from residual heavy-light interactions in
the corresponding two-meson threshold state can be
directly determined from the observed $B^*-B$, $B_s^*-B_s$,
$D^*-D$ and $D^*_s-D_s$ splittings, bearing in mind that
the correct two-meson threshold must be chosen. Thus,
for example, denoting by $V^\prime$ and $P^\prime$
the $\bar{b}^\prime\ell$ vector and pseudoscalar states,
one has that, for tetraquarks of the $ud\bar{b}\bar{b}^\prime$
type, the relevant threshold is $B^*P^\prime$ for 
$m_{b^\prime}<m_b$ but $BV^\prime$ for $m_{b^\prime}>m_b$.
We assume that the observed $1/m_h$ scaling of the 
bottom and charm vector-pseudoscalar splittings persists for
the variable-$b$-mass $V^\prime -P^\prime$ splittings.
The two-meson threshold contributions for a given
physical-to-variable $b$ quark mass ratio,
$r=m_b/m_{b^\prime}$, are then fixed by the observed 
charm/bottom meson splittings, and depend on 
$m_{h1}$ and $m_{h2}$ in a manner that varies
depending on the relation between these two masses. 
We use that the vector meson, $V^\prime$, and pseudoscalar 
meson, $P^\prime$, lie, respectively, 
${\frac{1}{4}} \left( m_{V^\prime}-m_{P^\prime}\right)$
above and ${\frac{3}{4}} \left( m_{V^\prime}-m_{P^\prime}\right)$
below the spin-average of the $V^\prime$ and $P^\prime$
masses.

Taking these expectations into account, and writing the results in
terms of the physical-to-variable mass ratio $r=m_b/m_{b^\prime}$,
one expects to obtain a good-quality fit to the binding energies 
of tetraquarks with at least one unphysical variable-mass antiquark
using an expression having the form 
\begin{eqnarray}
\Delta E &&={\frac{C_0}{2r}}\, +\, C_1^{ud} \, +\, 
C_2^{ud}\left( 2r\right)\, +\, 23\ {\rm MeV}\, r
\label{udbpbpform}\end{eqnarray}
for the $ud \bar{b}^\prime\bar{b}^\prime$ case, where
the first term represents the Coulomb binding contribution,
the second the good-$ud$-diquark attraction, the third
the residual heavy-light interactions in the tetraquark
state and the fourth the two-meson threshold contribution.
The numerical value appearing in the fourth term follows
from the observed meson splittings. Similarly, for 
the $ud \bar{b}^\prime\bar{b}$ case, one expects the form
\begin{eqnarray}
\Delta E &&={\frac{C_0}{1+r}}\, +\, C_1^{ud} \, +\, 
C_2^{ud}\left( 1+r\right)
\, +\, \left( 34\ {\rm MeV} -11\ {\rm MeV}\, r\right)
\label{udbpbmbpgtmbform}\end{eqnarray}
to provide a good representation for $m_{b^\prime}>m_b$ and
the form
\begin{eqnarray}
\Delta E &&={\frac{C_0}{1+r}}\, +\, C_1^{ud} \, +\, 
C_2^{ud}\left( 1+r\right)
\, +\, \left( 34\ {\rm MeV}\, r -11\ {\rm MeV}\right)
\label{udbpbmbpltmbform}\end{eqnarray}
to provide a good representation for $m_{b^\prime}<m_b$. The corresponding expectations for the cases involving an $\ell s$,
rather than $ud$, good-diquark are
\begin{eqnarray}
\Delta E &&={\frac{C_0}{2r}}\, +\, C_1^{\ell s} \, +\, 
C_2^{\ell s}\left( 2r\right)\, +\, 24\ {\rm MeV}\, r
\label{lsbpbpform}\end{eqnarray}
for $\ell s\bar{b}^\prime \bar{b}^\prime$,
\begin{eqnarray}
\Delta E &&={\frac{C_0}{1+r}}\, +\, C_1^{\ell s} \, +\, 
C_2^{\ell s}\left( 1+r\right)
\, +\, \left( 34\ {\rm MeV} -12\ {\rm MeV} r\right)
\label{lsbpbmbpgtmbform}\end{eqnarray}
for $\ell s\bar{b}^\prime \bar{b}$ with $m_{b^\prime}>m_b$ and
\begin{eqnarray}
\Delta E &&={\frac{C_0}{1+r}}\, +\, C_1^{\ell s} \, +\, 
C_2^{\ell s}\left( 1+r\right)
\, +\, \left( 36\ {\rm MeV}\, r -11\ {\rm MeV}\right)
\label{lsbpbmbpltmbform}\end{eqnarray}
for $\ell s\bar{b}^\prime \bar{b}$ with $m_{b^\prime}<m_b$.

For the variable-$b$-mass study just described, 
details of the setup and determination of the resulting tetraquark binding energies may be found in Appendices A and B. 
Fig.~\ref{fig:massdep1} and Tab.~\ref{tab:app_bbbind} display these results for $m_{b^\prime}$
running from $6.29$ to $0.60$ times $m_b$. The restriction
to $m_{b^\prime}\geq 0.60 m_b$ is designed to ensure that
the $b^\prime$ masses considered are all sufficiently heavy
that the NRQCD approximation is reliable. The same $2\times2$ 
GEVPs used in Ref.~\cite{Francis:2016nmj} are employed for 
the $ud\bar{b}^\prime\bar{b}^\prime$ and 
$\ell s\bar{b}^\prime\bar{b}^\prime$ channels, while new
$3\times3$ GEVPs, described in more detail in the next
section, are used for $ud\bar{b}^\prime\bar{b}$
and $\ell s\bar{b}^\prime\bar{b}$. The results of
a fit to this data using the forms detailed above
are shown in Fig.~\ref{fig:massdep1} and Tab.~\ref{tab:heavy_results}. The success of this fit
in describing tetraquark binding energies over a 
wide range of variable heavy quark masses confirms that 
the physical picture underlying those fit forms 
successfully captures the main features responsible for
the binding observed in these systems.

\begin{figure}[ht!]
\centering
\includegraphics[width=0.80\textwidth]{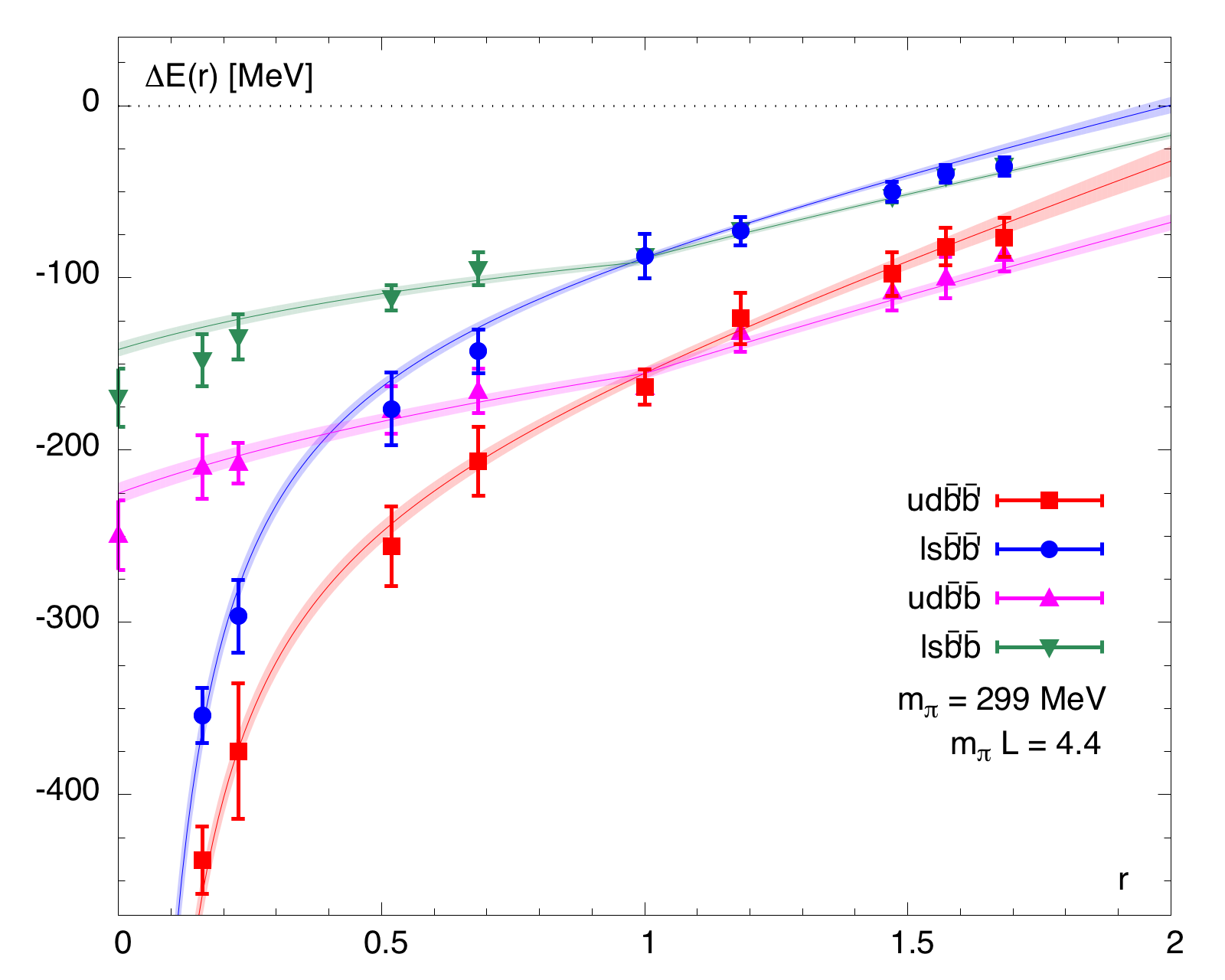}
\caption{The dependence on the heavy-quark mass ratio,
$r=m^b_{\rm bare}/m^{b'}_{\rm bare}$, of the binding energies for the $ud\bar b' \bar b$, $ud\bar b' \bar b'$, $\ell s b^\prime b$
and $\ell s b^\prime b^\prime$ channels.  
The results for each channel are separately fit to the phenomenologically motivated $\Delta E(m_Q)$ ansatze detailed in 
Eqs.~(\ref{udbpbpform}-\ref{lsbpbmbpltmbform})
of the text. 
}\label{fig:massdep1}
\end{figure}

While the use of NRQCD precludes extending the results of the
variable-$b$-mass study down to the charm mass, the pattern
of binding energies shown in Fig.~\ref{fig:massdep1} clearly
points to the 
\udcb channel as by far the
most likely among the four channels, 
\udcb, \lscb, \udcc, or \lscc,
in which one or both of the $\bar{b}$ antiquarks in 
$ud\bar{b}\bar{b}$ is replaced by $\bar{c}$, to support
a strong-interaction-stable bound state. A naive extrapolation
of the results for this channel, moreover,
produces a result very near $\bar{D}B^*$ threshold, strongly 
motivating direct \udcb 
simulations using a relativistic action for the charm quark. Similar naive
extrapolations of the variable-heavy-mass results suggest
none of the other three channels is likely to support a
strong-interaction-stable bound state. We thus focus,
in what follows, on the \udcb 
channel, leaving detailed simulations of the other channels for
a subsequent work. 

\begin{table}[ht!]
\centering
\begin{tabular}{c|cccc}
\toprule
& \multicolumn{4}{c}{$\Delta E$ [MeV]} \\
$m_{b'}/m_b$ & $ud\bar b' \bar b'$ & $ud\bar b' \bar b$ & $\ell s\bar b' \bar b'$ & $\ell s\bar b' \bar b$\\\hline
0.594 & -76(11) & -86(10) & -35(5) & -35(4) \\
0.636 & -82(11) & -100(12) & -39(5) & -41(3) \\
0.680 & -98(13) & -108(11) & -50(6) & -53(2) \\
0.846 & -123(15) & -131(12) & -73(8) & -72(3) \\
1.000 & -163(8)* & - & -94(9)* & - \\
1.463 & -206(20) & -166(13) & -143(13) & -95(10) \\
1.928 & -256(23) & -177(14) & -176(21) & -112(7) \\
4.935 & -375(39) & -208(12) & -296(21) & -134(13) \\
6.287 & -438(20) & -210(18) & -354(16) & -148(15)\\
$\infty$ & - & -249(30)  & - & -170(17)\\ \hline
\botrule
\end{tabular}
\caption{Table of binding energies determined in the non-relativistic regime. Note, the values denoted with ``*'' were calculated previously in \cite{Francis:2016nmj}.}
\label{tab:app_bbbind}
\end{table}

\begin{table}[ht!]
\begin{tabular}{ccccc}
\toprule
$C_0 $ & $C_1^{ud}$ & $C_2^{ud}$ & $C_1^{\ell s}$ & $C_2^{\ell s}$\\
\hline
{-82(6)}& {-217(14)} & {40(5)} & {-116(10)} & {22(3)} \\ 
\botrule
\end{tabular}
\caption{Fit results for the constants parametrizing the
heavy-quark mass dependence of tetraquark binding energies
on the ensemble $E_M$ with pion mass $m_\pi=299$ MeV.}\label{tab:heavy_results}
\end{table}

\section{Lattice correlators and operators}\label{sec:ops}

The generic form of a lattice QCD correlation function for a particle at rest, i.e. ${\bf p}=0$, in Euclidean time is given by
\begin{equation}
\begin{aligned}
C_{\mathcal{O}_1 \mathcal{O}_2}(t) &= \sum_{{\bf x}} \Big\langle \mathcal{O}_1({\bf x},t) \mathcal{O}_2({\bf 0},0)^\dagger   \Big\rangle \\
&=\sum_n \frac{\langle 0 | \mathcal{O}_1 | n \rangle\langle n| \mathcal{O}_2 | 0 \rangle}{2E_n} \,e^{-E_n t}~,
\end{aligned}
\end{equation}
with the interpolating operators $\mathcal{O}_i$ being chosen to have the quantum numbers of the continuum state to be studied. For example, the simplest local meson operator is
\begin{equation}\label{eq:mes_op}
\mathcal{O}_m(x) = \bar{q}^\alpha_a(x) \Gamma^{\alpha\beta} q'^{\beta}_a(x)~,
\end{equation}
where upper (Greek) indices denote Dirac spin and lower (Roman) color. The $q$ and $q'$ represent the constituent quark flavors.

In the case of the $\bar{3}_F$, $J^P=1^+$
tetraquarks, the relevant free-streaming two-meson thresholds are
given by the sums of the lowest-lying pseudoscalar (P) $\Gamma=\gamma_5$ and vector (V) $\Gamma=\gamma_i$ meson masses. 
Tab.~\ref{tab:mes_th} provides a list of these thresholds
for the cases of interest here.

\begin{table}[ht!]
\begin{tabular}{c|c|c|c}
\toprule
Tetraquark ~~ & ~~Threshold~~ & $\mathcal{O}_m=P(x)$ & $\mathcal{O}_m=V(x)$\\ 
\hline
\udbb & $BB^*$ & ~~ $\bar{b}^\alpha_a(x) \gamma_5^{\alpha\beta}u^{\beta}_a(x)$~~ & ~~$\bar{b}^\alpha_a(x) \gamma_i^{\alpha\beta} d^{\beta}_a(x)$~~\\
\lsbb & $B_sB^*$ & ~~$\bar{b}^\alpha_a(x) \gamma_5^{\alpha\beta}s^{\beta}_a(x)$~~ & ~~$\bar{b}^\alpha_a(x) \gamma_i^{\alpha\beta} d^{\beta}_a(x)$~~\\
\udcb & $\bar{D}B^*$ & ~~$\bar{c}^\alpha_a(x) \gamma_5^{\alpha\beta}u^{\beta}_a(x)$~~ & ~~ $\bar{b}^\alpha_a(x) \gamma_i^{\alpha\beta} d^{\beta}_a(x)$~~ \\
\botrule
\end{tabular}
\caption{Lowest two-meson thresholds for each of the flavor antisymmetric, $J^P=1^+$ tetraquark channels.}\label{tab:mes_th}
\end{table}

Given the phenomenological picture of the previous sections, a natural first choice of interpolating operator for 
$qq^\prime \bar{Q}\bar{Q}^\prime$-type tetraquarks 
is one with a diquark-anti-diquark structure. Using the epsilon identity $\epsilon_{abc}\epsilon_{dec}=\delta_{ad}\delta_{be}-\delta_{ae}\delta_{bd}$, the
local version of this operator takes the form
\begin{equation}\label{eq:di_antidi}
\begin{aligned}
D(x) = \Big(\; ( &q^\alpha_a(x) )^T ( C\gamma_5 )^{\alpha\beta} q'^{\beta}_b(x) \;\Big)\times \\
&\left[ \bar{Q}^\kappa_a(x) ( C\gamma_i )^{\kappa\rho} ( \bar{Q'}^\rho_b(x) )^T - \bar{Q}^\kappa_b(x) ( C\gamma_i )^{\kappa\rho} ( \bar{Q'}^\rho_a(x) )^T\right],
\end{aligned}
\end{equation}
where $C=i\gamma_y\gamma_t$ is the charge-conjugation matrix. This operator 
has $\bar{Q}\bar{Q'}$ color $3_c$, spin $1$ and light-quark 
flavor-spin-color $(\bar{3}_F,0,\bar{3}_c)$.

A second possible local operator is one whose discrete structure is meson-meson-like.
For tetraquark channels with $Q=Q'$ this could be:
\begin{equation}\label{eq:dimeson}
\begin{aligned}
M(x) = \, \Big(\bar{Q}^\alpha_a(x) \gamma_5^{\alpha\beta} q^{\beta}_a(x) \Big) 
\Big(\bar{Q}_b^{\kappa}(x) \gamma_i^{\kappa\rho} {q'}_b^{\rho}(x)\Big)
\, - \Big(\bar{Q}^\alpha_a(x) \gamma_5^{\alpha\beta} {q'}^{\beta}_a(x)\Big)
\Big(\bar{Q}_b^{\kappa}(x) \gamma_i^{\kappa\rho} q_b^{\rho}(x)\Big).
\end{aligned}
\end{equation}

When $Q\neq Q'$, a second local flavor antisymmetric, $J^P=1^+$ meson-meson-like combination can also be constructed. Suppressing the spin indices, 
the two possible ``meson-meson'' interpolating operators in this case are:
\begin{equation}
\begin{aligned}
M_1(x) &= ( \bar{Q}_a \gamma_5 q_a )( \bar{Q'}_b \gamma_i q'_b )- ( \bar{Q}_a \gamma_5 q'_a )( \bar{Q'}_b \gamma_i q_b )~,\\
M_2(x) &= ( \bar{Q'}_a \gamma_5 q_a )( \bar{Q}_b \gamma_i q'_b )- ( \bar{Q'}_a \gamma_5 q'_a )( \bar{Q}_b \gamma_i q_b )~.
\end{aligned}
\end{equation}

With these interpolating operators, there are several options to study the ground state energies of the proposed tetraquarks channels. One is to form the 
so-called binding correlator, the ratio of 
tetraquark correlation functions to the product of correlation functions,  $C_{PP}(t)$ and $C_{VV}(t)$, of the pseudoscalar and vector mesons making up the 
corresponding non-interacting two-meson threshold in the channel in question:
\begin{equation}\label{eq:bcorr}
G_{\mathcal{O}_1 \mathcal{O}_2}(t) = \frac{C_{\mathcal{O}_1 \mathcal{O}_2}(t)}{C_{PP}(t) C_{VV}(t)}~.
\end{equation}
This definition is beneficial in the present study as the additive mass renormalisation for NRQCD quarks explicitly cancels. 

The binding correlator behaves, for large $t$, as $\sim e^{-\Delta E t}$, with $\Delta E=m_0-m_V-m_P$, where $m_0$ is the ground
state mass in the channel. Thus, observing an exponentially-increasing binding correlator signals the existence
of a ground state lighter than the corresponding two-meson threshold and provides compelling evidence for a bound tetraquark state.

In the binding correlator one might also gain signal through cancellations between the two-meson and tetraquark fluctuations. At the same time, 
however, forming the binding correlator may introduce a difficult-to-control systematic through contamination of the tetraquark signal with residual 
excited state effects originating from the two-meson correlators. The potential problem is that ground state saturation might occur at different lattice 
times for each of the three different correlation functions entering the binding correlator. The binding correlator plateau will then depend on the 
slowest-plateauing of the three constituent correlators.

To handle (and quantify, if present) this effect, we also compute the individual correlation functions for the tetraquark candidates and 
two-meson threshold combination(s). This is especially important, since we expect 
a significantly smaller binding energy for \udcb tetraquarks than
was observed for the $ud\bar{b}\bar{b}$ channel. An unambiguous
determination of the ground state energies is thus essential.
To combine the two mesons we compute the product $C_{PP}(t) C_{VV}(t)$.
The resulting mass from a single-exponential fit to this combination for the case of the $\bar{D}B^*$ threshold is given in Tab.~\ref{tab:lat_params}. 
Note, however, that in this case the NRQCD additive mass renormalisation does not drop out and the results are given in lattice units with this shift 
included.

\subsection{Variational analysis}

With access to several operators with the same quantum numbers, a
variational analysis can be used to determine the ground and 
excited state energies. In the case $\bar Q=\bar Q'$, this analysis
involves the $2\times 2$ matrix
\begin{equation}
F(t) = 
\begin{pmatrix}
G_{DD}(t) & G_{DM}(t)  \\
G_{MD}(t) & G_{MM}(t) 
\end{pmatrix}.
\label{eq:gmatrix}
\end{equation}

When $\bar Q\neq \bar Q'$, there are now two meson-meson interpolating operators, allowing use of the $3\times 3$ matrix
\begin{equation}
F(t) = 
\begin{pmatrix}
G_{DD}(t) & G_{DM_1}(t) & G_{DM_2}(t)\\
G_{M_1D}(t) & G_{M_1M_1}(t) & G_{M_1M_2}(t)\\
G_{M_2D}(t) & G_{M_2M_1}(t) & G_{M_2M_2}(t)\\
\end{pmatrix}.
\label{eq:3x3matrix}
\end{equation}

Given the matrix $F(t)$, one can solve the generalised eigenvalue problem (GEVP) for some reference time $t_0$\footnote{We monitor $t_0$ to make sure our ground state mass evaluation is stable. We find $t_0/a=4$ to be a reasonable choice.},
\begin{equation}
F(t)\nu = \lambda_i(t) F(t_0) \nu\;,
\label{eq:gevp}
\end{equation}
where $\nu$ are the (generalised) eigenvectors and $\lambda_i(t)$
the eigenvalues. The solution of the GEVP 
gives independent eigenvalues corresponding to different states in the system,
\begin{equation}\label{eq:eigval}
\lambda_i(t) =  A_i e^{-\Delta E_i( t - t_0 ) }.
\end{equation}
In this work the aim will be to determine the ground state of the system ($\lambda_1$). As such, a $2\times2$ or $3\times3$ matrix should suffice, so long as the chosen operators have good overlap with the desired ground state.

\section{Numerical setup}\label{sec:numerical_setup}

\begin{table}[hbt!]
\centering
\begin{tabular}{c|ccc}
\toprule
& \multicolumn{3}{c}{Ensemble parameters} \\
Label & $E_H$ & $E_M$& $E_L$ \\ \hline
Extent & $\:32^3\times64\:$ & $\:32^3\times64\:$ & $\:32^3\times64\:$ \\
$a^{-1}\;\left[\text{GeV}\right]$\cite{Namekawa:2013vu} & 2.194(10) & 2.194(10) & 2.194(10) \\
$\kappa_l$ & $0.13754$ & $0.13770$ & $0.13781$ \\
$\kappa_s$ & $0.13640$ & $0.13640$ & $0.13640$ \\
$am_\pi$ & 0.18928(36) & 0.13618(46) & 0.07459(54) \\
$m_\pi L$ & 6.1 & 4.4 & 2.4 \\
$aE_{\bar{D}B^*}$ & 1.3588(17) & 1.3367(9) & 1.3095(12) \\
$M_{J/\Psi}\;\left[\text{GeV}\right]$ & 3.0862(2) & 3.0847(2) & 3.0685(11)\\
$M_{\Upsilon}\;\left[\text{GeV}\right]$ & 9.528(79) & 9.488(71) & 9.443(76)\\
Configurations & 400 & 800 & 195 \\
Measurements & 4000 & 6400 & 9360 \\
\botrule
\end{tabular}
\caption{Overview of our ensemble parameters (see also \cite{Hudspith:2017bbh,Francis:2016nmj,Francis:2016hui}). These configurations \cite{Aoki:2008sm} use the Iwasaki gauge action with $\beta=1.9$ and non-perturbatively tuned clover coefficient $c_{SW}=1.715$. The $\bar{D}B^*$ threshold is extracted from a single-exponential fit to the product of the two relevant $\bar{D}$- and $B^*$-meson correlators. Due to NRQCD's additive mass renormalisation, these values have not been converted into physical units. 
}\label{tab:lat_params}
\end{table}

Throughout this work the calculations are performed on three $n_f=2+1$, clover-improved \cite{Sheikholeslami:1985ij}, Iwasaki gauge \cite{Iwasaki:1985we}, PACS-CS ensembles introduced in \cite{Aoki:2008sm} and which we label by $E_H,E_M$, and $E_L$.
The lattice spacing is \cite{Namekawa:2013vu} $a^{-1}\,$=2.194(10) GeV for all three ensembles.
We use a partially-quenched strange quark tuned to the (connected) $\phi$-meson \cite{Lang:2014yfa}, which gives a near-physical kaon mass in the chiral limit. The labeling (and pion masses) of the ensembles are consistent with those used in our previous work \cite{Francis:2016hui,Francis:2016nmj,Hudspith:2017bbh}.
In the valence sector we use Coulomb gauge-fixed wall sources using the FACG algorithm \cite{Hudspith:2014oja}, as before. 
We put sources at multiple time positions and compute propagators for light, strange and charm
quarks using a modified deflated SAP-solver \cite{Luscher:2005rx}. 
An overview of the lattice parameters and ensemble properties can be found in Tab.~\ref{tab:lat_params}. 

To reliably handle charm quarks on these lattices we employ a relativistic heavy quark (RHQ) action \cite{ElKhadra:1996mp,Aoki:2001ra,Christ:2006us,Oktay:2008ex}, in particular, the ``Tsukuba'' formulation \cite{Aoki:2001ra}: 
\begin{equation}\label{eq:tsukuba}
\begin{aligned}
D_{x,y} =~ &\delta_{xy} - \kappa_f \Big[ (1-\gamma_t)U_{x,t}\delta_{x+\hat t,y} + (1+\gamma_t)U_{x,t}\delta_{x+\hat t,y}  \Big] \\
&-\kappa_f \sum_i \Big[ (r_s-\nu_s\gamma_i)U_{x,t}\delta_{x+\hat i,y} + (r_s+\nu_s\gamma_i)U_{x,t}\delta_{x+\hat i,y} \Big] \\
&-\kappa_f  \Big[ c_E \sum_{i} F_{it}(x)\sigma_{it} + c_B  \sum_{i,j} F_{ij}(x)\sigma_{ij}  \Big]~~.
\end{aligned}
\end{equation}
The common approach of RHQ actions is to re-interpret the discretisation effects and to re-tune the fully relativistic lattice action by introducing anisotropy in the valence sector to reproduce the correct dispersion relation, i.e., the equivalence between rest and kinetic mass, and physical spectrum.
For our RHQ action, the tuning parameters have been previously computed for the ensembles studied here. The values of the
parameters $\kappa_f,r_s,\nu_s,c_E$ and $c_B$ can be found in \cite{Namekawa:2011wt}.
Meson masses using quark propagators computed from our implementation of this action are seen to be within $\sim 1\%$ of the experimentally observed values, with splittings between, for example, $D$ and $D^*$ mesons equally well-behaved.

For bottom quarks, as noted above, we employ the NRQCD action, 
as detailed in Appendix A.
Overall, the NRQCD action is known to capture the relevant 
heavy-light quark physics and account for relativistic effects 
at the few percent level \cite{Lewis:2008fu,Gray:2005ur,Brown:2014ena}.

A list of the bare quark masses used in this study and the ratios of kinetic masses compared to the physical bottom quark is provided 
in Tab.~\ref{tab:heavymasses}.
\begin{table}[h!]
\centering
\begin{tabular}{ccc}
\toprule
~~$am_Q^\prime$~~ & ~~$m_Q^\prime/m_Q$~~ & ~~$m^{b'}/m^{b}$~~ \\\hline
0.9 & 0.466 & 0.594(3) \\
1.0 & 0.518 & 0.636(2) \\
1.2 & 0.622 & 0.680(5) \\
1.6 & 0.829 & 0.846(7) \\
3.0 & 1.554 & 1.463(12) \\
4.0 & 2.073 & 1.928(17) \\
8.0 & 4.145 & 4.395(35) \\
10.0 & 5.181 & 6.287(48) \\
\botrule
\end{tabular}
\caption{Unphysical heavy quark masses used in our variable-mass
heavy anti-diquark investigation. Values are 
given divided by the physical bottom quark mass. 
For the definition of the entries in
the two columns see Appendix A. The calculations were 
performed using on the ensemble $E_M$ with one source position.}
\label{tab:heavymasses}
\end{table}

\section{Indications of a bound \texorpdfstring{$ud\bar c\bar b$}{udcb} tetraquark in nature}\label{sec:udcb}

The successful phenomenological description of the
heavy-quark mass dependence of $J^P=1^+$ tetraquark 
binding energies described earlier identifies the 
$ud\bar{c}\bar{b}$ channel as the only such channel
containing at least one charm antiquark likely
to support a strong-interaction-stable bound tetraquark state.
Given this insight, in this section, we focus our attention 
and resources on this channel and present 
results for the direct calculation of the 
mass of an $I(J^P)=0(1^+)$ \udcb tetraquark. 

\begin{figure}
\centering
\subfloat
{\includegraphics[width=0.48\textwidth]{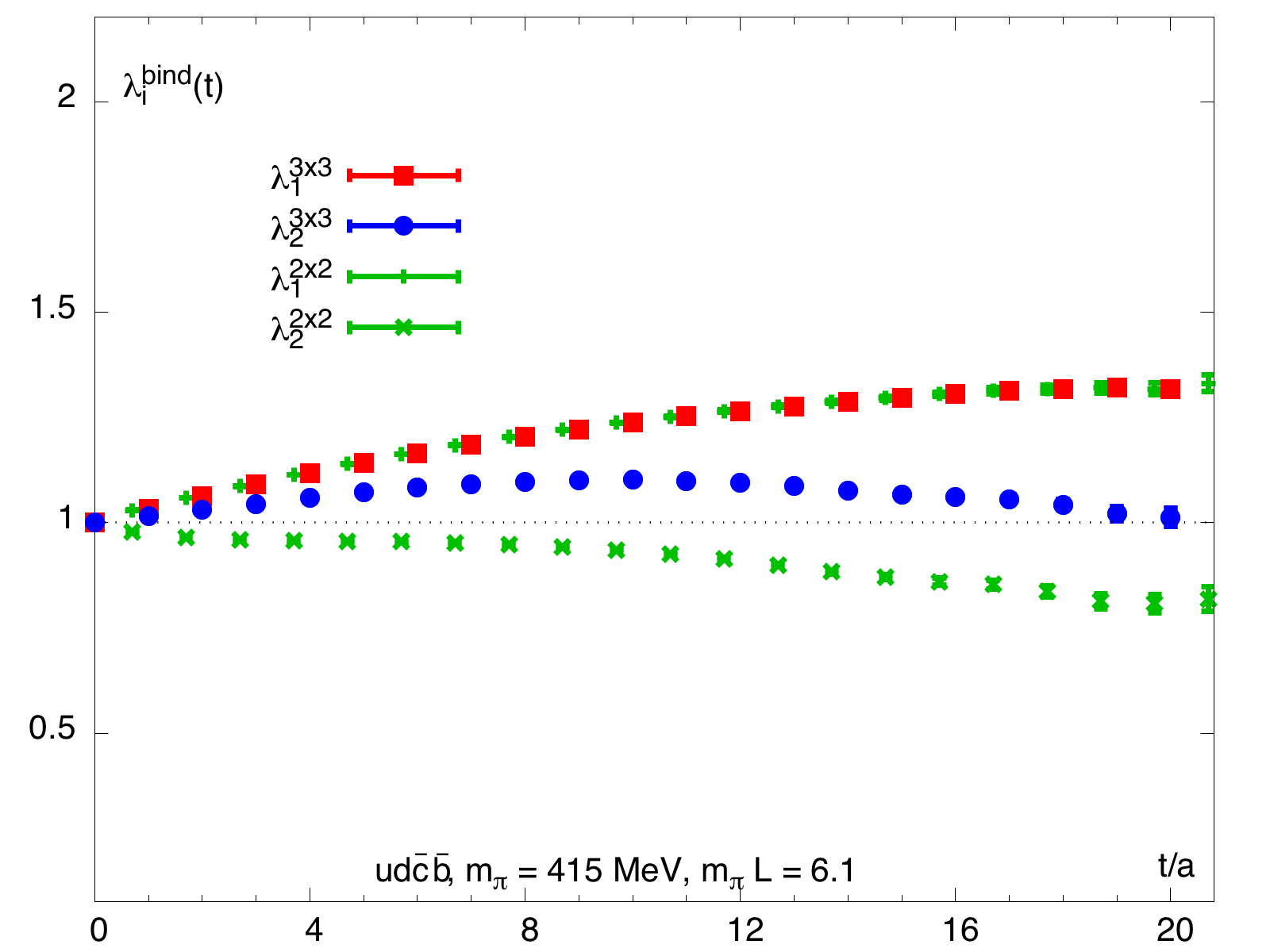}
}
\subfloat
{\includegraphics[width=0.48\textwidth]{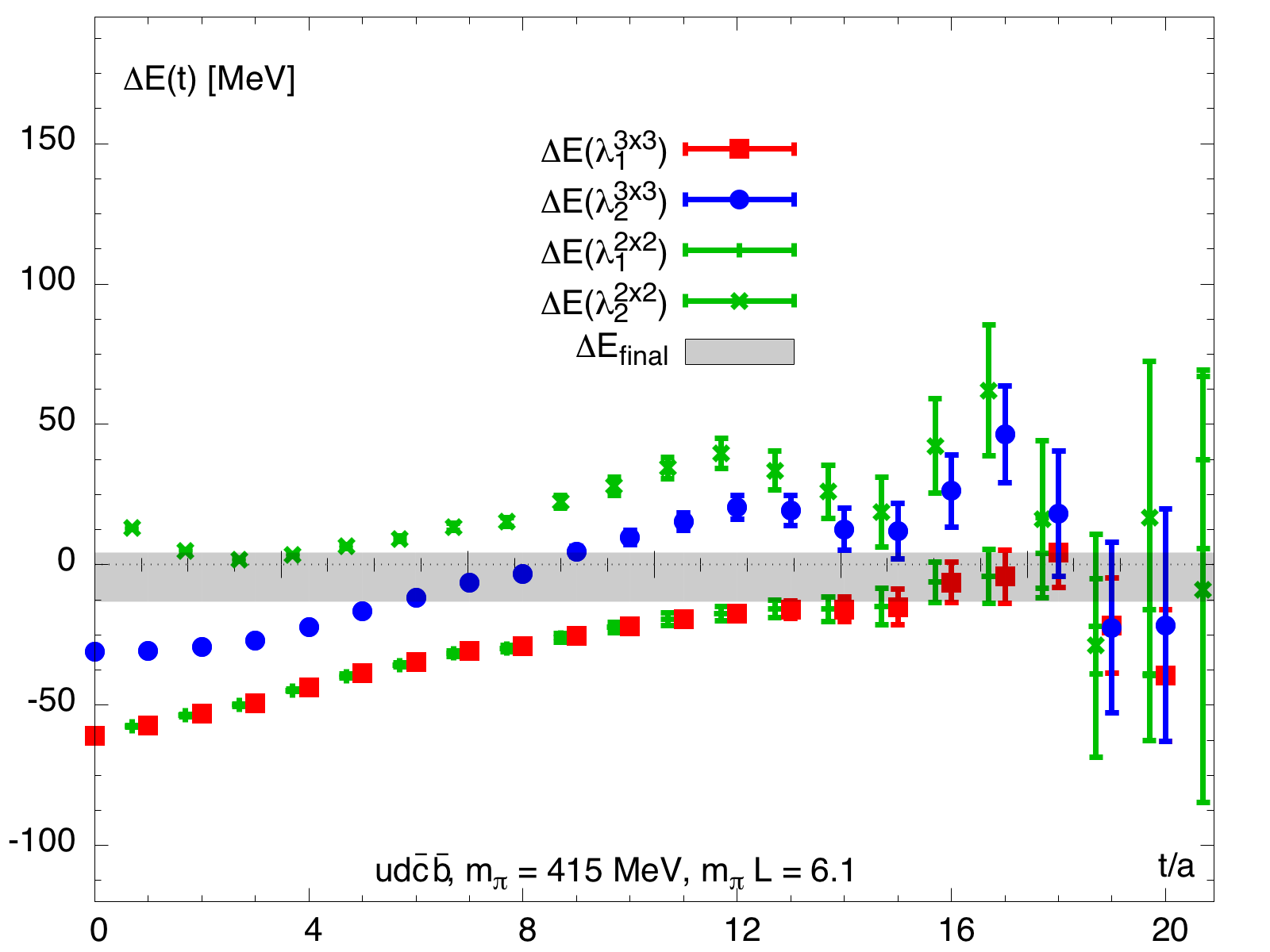}
}
\vspace{2pt}
\subfloat
{
\includegraphics[width=0.48\textwidth]{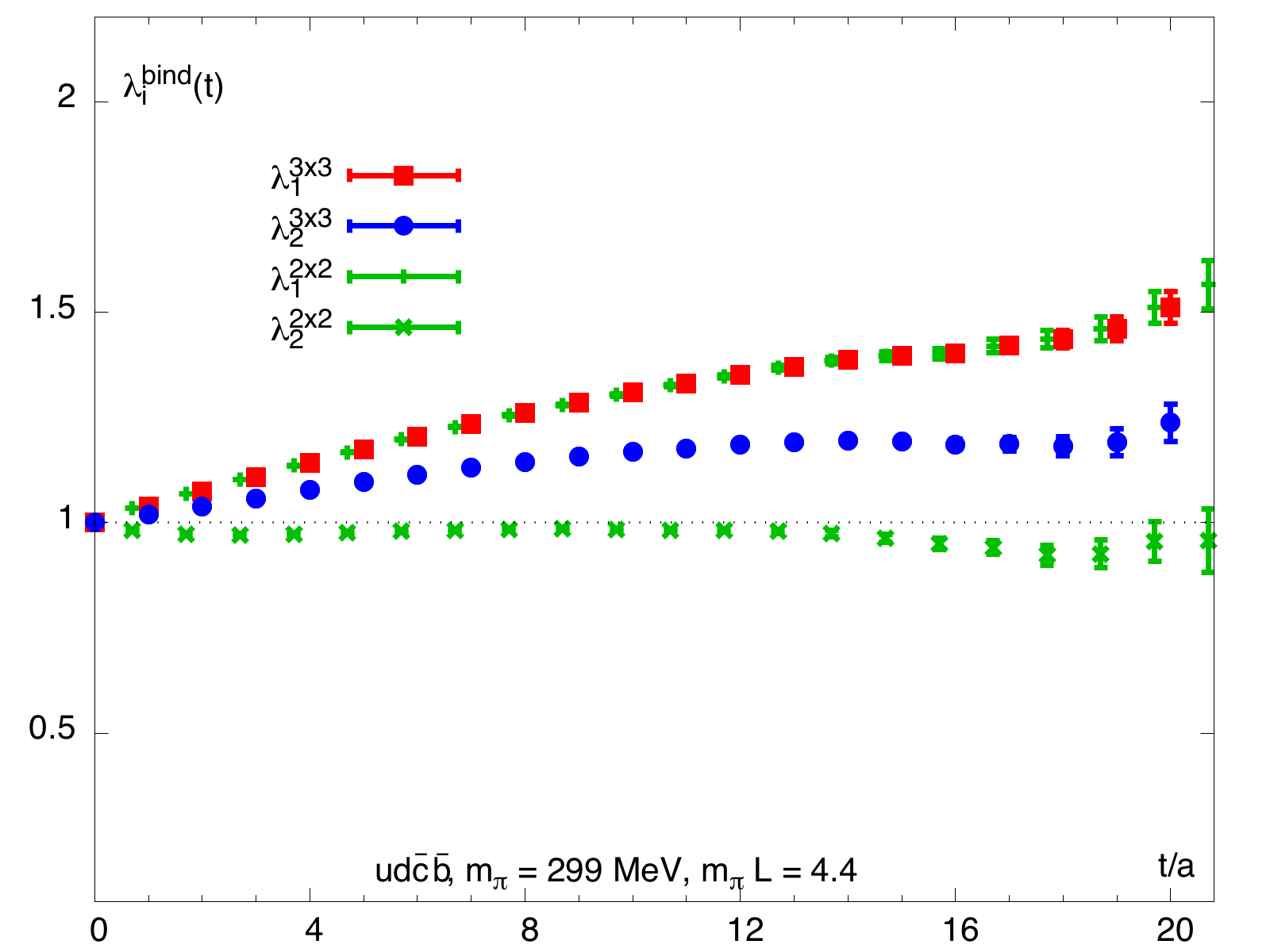}
}
\subfloat
{
\includegraphics[width=0.48\textwidth]{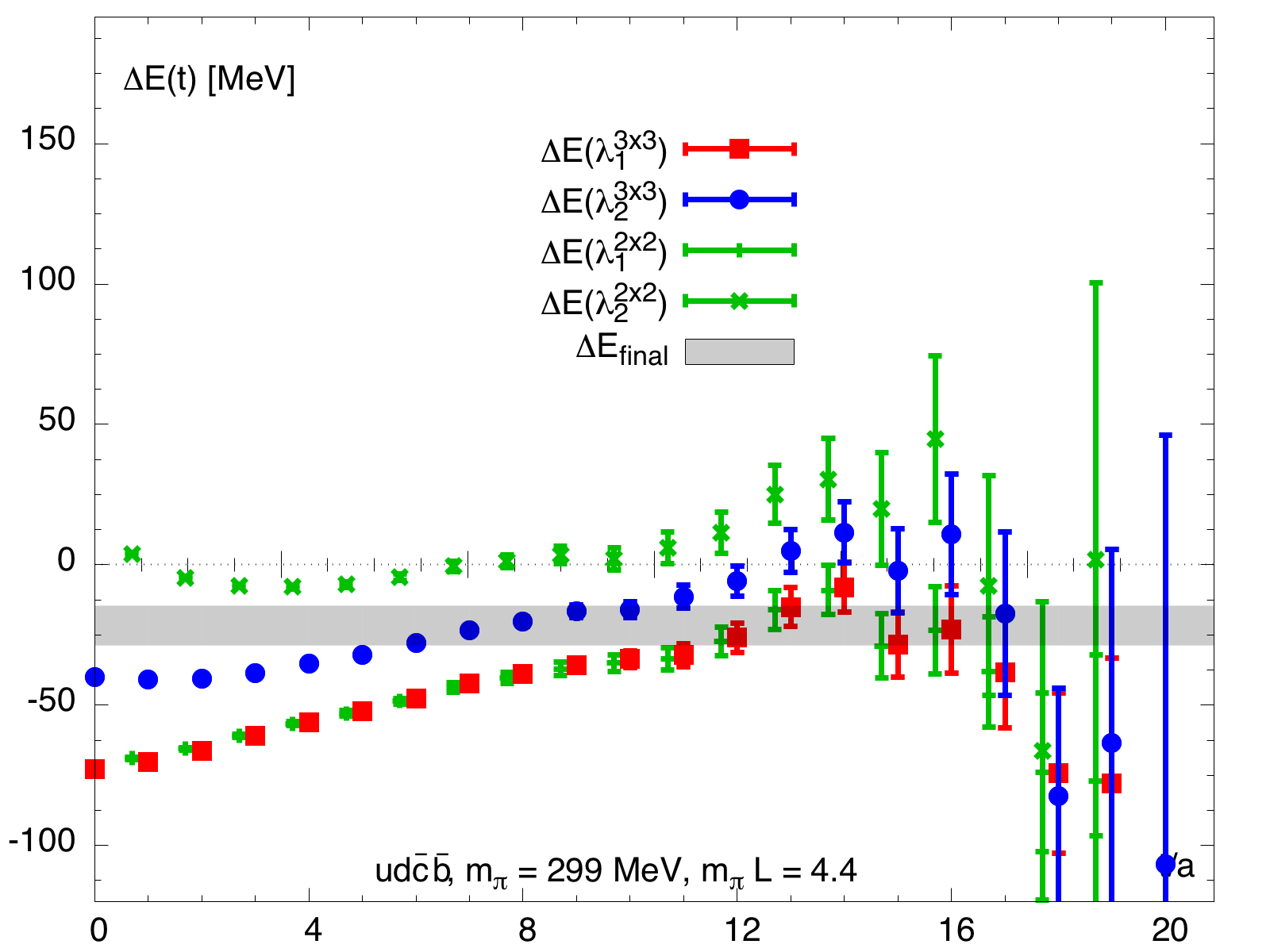}
}
\vspace{2pt}
\subfloat
{
\includegraphics[width=0.48\textwidth]{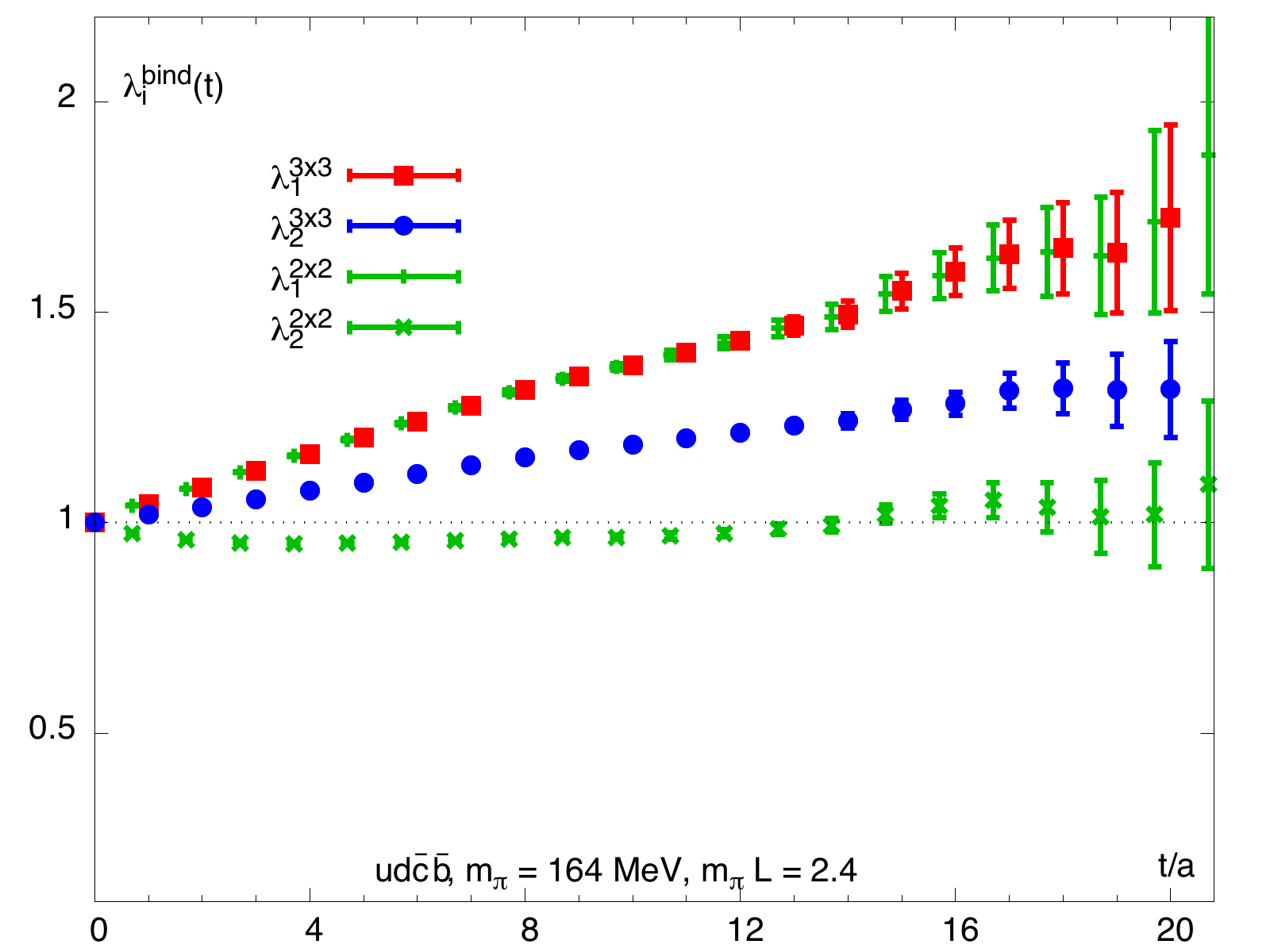}
}
\subfloat
{\includegraphics[width=0.48\textwidth]{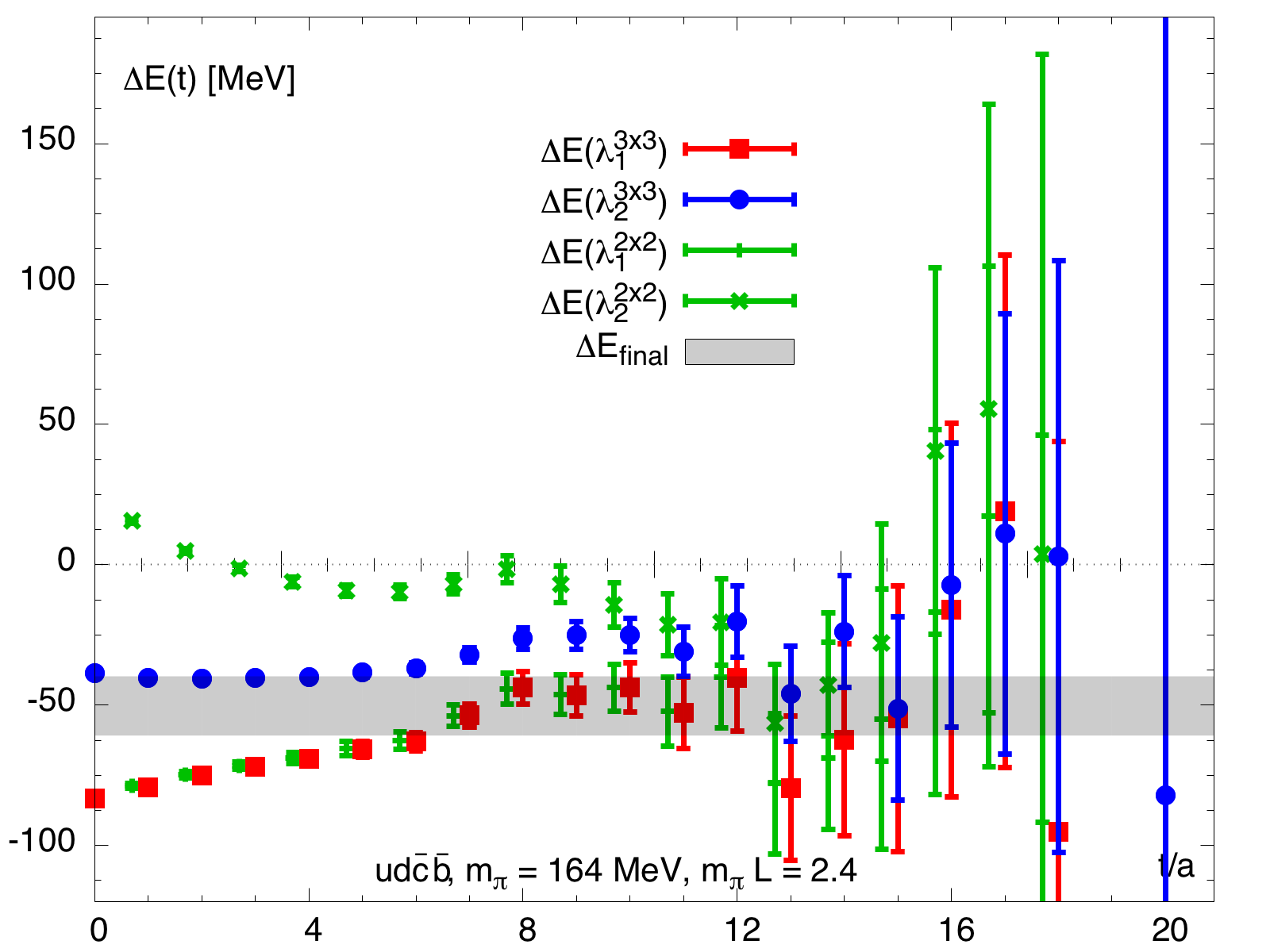}
}
\caption{$ud\bar{c}\bar{b}$ tetraquark results for binding
correlator eigenvalues (left panels) and binding energies
(right panels) on $E_H$ (top), $E_M$ (center) and $E_L$ (bottom).
Red squares and blue circles denote $3\times 3$ GEVP ground 
and first excited state results, respectively, green vertical
dashes and green diagonal crosses ground and first excited state $2\times 2$ GEVP results. The grey bands depict the final binding energies, derived from single-exponential fits to the first eigenvalues. Further details may be found in the text, and supplementary material in 
App.~\ref{sec:app_fits}. $2\times 2$ GEVP results are offset 
slightly in $t$ for visual clarity.}
\label{fig:udcb}
\end{figure}

Our lattice results for the ground (red) and first excited (blue) state binding energies, obtained from the $3\times 3$ GEVPs, 
are shown in Fig.~\ref{fig:udcb}.
Results obtained using the corresponding $2\times 2$ GEVPs,
in which the local operator with $\bar{D}^*B$ discrete
structure has been omitted, are shown for comparison in green, 
offset slightly in $t$ for presentational clarity.

Results shown in the left-hand panels are those obtained using 
the binding correlators Eq.~(\ref{eq:bcorr}). 
For the $E_M$ and $E_L$ ensembles, one observes a 
clear rising exponential behavior in both the
$3\times 3$ and $2\times 2$ GEVP results, indicating the 
presence of a bound \udcb tetraquark ground
state for these ensembles.

The right-hand panels of Fig.~\ref{fig:udcb}, in contrast, focus 
on the binding energies themselves, which are shown in
log-effective form, $E_{\rm eff}(t)$, and given in physical units. 
The choice of log-effective form is for presentational
purposes only; our final binding energies are obtained
through fits to the eigenvalues shown in the left-hand panels. 
The right-hand panels also provide a direct comparison of the 
eigenvalue-fit results (shown by the shaded bands) with 
the corresponding log-effective-energies. 

The data allow the first two eigenvalues to be clearly resolved 
for each ensemble, with a signal that can be followed to 
$t/a \sim 16$. The overlaps with the chosen multi-quark 
interpolating operators therefore seem good. Log-effective
energies of the states corresponding to these eigenvalues
are shown in the right panels of Fig.~\ref{fig:udcb2} in 
Appendix C. These results confirm that the signal is strong 
enough to reach the plateau regions of $aE_{\rm eff}(t)$. In 
the case of the binding correlator, the plateaus are in 
general reached later and are therefore shorter since the 
point of signal deterioration is the same in both approaches. 
This is a sign of unwanted contamination from excited states 
in the meson denominator of the binding correlator ratio, as 
discussed in Sec.~\ref{sec:ops}.

In all cases the third eigenvalue of the $3\times 3$ GEVP (not
shown here) is found to lie significantly higher than 
the first two, a fact confirmed by the eigenvalue plots in 
the left-hand panels of Fig.~\ref{fig:udcb2} in Appendix C.

Note that the results for the ground state eigenvalues obtained from the smaller $2\times 2$ GEVP defined in Eq.~(\ref{eq:gmatrix}), 
shown by the green vertical dashes in Fig.~\ref{fig:udcb}, agree 
very well with those obtained from the corresponding $3\times 3$ 
GEVP. In contrast, the $2\times 2$ results for the second eigenvalues, denoted by the green diagonal crosses, lie lower
than those of the $3\times 3$ analysis, and hence correspond
to higher energies. This is as expected for analyses
employing interpolating fields having a good overlap with 
the ground state, where both the $2\times 2$ and $3\times 3$ 
analyses should provide good access to the ground state energy, 
but the $2\times 2$ analysis, where the second eigenvalue 
will have to represent, not just the effects of the actual 
first excited state, but also those of all higher excited 
states, should yield results corresponding to higher energies
than those obtained from the $3\times 3$ analysis.

For the heavy pion ensemble, $E_H$, with 
$m_\pi=415\text{ MeV}$, the ground state 
plateaus at the $\bar{D}B^*$ threshold, whereas already
for $E_M$, with $m_\pi=299\text{ MeV}$, the second eigenvalue is 
consistent with the threshold and the first eigenvalue with a 
bound tetraquark interpretation. Finally, for $E_L$, with 
$m_\pi=164\text{ MeV}$, both the first and second eigenvalues 
have central values corresponding to states below threshold. 
In light of the error bars, we suspect that, in this case, 
the second eigenvalue corresponds to a scattering state affected 
by the finite lattice volume rather than a genuine bound state. 
A more detailed investigation of this question must, however, 
await a future study involving sufficiently high-statistics
ensembles with larger volumes for near-physical $m_\pi$.

To estimate the binding energies we have performed (uncorrelated) 
single-exponential fits Eq.~(\ref{eq:eigval}) to the individual
eigenvalues and accepted results satisfying $\chi^2/d.o.f. 
\lesssim 1$. We perform fits to the eigenvalues of both 
the tetraquark correlators and binding correlators. In the 
former case, we subtract the sum of the threshold two-meson 
state masses to obtain the binding energy. All accepted 
results are compared and a result that is representative of both 
procedures is chosen. The final fit ranges are $t/a\in[10:23]$ 
for $E_L$, $t/a\in[14:20]$ for $E_M$, and $t/a\in[17:21]$ 
for $E_H$. Further details of this procedure, as well as the 
resulting fit stabilities, are given in Appendix~\ref{sec:app_fits}.
The final results obtained in this manner are listed in
Tab.~\ref{tab:Eudcb} and plotted as the grey shaded bands 
in the right-hand panels of Fig.~\ref{fig:udcb}.

\begin{table}
\centering
\begin{tabular}{c|ccc}
\toprule
Ensemble ~~ & $m_\pi[\rm{MeV}]$ & $m_\pi L$ & ~~$\Delta E_{ud\bar c \bar b}$[MeV] \\
\noalign{\smallskip}\hline\noalign{\smallskip}
$E_H$ & 415 & 6.1 & -4(9)  \\
$E_M$ & 299 & 4.4 & -22(7)  \\
$E_L$ & 164 & 2.4 & -50(11) \\ 
\botrule
\end{tabular}
\caption{
Final estimated $ud\bar{c}\bar{b}$ binding energies. A negative
value signals binding with respect to $\bar{D}B^*$ threshold.}
\label{tab:Eudcb}      
\end{table}

%

Although further control of the dominant systematic uncertainties
is necessary before an accurate prediction can be made, the 
results of our direct simulation provide good evidence for the existence of a strong-interaction-stable 
$J^P=1^+$ $ud\bar{c}\bar{b}$ tetraquark 
state at physical $m_\pi$. We take the upper bound of the $E_M$ 
result and lower bound of the $E_L$ result 
as providing the best assessment of the
likely range of binding, leading to the expectation  
\begin{equation}
-61 \text{ MeV}<\Delta E_{ud\bar{c}\bar{b}} < - 15 \text{ MeV}
\label{udbcberange}\end{equation}
for the binding energy of the $ud\bar{c}\bar{b}$ tetraquark
ground state relative to the $\bar{D}B^*$ threshold.
For presentational simplicity in what follows, we will also
quote this result in the equivalent compact form 
$\Delta E_{ud\bar c\bar b}= -38(23) \text{ MeV}$, using the 
midpoint of the range in Eq.~(\ref{udbcberange}) as the central value, 
and half the width as the error estimate. This strategy for 
estimating the binding energy implied by our results appears 
plausible, given the ensembles currently available to us; 
since we expect both some deepening of the binding in going 
from $E_M$ to $E_L$ (as a consequence of the increasing 
spin-dependent, good-light-diquark attraction with decreasing
light quark mass) but also potentially non-negligible 
finite-volume effects due to the small $m_\pi L$ of 
the $E_L$ ensemble. It thus seems reasonable to expect the 
true, infinite volume binding for physical $m_\pi$ to lie 
somewhere between the two bounds noted above. Our finite-volume 
systematic uncertainty, at present, clearly dwarfs the
statistical errors on the individual measurements and only a careful finite-volume study will allow for a more accurate prediction. We are currently in the process of generating
ensembles with larger volumes at near-physical $m_\pi$,
and, once this is completed, we will report on the results 
obtained using these ensembles to more fully quantify the 
impact of finite-volume effects on the estimated
physical-$m_\pi$ binding in a future work.

Taking PDG \cite{Patrignani:2016xqp} values for the physical 
$D$ and $B^*$ masses,\footnote{As the ensembles we use have identical
$u$ and $d$ quark masses, and hence exact isospin symmetry, 
we use the average of the $D^+$ and $D^0$ masses for $m_D$.} 
the binding energy estimate obtained above corresponds to 
a tetraquark mass of $\approx 7154(23)$ MeV.

%
For completeness, we quote here also the energy differences 
between the first excited state (corresponding to the second 
eigenvalue) and $\bar{D}B^*$ 
threshold, which are $18(6)~{\rm MeV}$ 
for $E_H$, $8(8)~{\rm MeV}$ for $E_M$, and $-26(7)~{\rm MeV}$ 
for $E_L$. These are compatible with our interpretation of the
second eigenvalue as corresponding to the 
$\bar{D}B^*$ threshold for 
the $E_M$ and $E_L$ ensembles, with potentially non-negligible
finite-volume effects present in the $E_L$ case. One should,
however, also bear in mind that the limited size of our basis 
of operators may impact how reliably we can extract the energy 
of the first excited state, particularly if this state 
corresponds to an infinite-volume scattering state and, as 
is likely, finite-volume effects are not negligible for 
the $E_L$ ensemble. A larger basis of operators and 
finite-volume analysis are thus desirable for a more robust 
study of the nature of this state.

\subsection{Discussion of systematic uncertainties}

In this calculation, the systematic error originating from the uncertainty of the lattice spacing in Tab.~\ref{tab:lat_params} is negligible in comparison to the statistical error of the final result.

As noted above, we believe the dominant source of uncertainty 
in our result comes from finite-volume effects. The 
light-quark-mass dependence of the \udcb masses covers 
multiple $m_\pi L$ and appears to support the interpretation 
of the ground states for
the $E_M$ and $E_L$ ensembles as corresponding to genuine bound 
tetraquark states, and the expectation that such a bound tetraquark 
state will therefore also exist at physical $m_\pi$. However, without additional ensembles with larger lattice volumes, a 
direct study of the scaling of the binding with $m_\pi L$, and 
a full extrapolation to infinite volume and physical $m_\pi$, 
is not feasible at present, and must be left for future study.

Aside from general terms $\sim\exp(-m_\pi L)$, finite lattice volume may affect particle energies in two ways. First, a 
bound state receives corrections proportional to $\exp(-|p| L)$, where $|p|$ is the binding momentum, defined via the 
energy-momentum relation $-\Delta E = \sqrt{E_1^2 + p^2} + 
\sqrt{E_2^2 +p^2} -E_1 -E_2$, where $E_1$ and $E_2$ are the energies of the threshold particles. Second, a state which 
becomes a scattering state in the infinite-volume limit, but
which lies below threshold at finite volume, may receive 
power-law corrections in $1/L$, which for the $n=0$ states 
depend on $a_0/L$, where $a_0$ is the scattering length of 
the particles that define the two-meson threshold in the 
channel in question \cite{Luscher:1986pf,Luscher:1990ux}.

Unlike the case of the \udbb and \lsbb channels, where  
binding energies much larger than expected finite-volume
effects were found \cite{Francis:2016hui}, 
finite-volume effects may play a more important qualitative
role in the \udcb channel, where the expected binding and 
overall energy scale are lower. Even though a rigorous 
finite-volume study must be left for future work, we observe 
that, if the ground state energies for the $E_M$ and $E_L$ 
ensembles do, indeed, as the evidence above suggests, 
correspond to bound $ud\bar{c}\bar{b}$ tetraquark states, 
the binding momenta for these states would be
$(|p|)_L=373(171)\,$MeV and $(|p|)_M=245(140)\,$MeV, 
respectively, implying a strong suppression of the 
finite-volume effects on the determined bound-state energies.
Finite-volume studies with larger lattice volumes, however, 
remain desirable to more strongly test the bound-state
interpretation of the ground states for the $E_M$ and $E_L$
ensembles.

\section{Possible experimental detection}\label{sec:experiment}

With a predicted binding of between $15$ and $61$ MeV below the $\bar{D}B^*$ threshold, it is unclear whether the $J^P=1^+$ 
$ud\bar{c}\bar{b}$ tetraquark predicted by our results should,
like its $ud\bar{b}\bar{b}$ and $\ell s\bar{b}\bar{b}$
analogues, be both strong- and electromagnetic-interaction
stable, or be able to decay electromagnetically, 
to $\bar{D}B\gamma$, whose threshold lies $\sim 45$ MeV 
below $\bar{D}B^*$. In either 
case, such a state, with a mass $m_{udcb}\simeq 7154(23)$ MeV,
should be much more accessible to current experimental programs
than are the corresponding doubly bottom states. 
\footnote{It is
worth mentioning that, even for the more experimentally
challenging doubly bottom tetraquarks, a recent study of 
possible branching fractions came to a positive conclusion
regarding their discovery potential in current experiments \cite{Ali:2018xfq}.} 

\begin{figure}[t!]
\centering
\includegraphics[width=0.96\textwidth,trim =0 610 0 0,clip=true]{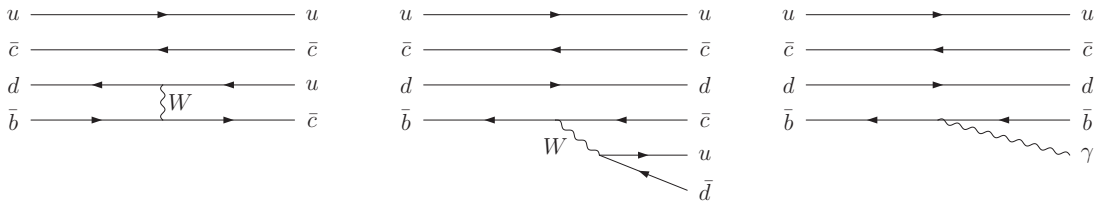}
\caption{Examples of potentially observable \udcb 
tetraquark decay channels. Channels
accessible if only weak decays are possible: 
(left) into two mesons (e.g to $\bar D^0 \bar D^0$), and 
(center) into three mesons, (e.g to $\pi^+\bar D^0 D^-$). Right: decay to $\bar{D}B\gamma$, expected
to dominate if electromagnetic decay is possible.}
\label{fig:decays}
\end{figure}

Should the binding be greater than $\sim 45$ MeV, the 
\udcb tetraquark will decay only weakly, and suffer from
the experimental disadvantage of having a large number of 
exclusive decay modes, each with a relatively small 
branching fraction. Such decay modes would, however, have
the compensating advantage of being accompanied by
secondary heavy-meson decays with displaced vertices. The left
and center panels of Fig.~\ref{fig:decays} show examples of
such decays. Two- or three-meson decays such as those to
$(\bar D^0\bar D^0)$ or $\pi^+ \bar{D}^0 D^-$, 
might then serve as useful potential experimental signatures.
Another three-body mode potentially suitable for detection, 
produced if the $W^+$ materializes as a $c\bar{s}$ rather 
than a $u\bar{d}$ pair, is $J/\psi\, \bar{D}^0 K^0$.

In contrast, should the \udcb binding energy be less than 
$\sim 45$ MeV, the state should decay electromagnetically
essentially 100\% of the time to $\bar{D}B\gamma$, as in the rightmost panel of Fig.\ref{fig:decays}.

With (given the predicted binding energy) less than 
$\sim 20$ MeV of phase space available to the electromagnetic 
decay, such a tetraquark, produced with a reasonable 
momentum in the lab frame, will decay to produce a 
mixed-heavy-flavour $D^+B^-$ or $\bar{D}^0B^0$ meson pair
highly collimated in the lab frame.

\section{Conclusions}

The study of heavy tetraquarks of the $qq'\bar Q'\bar Q$ 
type continues to yield insight into the binding of exotic
hadrons. This paper provides evidence of a 
strong-interaction-stable $I(J^P)=0(1^+)$ \udcb tetraquark 
which we estimate to lie $15-61$ MeV below the 
corresponding free, two-meson ($\bar{D}B^*$) threshold. 
The predicted mass range, $\approx 7154(23)~$MeV, 
thus straddles the electromagnetic $\bar{D}B\gamma$ 
decay threshold, making it impossible, at present, 
to predict whether it will decay electromagnetically 
or only weakly. With $B_c$ production, and hence
the simultaneous production of $b\bar{b}$ and
$c\bar{c}$ pairs, already established experimentally,
these results clearly motivate a search for this
state at LHCb.

The results of the direct computation above further 
strengthen the argument that this class of tetraquarks 
becomes more strongly bound as the light-quark component 
becomes lighter. The variable, unphysical heavy-quark mass
study also confirms the expectation that such 
tetraquarks become less bound as the heavy anti-diquark 
reduced mass decreases.

Complementary to the direct computation of the binding
of the $I(J^P)=0(1^+)$ \udcb tetraquark, the NRQCD study 
of the $ud\bar{b}^\prime \bar{b}^\prime$, 
$ud\bar{b}^\prime \bar{b}$, 
$\ell s\bar{b}^\prime \bar{b}^\prime$ and 
$\ell s\bar{b}^\prime \bar{b}$ channels with variable
heavy quark mass, $m_{b^\prime}$, extending down to
$0.60\, m_b$ confirms our qualitative understanding of
the basic physics responsible for the observed
tetraquark binding. This study also suggests that 
strong-interaction-stable \udcc, \lscb, and \lscc 
tetraquarks are unlikely to exist, see also \cite{Cheung:2017tnt}, though direct 
computations using a relativistic charm action at almost physical light quark masses have
yet to be carried out for these channels.

Our prediction for the \udcb binding has an error dominated 
by what we believe to be a conservative estimate of the 
finite-volume systematic. While statistics does not 
appear to be a problem, having more operators and/or 
crafting sources and sinks that better overlap with our desired ground states should provide longer plateaus. In future, 
futher investigations of the \udcb channel using 
ensembles with additional near-physical light pion 
masses and larger spatial volumes will allow us to
obtain better quantitative control of the extrapolation
to physical $m_\pi$ and finite-volume effects.

\begin{acknowledgments}
Propagator inversions for this work were performed on the compute cluster ``GPC'' at SciNet, Toronto. This work was supported in part by the Natural Sciences and Engineering Research Council (NSERC) of Canada. The calculations were performed as part of an RAC allocation under the Compute Canada initiative.
\end{acknowledgments}
\appendix
\begin{figure}[ht!]
\centering
\subfloat{
\includegraphics[width=0.38\textwidth]{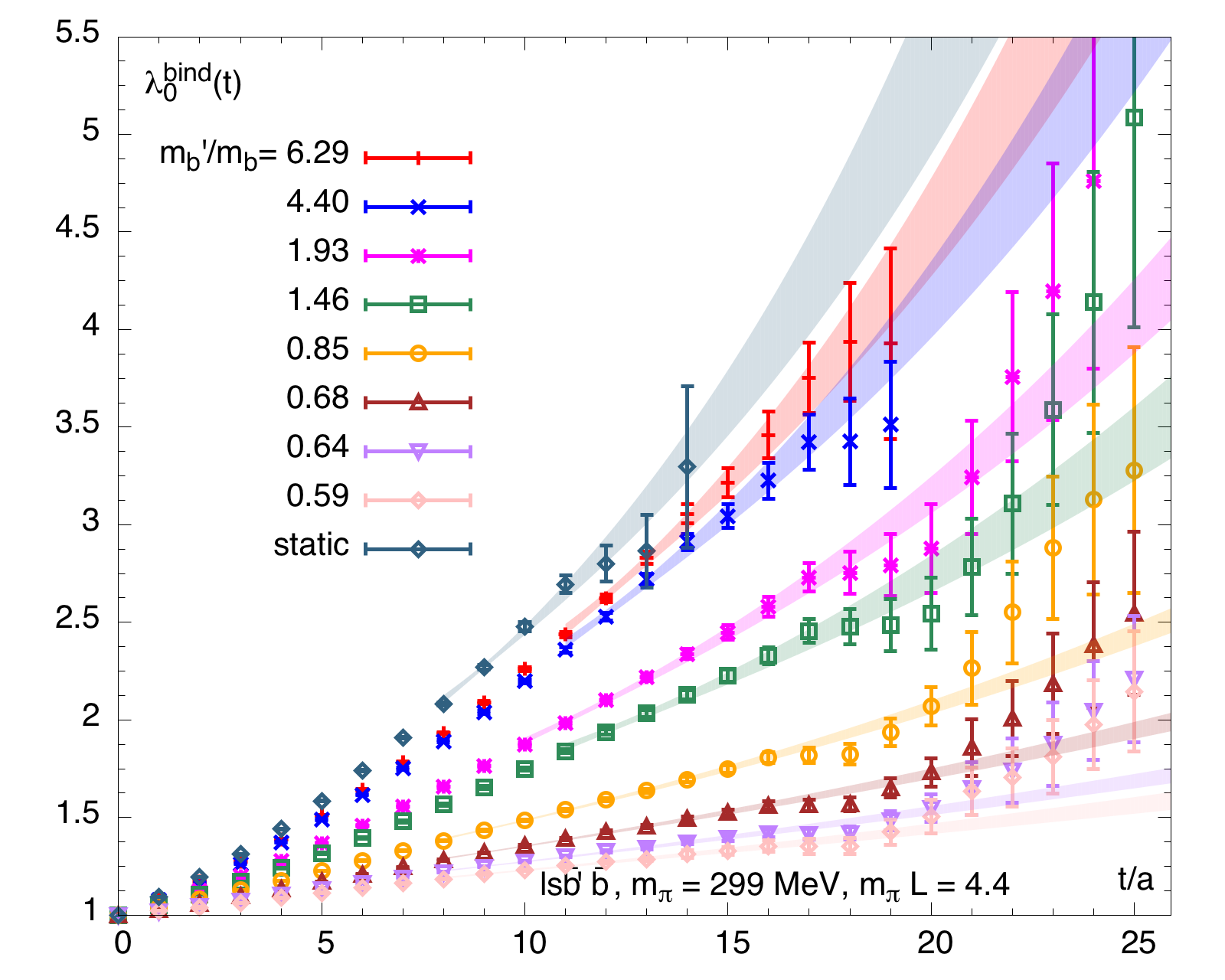}
}
\hspace{4pt}
\subfloat{
\includegraphics[width=0.38\textwidth]{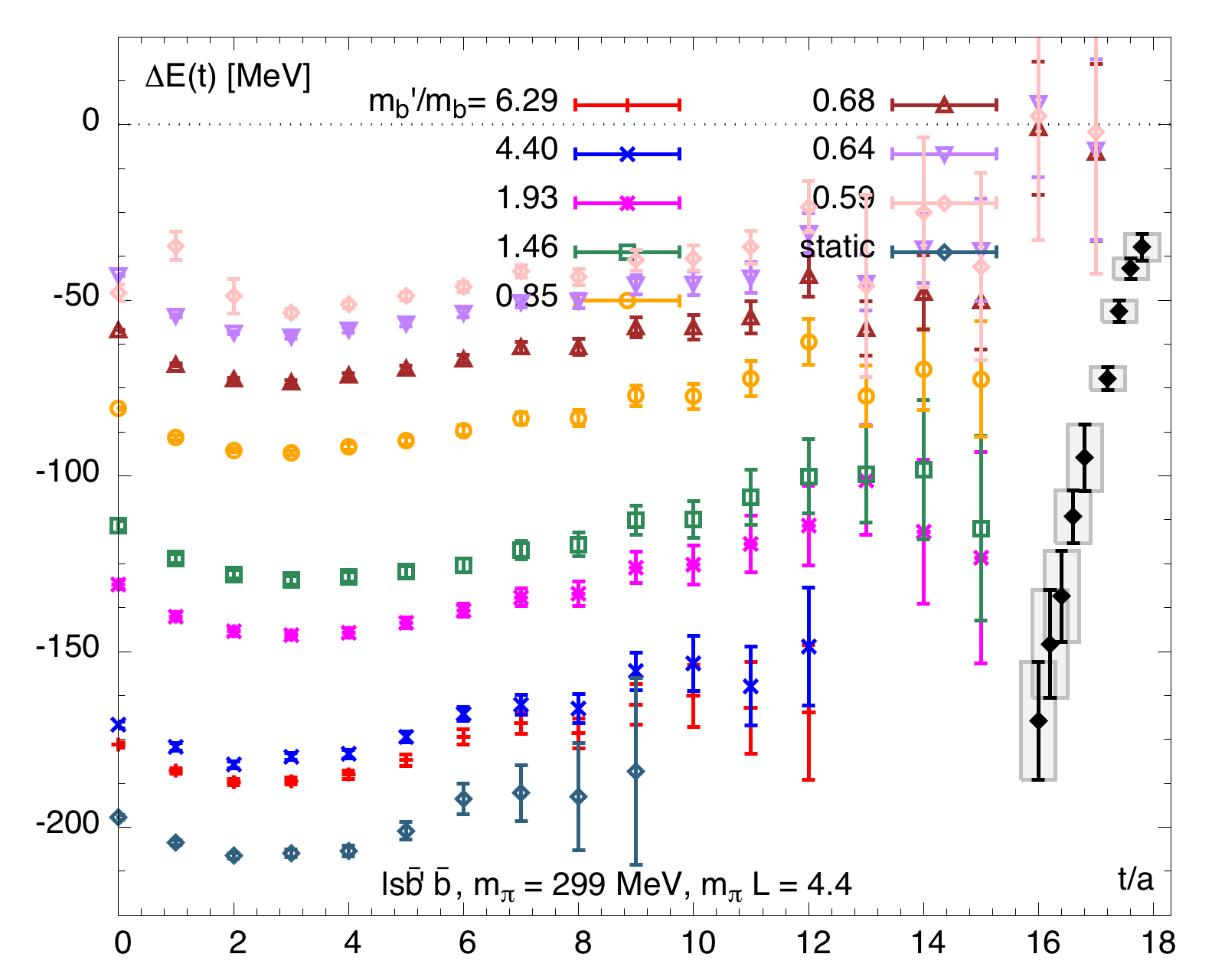}
}
\vspace{1pt}
\subfloat{
\includegraphics[width=0.38\textwidth]{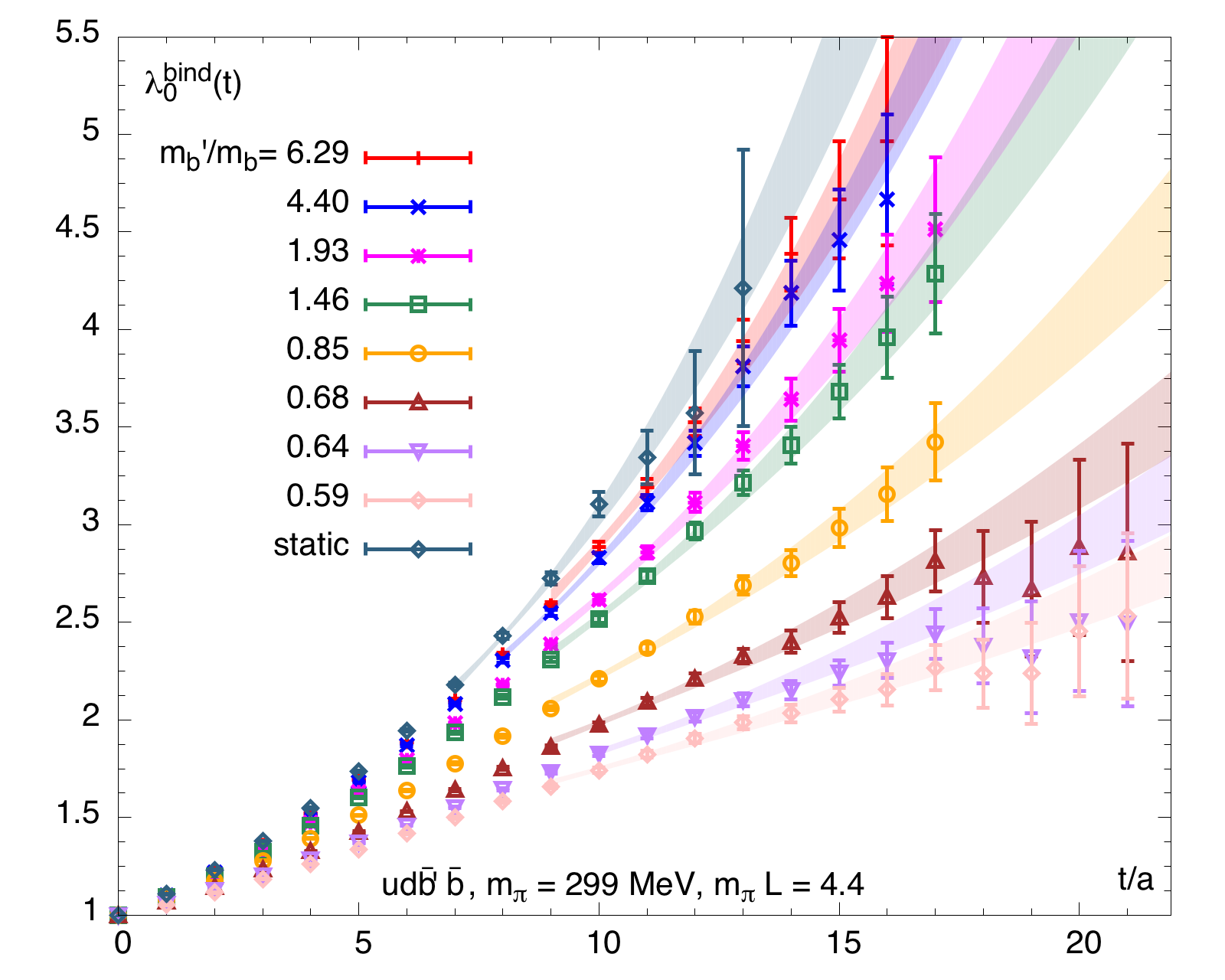}
}
\hspace{4pt}
\subfloat{
\includegraphics[width=0.38\textwidth]{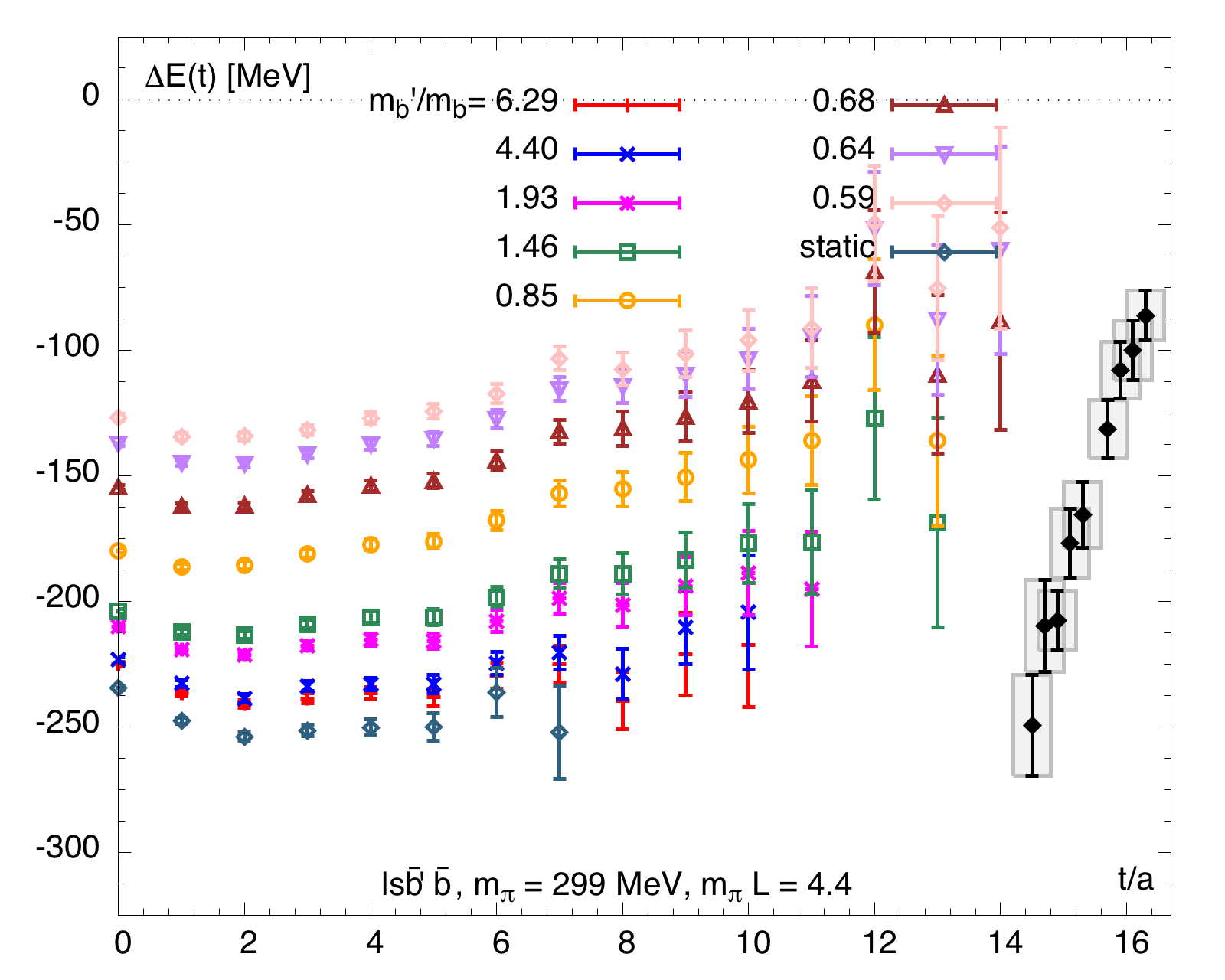}
}
\vspace{1pt}
\subfloat{
\includegraphics[width=0.38\textwidth]{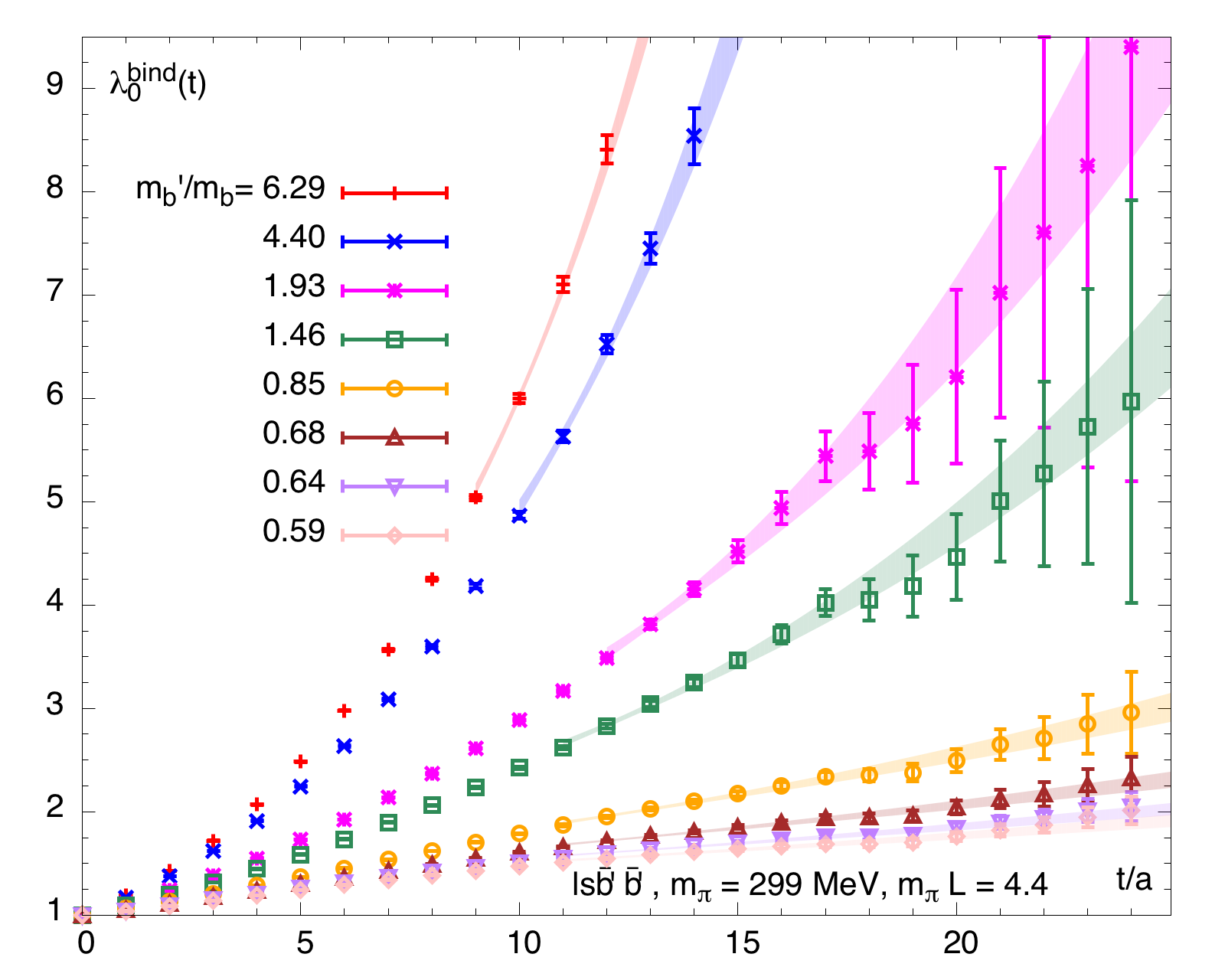}
}
\hspace{4pt}
\subfloat{
\includegraphics[width=0.38\textwidth]{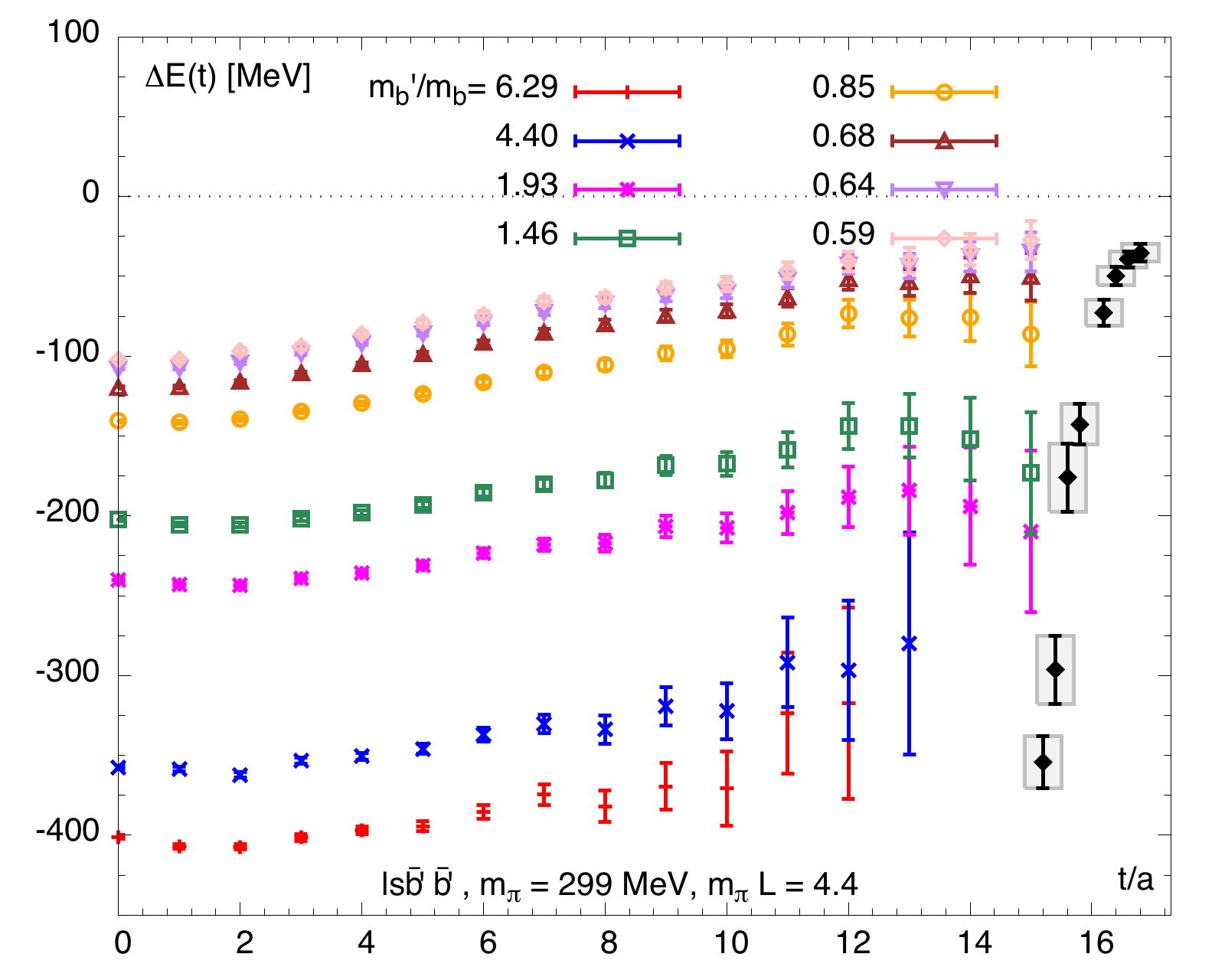}
}
\vspace{1pt}
\subfloat{
\includegraphics[width=0.38\textwidth]{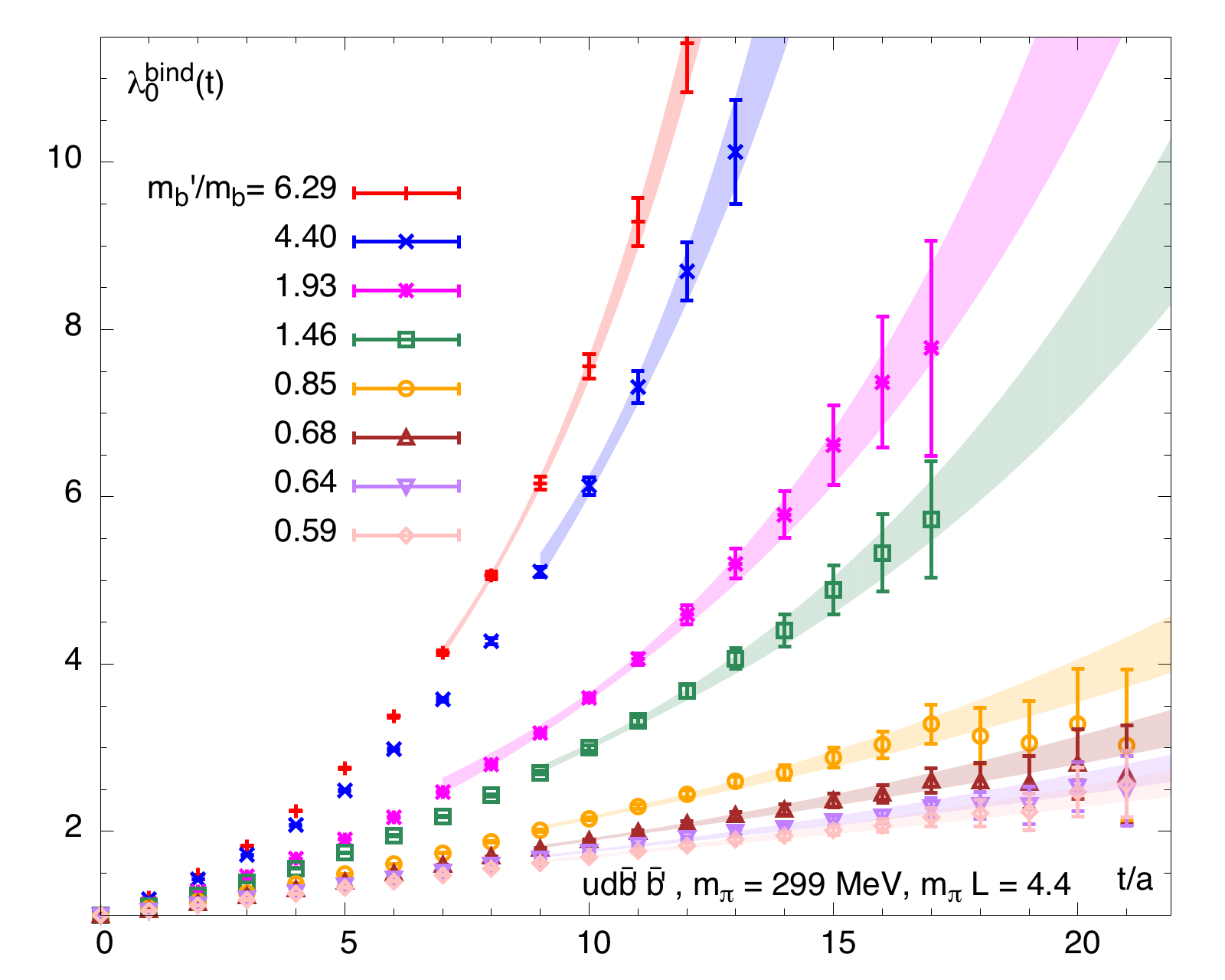}
}
\hspace{4pt}
\subfloat{
\includegraphics[width=0.38\textwidth]{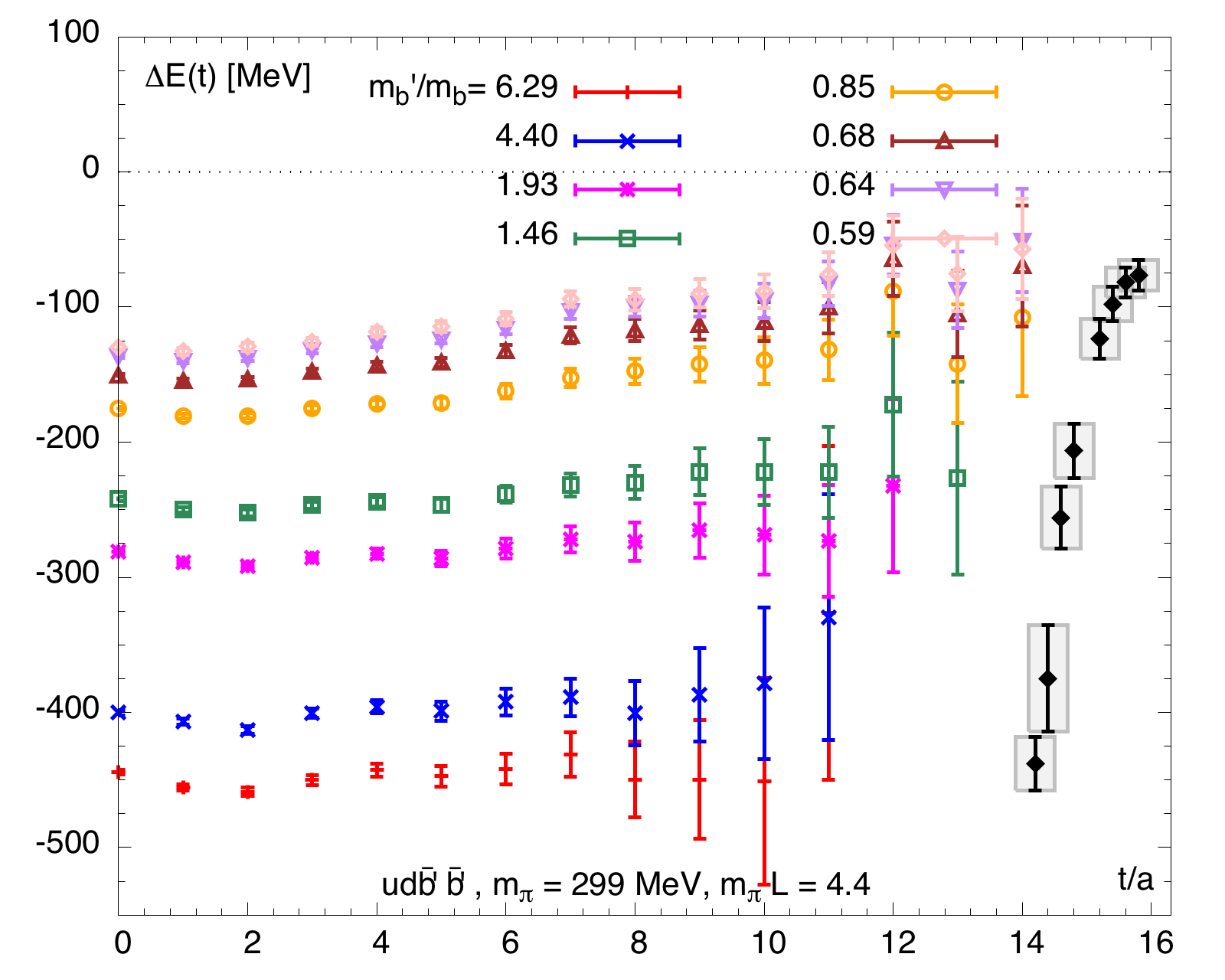}
}
\caption{ Binding correlators (left) and effective energies (right) for the four channels $\ell s \bar{b}^\prime\bar{b}$, $ud \bar{b}^\prime\bar{b}$, $\ell s \bar{b}^\prime\bar{b}^\prime$ and $ud \bar{b}^\prime\bar{b}^\prime$ (top to bottom). Fit results are given as shaded bands (left) or boxes (right).
}
\label{fig:heavies}
\end{figure}

\section{Our Implementation of the NRQCD Framework}

We use the NRQCD Hamiltonian \cite{Thacker:1990bm,Lepage:1992tx,Davies:1994mp}
\begin{equation}\label{eq:nrqcd}
\begin{aligned}
 H_0 = &-c_0\frac{\Delta^{(2)}}{2m_Q}\\
 \delta H = &-c_1\frac{(\Delta^{(2)})^2}{8m_Q^3}
 + \frac{c_2}{U_0^4}\frac{ig}{8m_Q^2}(\bm{\tilde\Delta\cdot\tilde{E}}-\bm{\tilde{E}\cdot\tilde\Delta}) 
 \\
 &- \frac{c_3}{U_0^4}\frac{g}{8m_Q^2}\bm{\sigma\cdot}(\bm{\tilde\Delta\times\tilde{E}}-\bm{\tilde{E}\times\tilde\Delta})\\
 &-\frac{c_4}{U_0^4}\frac{g}{2m_Q}\bm{\sigma\cdot\tilde{B}}
 + c_5\frac{a^2\Delta^{(4)}}{24m_Q}
 - c_6\frac{a(\Delta^{(2)})^2}{16nm_Q^2}\; .
\end{aligned}
\end{equation}
where $\bm{\tilde{E}}$, $\bm{\tilde{B}}$ and $\bm{\tilde\Delta}$ denote the $O(a)$-improved color electric field, color magnetic field, and spatial lattice derivative respectively.
$\Delta^{(2)}$ is the lattice Laplacian, $g$ is the bare gauge coupling and $n$ is a mode number used in the evolution equation (see Eq.~(\ref{evolution}) below).
We find $n=4$ to be a reasonable value for this stability parameter.

Here, the tadpole-improvement coefficient $U_0$ is set via the fourth root of the plaquette and the coefficients $c_i$ are
assigned their tree-level values of $1$. This leaves one 
free parameter, $m_Q$, to be tuned to give the desired 
heavy-quark physics. This tuning is achieved by measuring 
the slope of the dispersion relation, and hence the
``kinetic'' mass, of the spin-averaged $\Upsilon$ and $\eta_b$ 
on Fourier-transformed local-local current correlators. 
Explicitly, with
\begin{equation}
\begin{gathered}
\eta_b(M_0,p^2) = \eta_b(M_0) + \frac{p^2}{2m^b_\eta}, \qquad \Upsilon(M_0,p^2) = \Upsilon(M_0) + \frac{p^2}{2m^b_\Upsilon},
\end{gathered}
\end{equation}
we take
\begin{equation}
m^b = \frac{1}{4}\left( m^b_\eta + 3 m^b_\Upsilon \right).
\end{equation}

With the parameters of the NRQCD action all fixed, 
the evolution equation,
\begin{equation}\label{evolution}
G(x,t+1) = \left(1-\frac{H_0}{2n}\right)^n \left( 1 - \frac{\delta H}{2} \right)  \frac{U_{t}^\dagger(x)}{U_0} \left( 1 - \frac{\delta H}{2} \right) \left(1-\frac{H_0}{2n}\right)^n G(x,t),
\end{equation}
where $U_t$ denotes a gauge link in the temporal direction,
is used to compute our heavy quark propagators. 
Static heavy quark propagators can be realised by setting all the coefficients $c_i$ in this action to 0.

The NRQCD action can be organized as an expansion in $1/m_Q$ and naively its regime of validity is limited to the region $m_Q\gtrsim 1$ in lattice units.

\section{Details of the variable-heavy-mass study}\label{sec:heavies}

This study focuses on unphysical bottom quarks using the ensemble $E_M$ of Tab.~\ref{tab:lat_params} with a single source position.
Visible effects due to violations of the condition 
$am_Q\gtrsim 1$ are e.g. the increase of the hyperfine splitting. While tunings of the 
parameters $c_i$ beyond tree level, as well as the inclusion of additional terms in the action, and/or an increase in the stability parameter $n$, could be considered to expand the region
of validity of the NRQCD approximation, here we choose to work with tree-level values and the form of the NRQCD Hamiltonian
detailed in Appendix A. Monitoring the hyperfine splitting, we find that $m_Q\simeq 0.594\, m_b$ appears to be the lowest acceptable 
value for our calculation.
From extrapolation of our heavy-quark data, the charm quark lies at around $m_c\simeq 0.33\, m_b$ and is thus beyond our reach, in
the NRQCD approach. 

The binding energies of the $ud\bar b' \bar b'$, $ud\bar b' \bar b$, $\ell s\bar b' \bar b'$ and $\ell s\bar b' \bar b$ tetraquark states on the $E_M$ ensemble
are extracted by fitting the smallest eigenvalue of the respective GEVPs to a single-exponential ansatz. 
All fit results with $\chi^2/d.o.f. \leq 1$ are kept for further analysis. The fit procedure does not 
take into account correlations in the data. The final numbers are chosen as representative results with 
the longest fit window possible.

The resulting data and fits are shown in the left panels of Fig.~\ref{fig:heavies}. 
We observe positive exponential growth of the first eigenvalues determined from the binding 
correlator with Euclidean time $t$. This growth becomes steeper as the heavy quark mass 
is increased in all four tetraquark channels (especially so for the equal-heavy-mass
case, where the binding energy is unbounded as $m_{b^\prime}\rightarrow\infty$).
The shaded bands denote the corresponding fits and are seen to describe the data well. We 
also determine the effective binding energies and show these results (data), together with the
corresponding binding energy fits (shaded boxes) in the right panels of Fig.~\ref{fig:heavies}.
The success of the phenomenological fit forms detailed in Sec.~\ref{sec:pheno}
in reproducing this lattice data confirms our understanding of the basic binding mechanism. 

\begin{figure}[ht!]
\centering
\subfloat
{\includegraphics[width=0.43\textwidth]{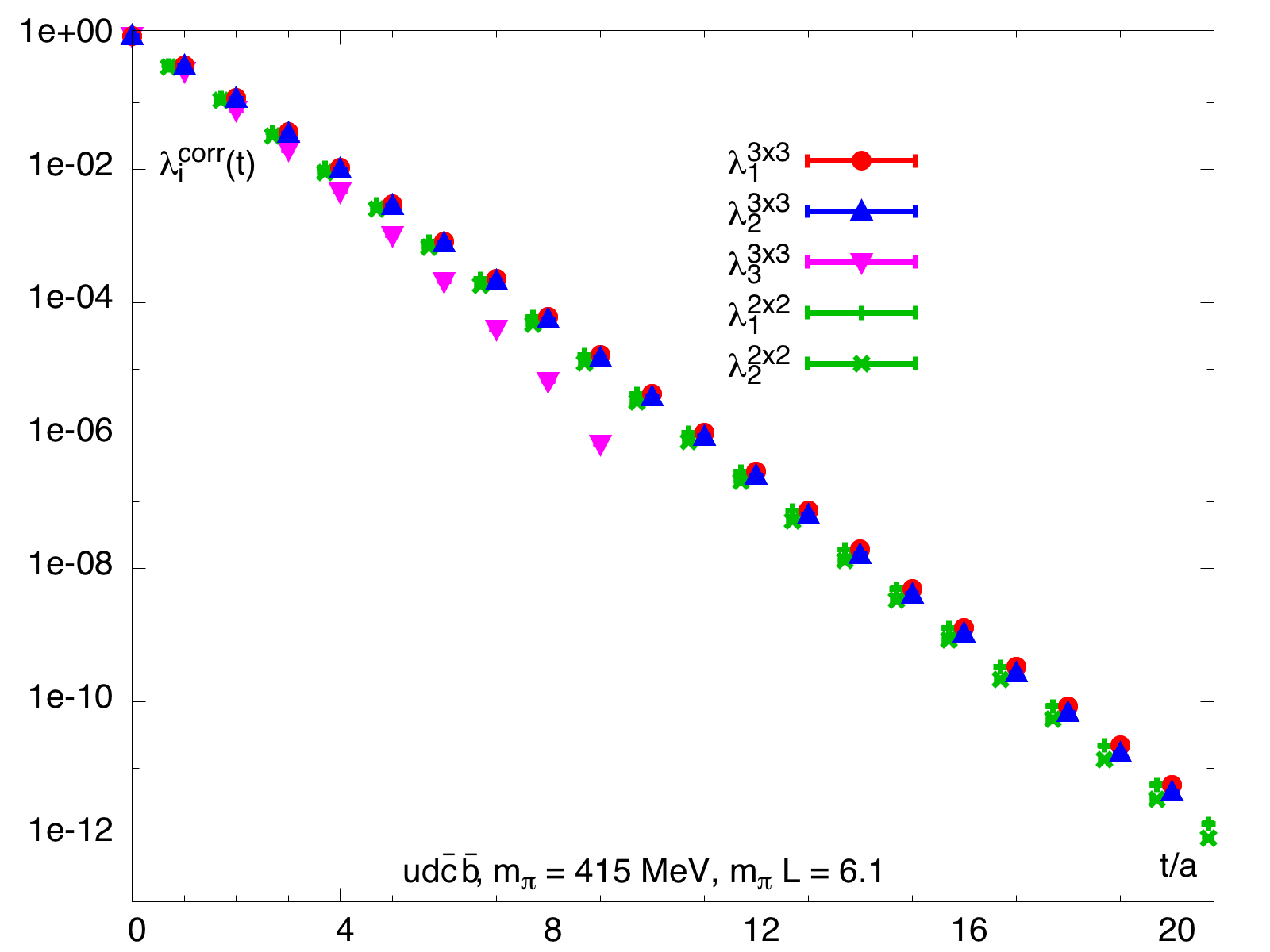}
}
\subfloat
{\includegraphics[width=0.43\textwidth]{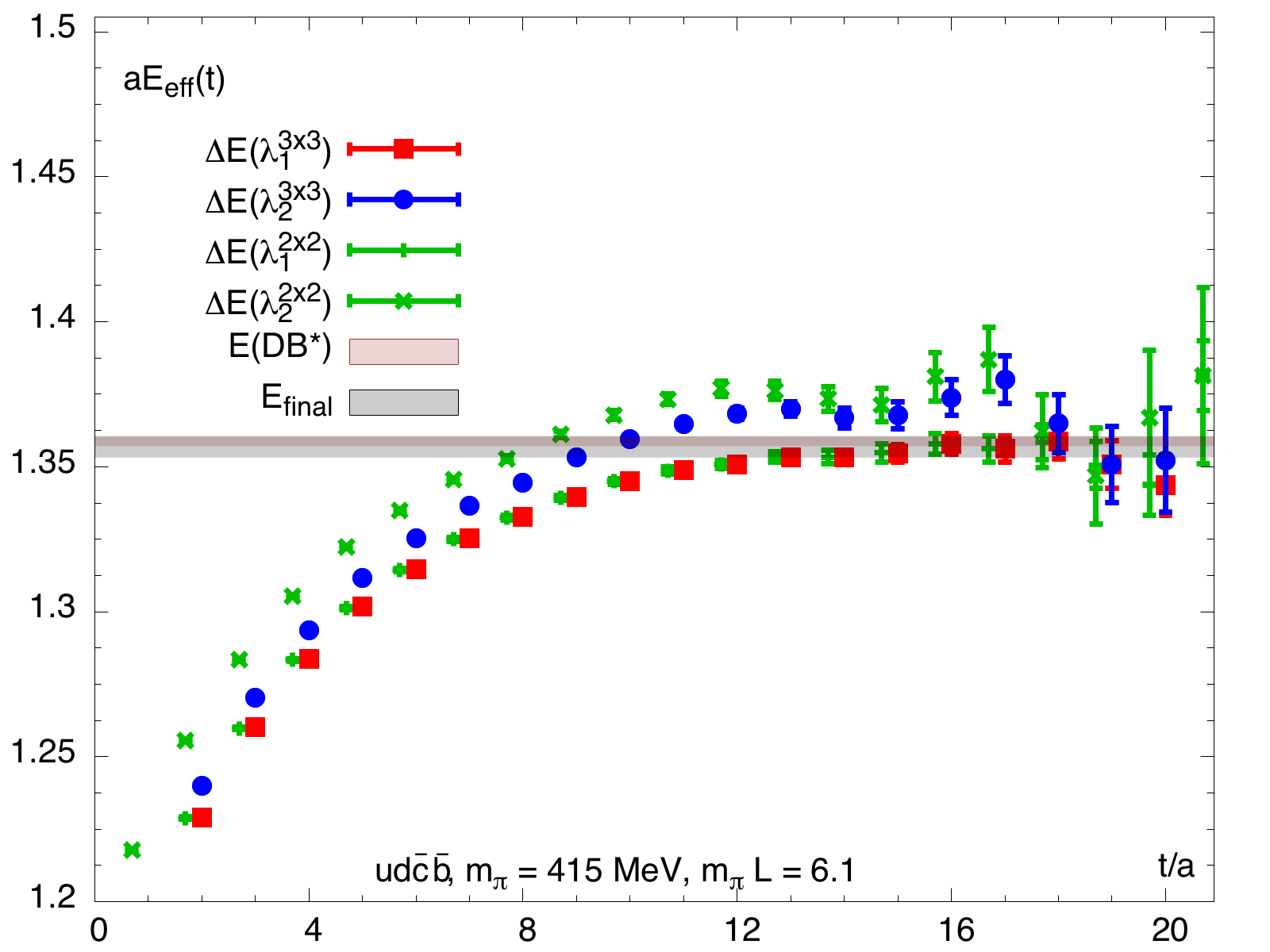}
}
\vspace{2pt}
\subfloat
{
\includegraphics[width=0.43\textwidth]{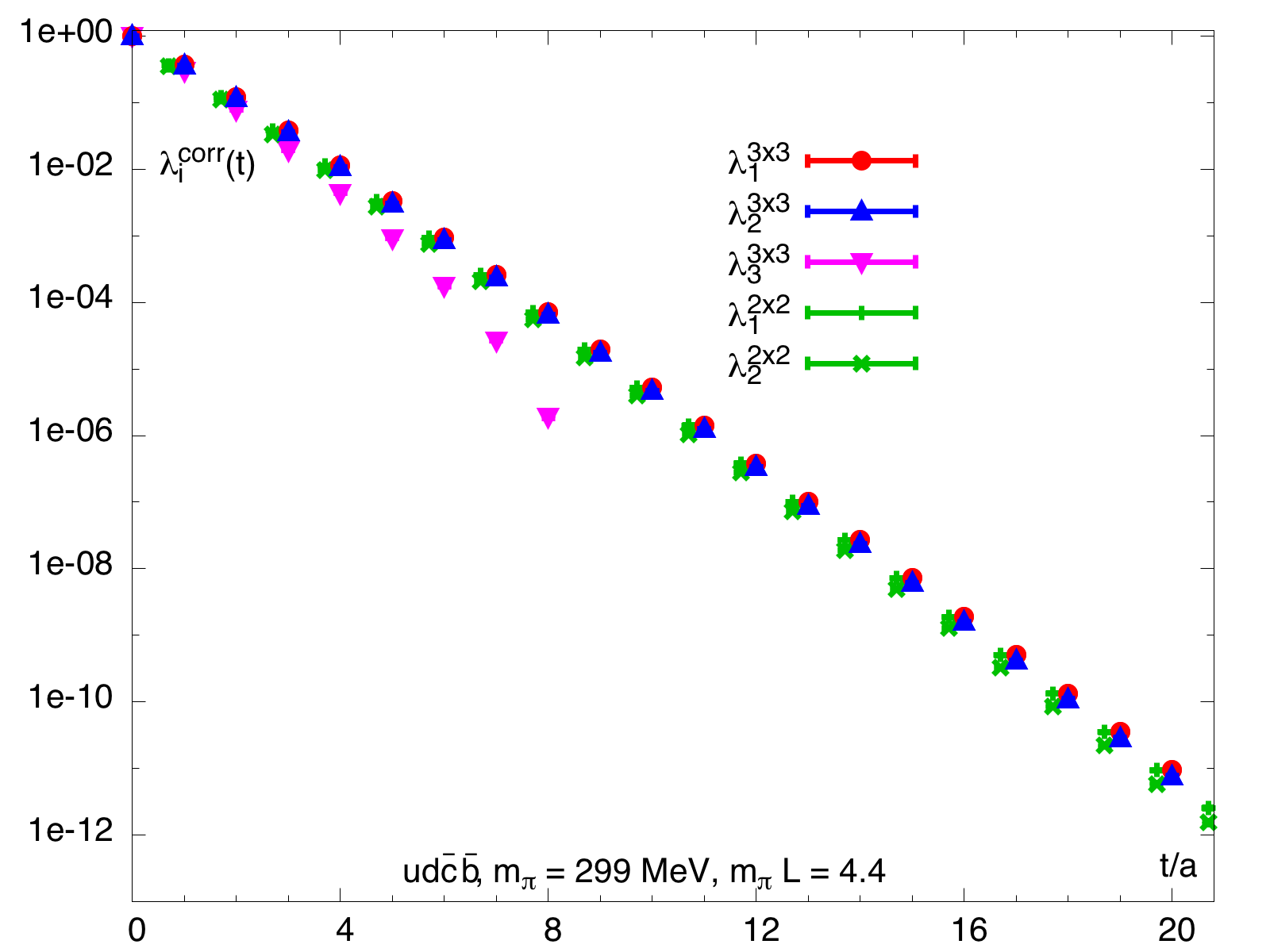}
}
\subfloat
{
\includegraphics[width=0.43\textwidth]{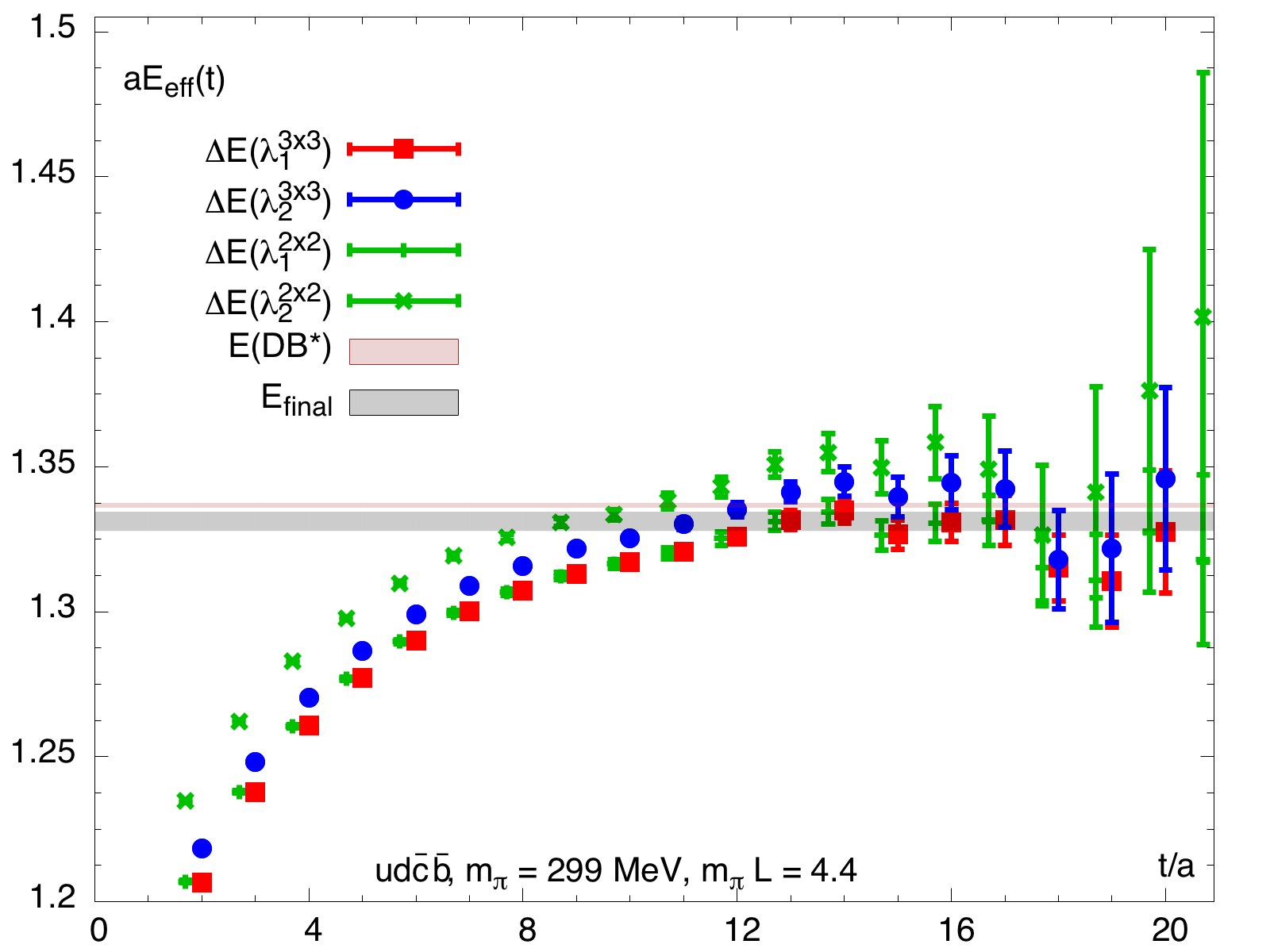}
}
\vspace{2pt}
\subfloat
{
\includegraphics[width=0.43\textwidth]{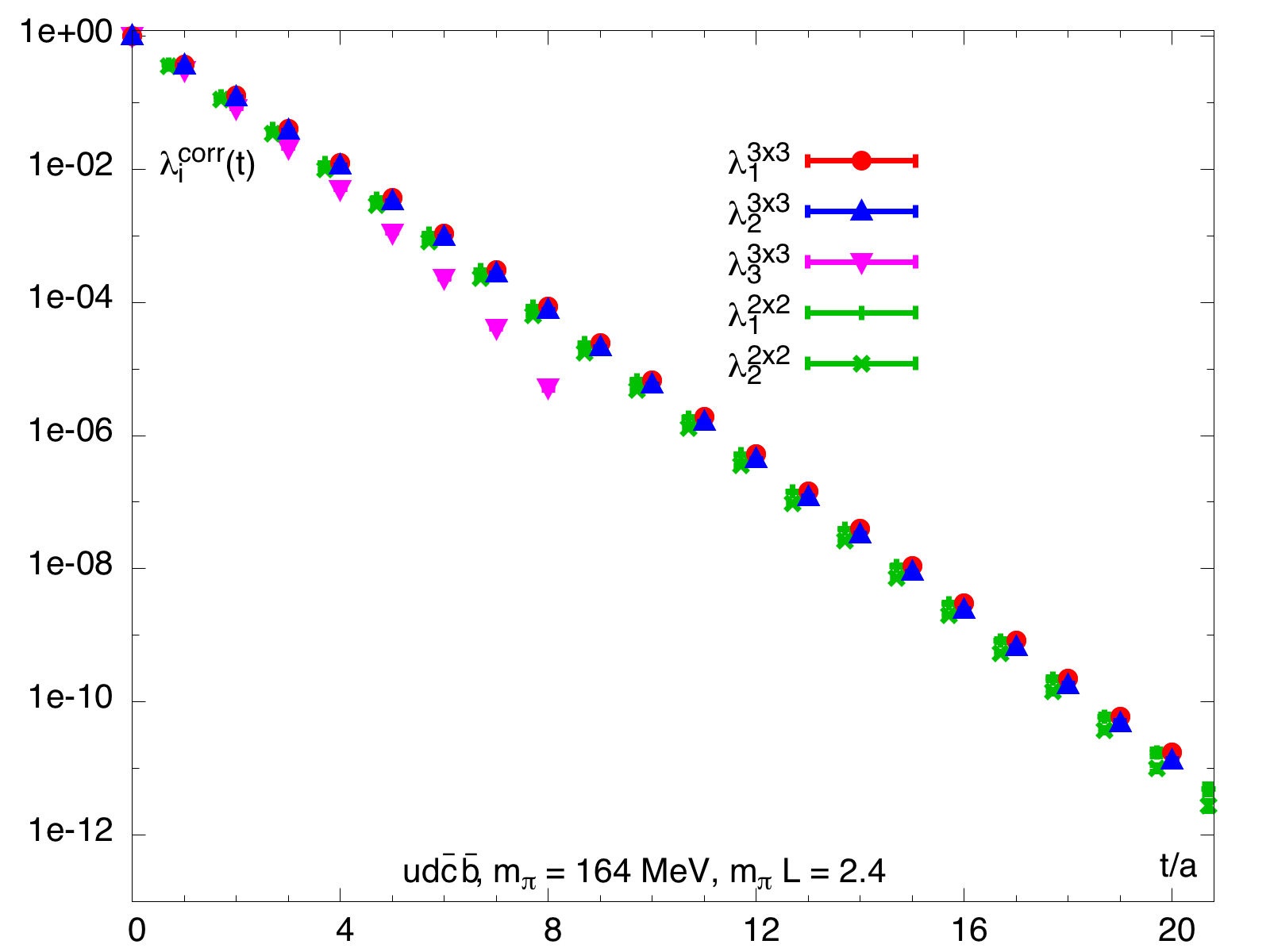}
}
\subfloat
{\includegraphics[width=0.43\textwidth]{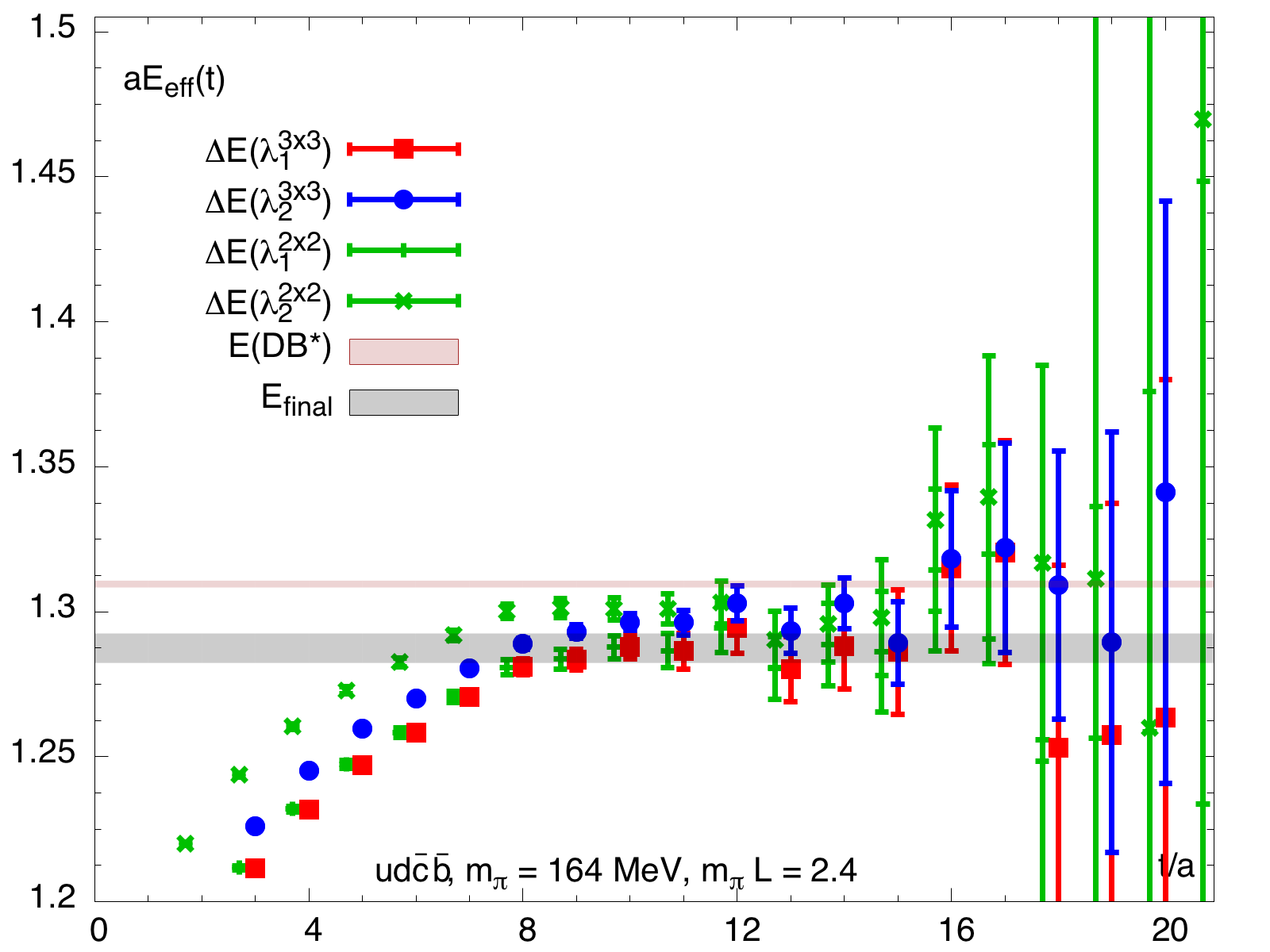}
}
\caption{$ud\bar{c}\bar{b}$ tetraquark results on $E_H$ (top), $E_M$ (center) and $E_L$ (bottom). The left panels 
show the GEVP eigenvalues, the right panels the 
corresponding energies.
The red and blue symbols show,
respectively, the ground and first excited state results 
obtained from the $3\times 3$ GEVP analyses. The green 
vertical dashes and green diagonal crosses similarly
denote the ground and first excited state results 
obtained from the $2\times 2$ GEVP analyses.
The two-meson $\bar{D}B^*$ thresholds, shown for
comparison, are given by the brown bands in the right-hand
panels. The grey bands in those same panels depict the final 
energies obtained from the single exponential fits
to the eigenvalues described above. Further details are 
given in the text. $2\times 2$ GEVP results have been 
offset in $t$ for visual clarity.
}
\label{fig:udcb2}
\end{figure}

\begin{figure}[ht!]
\centering
\subfloat{\includegraphics[width=0.99\textwidth]{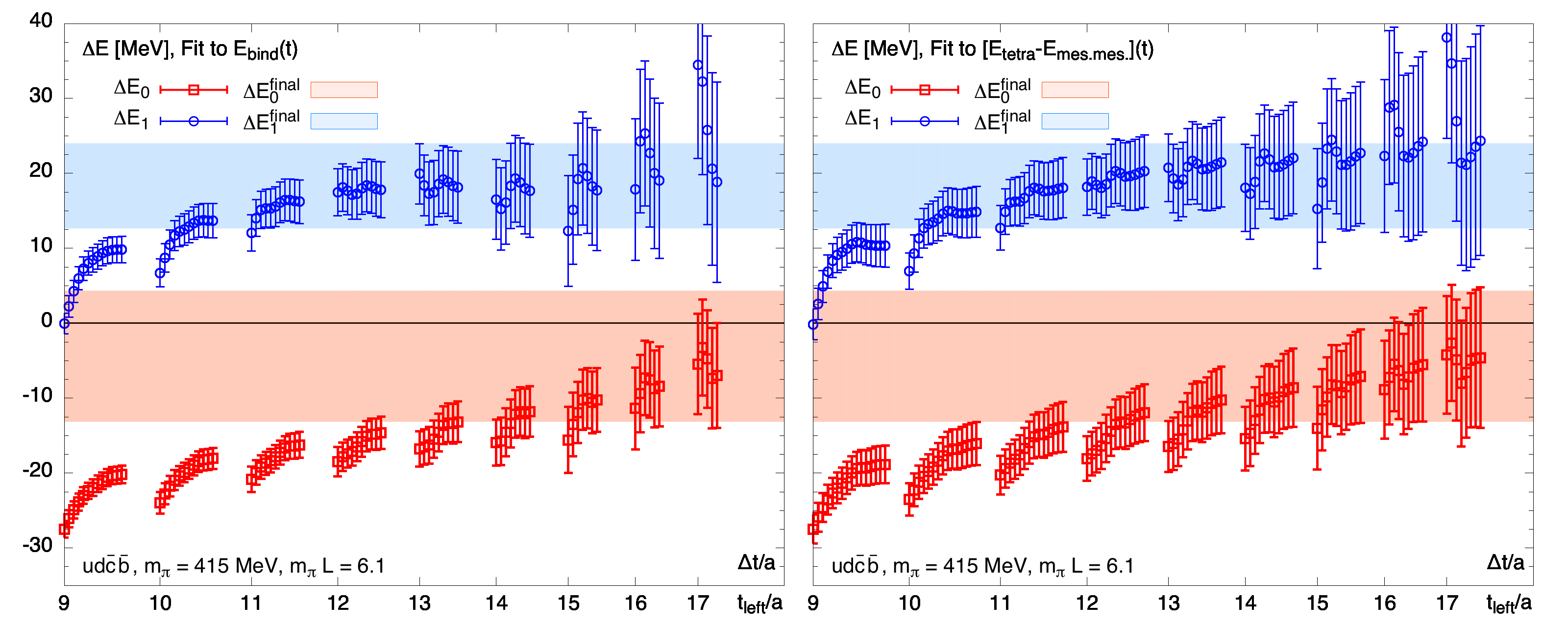}}
\vspace{4pt}
\subfloat{\includegraphics[width=0.99\textwidth]{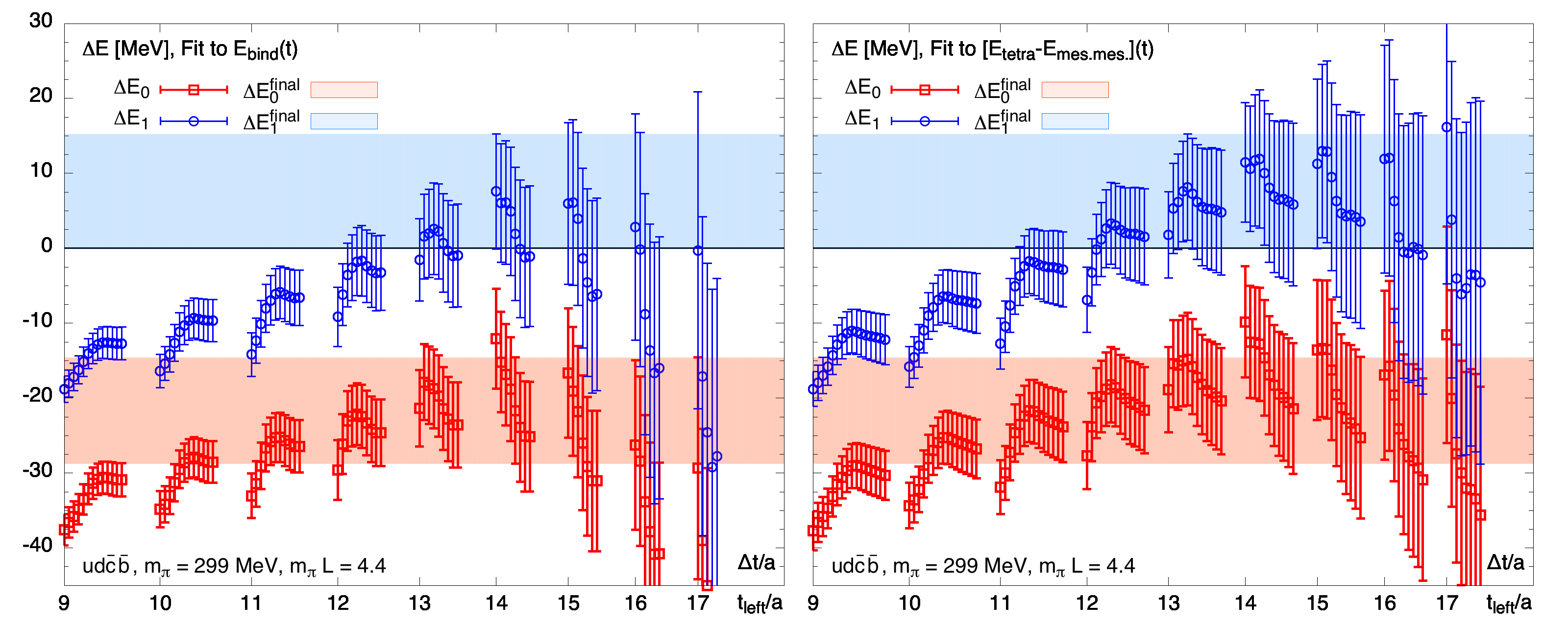}}
\vspace{4pt}
\subfloat{\includegraphics[width=0.99\textwidth]{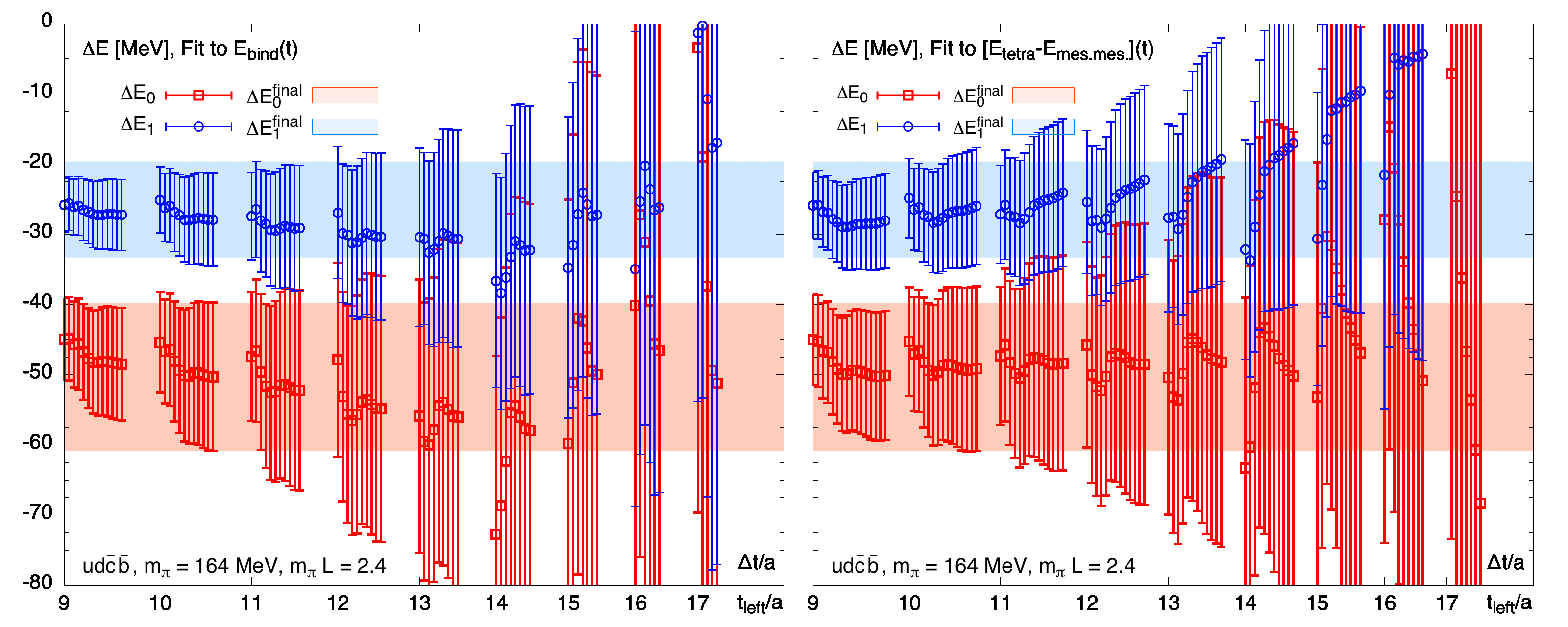}}
\caption{Fit window dependence for fitting the binding correlator GEVP solutions and those obtained from the particle correlators for the ensembles $E_H$ (top), $E_M$ (center), and $E_L$ (bottom). 
We obtain our final results from this selection of accepted fits in the stable region that is representative of both procedures.}
\label{fig:app_fits}
\end{figure}

\section{Further details of the \texorpdfstring{$ud\bar{c}\bar{b}$}{udcb} analysis}
\label{sec:app_fits}

We estimate the binding energies from (uncorrelated) fits to the eigenvalue data with a single exponential ansatz.
This data is shown in Fig.~\ref{fig:udcb2}. The eigenvalues
are shown in the left panels 
and the corresponding effective energies in the right panels. Throughout, the first, second and third eigenvalues 
obtained from the $3\times 3$ GEVP are given in red, blue and magenta, respectively, while the results obtained from the 
$2\times 2$ GEVP are specified using green 
vertical dashes and green diagonal crosses.

To extract the energies from the lattice data, we select fit windows in Euclidean $t$ extending from 
$t_{left}$ to $t_{right}=t_{left}+\Delta t$ whose results satisfy $\chi^2/d.o.f. \lesssim 1$. 
In the case of the binding correlator, the fit immediately gives the binding energy. In the case of the four-quark
``diquark-diquark'' and ``meson-meson'' correlators, the two meson threshold mass sum is subtracted from the 
resulting tetraquark mass. The accepted results obtained in this way are shown as a function of
$t_{left}$ in the right panels of Fig.~\ref{fig:app_fits}, and the corresponding results obtained using the 
binding correlator GEVPs in the corresponding left panels. The multiple points 
plotted for each $t_{left}$ are those corresponding to different choices of $\Delta t$.
From this selection of results the final, quoted numbers 
(corresponding to the grey bands in Fig.~\ref{fig:udcb2}) 
are chosen in such a way that they lie in the 
stable regions of the plots, and are representative of both procedures.
The final fit ranges used are 
$t/a\in[10:23]$ for $E_L$, $t/a\in[14:20]$ for $E_M$, obtained from the left panels, and $t/a\in[17:21]$ for $E_H$ chosen from the right panels. 

\bibliography{cbTetra}{}

\begin{thebibliography}{74}%
\makeatletter
\providecommand \@ifxundefined [1]{%
 \@ifx{#1\undefined}
}%
\providecommand \@ifnum [1]{%
 \ifnum #1\expandafter \@firstoftwo
 \else \expandafter \@secondoftwo
 \fi
}%
\providecommand \@ifx [1]{%
 \ifx #1\expandafter \@firstoftwo
 \else \expandafter \@secondoftwo
 \fi
}%
\providecommand \natexlab [1]{#1}%
\providecommand \enquote  [1]{``#1''}%
\providecommand \bibnamefont  [1]{#1}%
\providecommand \bibfnamefont [1]{#1}%
\providecommand \citenamefont [1]{#1}%
\providecommand \href@noop [0]{\@secondoftwo}%
\providecommand \href [0]{\begingroup \@sanitize@url \@href}%
\providecommand \@href[1]{\@@startlink{#1}\@@href}%
\providecommand \@@href[1]{\endgroup#1\@@endlink}%
\providecommand \@sanitize@url [0]{\catcode `\\12\catcode `\$12\catcode
  `\&12\catcode `\#12\catcode `\^12\catcode `\_12\catcode `\%12\relax}%
\providecommand \@@startlink[1]{}%
\providecommand \@@endlink[0]{}%
\providecommand \url  [0]{\begingroup\@sanitize@url \@url }%
\providecommand \@url [1]{\endgroup\@href {#1}{\urlprefix }}%
\providecommand \urlprefix  [0]{URL }%
\providecommand \Eprint [0]{\href }%
\providecommand \doibase [0]{http://dx.doi.org/}%
\providecommand \selectlanguage [0]{\@gobble}%
\providecommand \bibinfo  [0]{\@secondoftwo}%
\providecommand \bibfield  [0]{\@secondoftwo}%
\providecommand \translation [1]{[#1]}%
\providecommand \BibitemOpen [0]{}%
\providecommand \bibitemStop [0]{}%
\providecommand \bibitemNoStop [0]{.\EOS\space}%
\providecommand \EOS [0]{\spacefactor3000\relax}%
\providecommand \BibitemShut  [1]{\csname bibitem#1\endcsname}%
\let\auto@bib@innerbib\@empty
\bibitem [{\citenamefont {Ali}\ \emph {et~al.}(2017)\citenamefont {Ali},
  \citenamefont {Lange},\ and\ \citenamefont {Stone}}]{Ali:2017jda}%
  \BibitemOpen
  \bibfield  {author} {\bibinfo {author} {\bibfnamefont {A.}~\bibnamefont
  {Ali}}, \bibinfo {author} {\bibfnamefont {J.~S.}\ \bibnamefont {Lange}}, \
  and\ \bibinfo {author} {\bibfnamefont {S.}~\bibnamefont {Stone}},\ }\href
  {\doibase 10.1016/j.ppnp.2017.08.003} {\bibfield  {journal} {\bibinfo
  {journal} {Prog. Part. Nucl. Phys.}\ }\textbf {\bibinfo {volume} {97}},\
  \bibinfo {pages} {123} (\bibinfo {year} {2017})},\ \Eprint
  {http://arxiv.org/abs/1706.00610} {arXiv:1706.00610 [hep-ph]} \BibitemShut
  {NoStop}%
\bibitem [{\citenamefont {Aaij}\ \emph {et~al.}(2015)\citenamefont {Aaij} \emph
  {et~al.}}]{Aaij:2015tga}%
  \BibitemOpen
  \bibfield  {author} {\bibinfo {author} {\bibfnamefont {R.}~\bibnamefont
  {Aaij}} \emph {et~al.} (\bibinfo {collaboration} {LHCb}),\ }\href {\doibase
  10.1103/PhysRevLett.115.072001} {\bibfield  {journal} {\bibinfo  {journal}
  {Phys. Rev. Lett.}\ }\textbf {\bibinfo {volume} {115}},\ \bibinfo {pages}
  {072001} (\bibinfo {year} {2015})},\ \Eprint
  {http://arxiv.org/abs/1507.03414} {arXiv:1507.03414 [hep-ex]} \BibitemShut
  {NoStop}%
\bibitem [{\citenamefont {Zouzou}\ \emph {et~al.}(1986)\citenamefont {Zouzou},
  \citenamefont {Silvestre-Brac}, \citenamefont {Gignoux},\ and\ \citenamefont
  {Richard}}]{Zouzou:1986qh}%
  \BibitemOpen
  \bibfield  {author} {\bibinfo {author} {\bibfnamefont {S.}~\bibnamefont
  {Zouzou}}, \bibinfo {author} {\bibfnamefont {B.}~\bibnamefont
  {Silvestre-Brac}}, \bibinfo {author} {\bibfnamefont {C.}~\bibnamefont
  {Gignoux}}, \ and\ \bibinfo {author} {\bibfnamefont {J.~M.}\ \bibnamefont
  {Richard}},\ }\href {\doibase 10.1007/BF01557611} {\bibfield  {journal}
  {\bibinfo  {journal} {Z. Phys.}\ }\textbf {\bibinfo {volume} {C30}},\
  \bibinfo {pages} {457} (\bibinfo {year} {1986})}\BibitemShut {NoStop}%
\bibitem [{\citenamefont {Lipkin}(1986)}]{Lipkin:1986dw}%
  \BibitemOpen
  \bibfield  {author} {\bibinfo {author} {\bibfnamefont {H.~J.}\ \bibnamefont
  {Lipkin}},\ }\href {\doibase 10.1016/0370-2693(86)90843-9} {\bibfield
  {journal} {\bibinfo  {journal} {Phys. Lett.}\ }\textbf {\bibinfo {volume}
  {B172}},\ \bibinfo {pages} {242} (\bibinfo {year} {1986})}\BibitemShut
  {NoStop}%
\bibitem [{\citenamefont {Silvestre-Brac}\ and\ \citenamefont
  {Semay}(1993)}]{SilvestreBrac:1993ss}%
  \BibitemOpen
  \bibfield  {author} {\bibinfo {author} {\bibfnamefont {B.}~\bibnamefont
  {Silvestre-Brac}}\ and\ \bibinfo {author} {\bibfnamefont {C.}~\bibnamefont
  {Semay}},\ }\href {\doibase 10.1007/BF01565058} {\bibfield  {journal}
  {\bibinfo  {journal} {Z. Phys.}\ }\textbf {\bibinfo {volume} {C57}},\
  \bibinfo {pages} {273} (\bibinfo {year} {1993})}\BibitemShut {NoStop}%
\bibitem [{\citenamefont {Semay}\ and\ \citenamefont
  {Silvestre-Brac}(1994)}]{Semay:1994ht}%
  \BibitemOpen
  \bibfield  {author} {\bibinfo {author} {\bibfnamefont {C.}~\bibnamefont
  {Semay}}\ and\ \bibinfo {author} {\bibfnamefont {B.}~\bibnamefont
  {Silvestre-Brac}},\ }\href {\doibase 10.1007/BF01413104} {\bibfield
  {journal} {\bibinfo  {journal} {Z. Phys.}\ }\textbf {\bibinfo {volume}
  {C61}},\ \bibinfo {pages} {271} (\bibinfo {year} {1994})}\BibitemShut
  {NoStop}%
\bibitem [{\citenamefont {Pepin}\ \emph {et~al.}(1997)\citenamefont {Pepin},
  \citenamefont {Stancu}, \citenamefont {Genovese},\ and\ \citenamefont
  {Richard}}]{Pepin:1996id}%
  \BibitemOpen
  \bibfield  {author} {\bibinfo {author} {\bibfnamefont {S.}~\bibnamefont
  {Pepin}}, \bibinfo {author} {\bibfnamefont {F.}~\bibnamefont {Stancu}},
  \bibinfo {author} {\bibfnamefont {M.}~\bibnamefont {Genovese}}, \ and\
  \bibinfo {author} {\bibfnamefont {J.~M.}\ \bibnamefont {Richard}},\ }\href
  {\doibase 10.1016/S0370-2693(96)01597-3} {\bibfield  {journal} {\bibinfo
  {journal} {Phys. Lett.}\ }\textbf {\bibinfo {volume} {B393}},\ \bibinfo
  {pages} {119} (\bibinfo {year} {1997})},\ \Eprint
  {http://arxiv.org/abs/hep-ph/9609348} {arXiv:hep-ph/9609348 [hep-ph]}
  \BibitemShut {NoStop}%
\bibitem [{\citenamefont {Brink}\ and\ \citenamefont
  {Stancu}(1998)}]{Brink:1998as}%
  \BibitemOpen
  \bibfield  {author} {\bibinfo {author} {\bibfnamefont {D.~M.}\ \bibnamefont
  {Brink}}\ and\ \bibinfo {author} {\bibfnamefont {F.}~\bibnamefont {Stancu}},\
  }\href {\doibase 10.1103/PhysRevD.57.6778} {\bibfield  {journal} {\bibinfo
  {journal} {Phys. Rev.}\ }\textbf {\bibinfo {volume} {D57}},\ \bibinfo {pages}
  {6778} (\bibinfo {year} {1998})}\BibitemShut {NoStop}%
\bibitem [{\citenamefont {Barnes}\ \emph {et~al.}(1999)\citenamefont {Barnes},
  \citenamefont {Black}, \citenamefont {Dean},\ and\ \citenamefont
  {Swanson}}]{Barnes:1999hs}%
  \BibitemOpen
  \bibfield  {author} {\bibinfo {author} {\bibfnamefont {T.}~\bibnamefont
  {Barnes}}, \bibinfo {author} {\bibfnamefont {N.}~\bibnamefont {Black}},
  \bibinfo {author} {\bibfnamefont {D.~J.}\ \bibnamefont {Dean}}, \ and\
  \bibinfo {author} {\bibfnamefont {E.~S.}\ \bibnamefont {Swanson}},\ }\href
  {\doibase 10.1103/PhysRevC.60.045202} {\bibfield  {journal} {\bibinfo
  {journal} {Phys. Rev.}\ }\textbf {\bibinfo {volume} {C60}},\ \bibinfo {pages}
  {045202} (\bibinfo {year} {1999})},\ \Eprint
  {http://arxiv.org/abs/nucl-th/9902068} {arXiv:nucl-th/9902068 [nucl-th]}
  \BibitemShut {NoStop}%
\bibitem [{\citenamefont {Gelman}\ and\ \citenamefont
  {Nussinov}(2003)}]{Gelman:2002wf}%
  \BibitemOpen
  \bibfield  {author} {\bibinfo {author} {\bibfnamefont {B.~A.}\ \bibnamefont
  {Gelman}}\ and\ \bibinfo {author} {\bibfnamefont {S.}~\bibnamefont
  {Nussinov}},\ }\href {\doibase 10.1016/S0370-2693(02)03069-1} {\bibfield
  {journal} {\bibinfo  {journal} {Phys. Lett.}\ }\textbf {\bibinfo {volume}
  {B551}},\ \bibinfo {pages} {296} (\bibinfo {year} {2003})},\ \Eprint
  {http://arxiv.org/abs/hep-ph/0209095} {arXiv:hep-ph/0209095 [hep-ph]}
  \BibitemShut {NoStop}%
\bibitem [{\citenamefont {Vijande}\ \emph {et~al.}(2004)\citenamefont
  {Vijande}, \citenamefont {Fernandez}, \citenamefont {Valcarce},\ and\
  \citenamefont {Silvestre-Brac}}]{Vijande:2003ki}%
  \BibitemOpen
  \bibfield  {author} {\bibinfo {author} {\bibfnamefont {J.}~\bibnamefont
  {Vijande}}, \bibinfo {author} {\bibfnamefont {F.}~\bibnamefont {Fernandez}},
  \bibinfo {author} {\bibfnamefont {A.}~\bibnamefont {Valcarce}}, \ and\
  \bibinfo {author} {\bibfnamefont {B.}~\bibnamefont {Silvestre-Brac}},\ }\href
  {\doibase 10.1140/epja/i2003-10128-9} {\bibfield  {journal} {\bibinfo
  {journal} {Eur. Phys. J.}\ }\textbf {\bibinfo {volume} {A19}},\ \bibinfo
  {pages} {383} (\bibinfo {year} {2004})},\ \Eprint
  {http://arxiv.org/abs/hep-ph/0310007} {arXiv:hep-ph/0310007 [hep-ph]}
  \BibitemShut {NoStop}%
\bibitem [{\citenamefont {Janc}\ and\ \citenamefont
  {Rosina}(2004)}]{Janc:2004qn}%
  \BibitemOpen
  \bibfield  {author} {\bibinfo {author} {\bibfnamefont {D.}~\bibnamefont
  {Janc}}\ and\ \bibinfo {author} {\bibfnamefont {M.}~\bibnamefont {Rosina}},\
  }\href {\doibase 10.1007/s00601-004-0068-9} {\bibfield  {journal} {\bibinfo
  {journal} {Few Body Syst.}\ }\textbf {\bibinfo {volume} {35}},\ \bibinfo
  {pages} {175} (\bibinfo {year} {2004})},\ \Eprint
  {http://arxiv.org/abs/hep-ph/0405208} {arXiv:hep-ph/0405208 [hep-ph]}
  \BibitemShut {NoStop}%
\bibitem [{\citenamefont {Ebert}\ \emph {et~al.}(2007)\citenamefont {Ebert},
  \citenamefont {Faustov}, \citenamefont {Galkin},\ and\ \citenamefont
  {Lucha}}]{Ebert:2007rn}%
  \BibitemOpen
  \bibfield  {author} {\bibinfo {author} {\bibfnamefont {D.}~\bibnamefont
  {Ebert}}, \bibinfo {author} {\bibfnamefont {R.~N.}\ \bibnamefont {Faustov}},
  \bibinfo {author} {\bibfnamefont {V.~O.}\ \bibnamefont {Galkin}}, \ and\
  \bibinfo {author} {\bibfnamefont {W.}~\bibnamefont {Lucha}},\ }\href
  {\doibase 10.1103/PhysRevD.76.114015} {\bibfield  {journal} {\bibinfo
  {journal} {Phys. Rev.}\ }\textbf {\bibinfo {volume} {D76}},\ \bibinfo {pages}
  {114015} (\bibinfo {year} {2007})},\ \Eprint {http://arxiv.org/abs/0706.3853}
  {arXiv:0706.3853 [hep-ph]} \BibitemShut {NoStop}%
\bibitem [{\citenamefont {Vijande}\ \emph {et~al.}(2007)\citenamefont
  {Vijande}, \citenamefont {Weissman}, \citenamefont {Valcarce},\ and\
  \citenamefont {Barnea}}]{Vijande:2007rf}%
  \BibitemOpen
  \bibfield  {author} {\bibinfo {author} {\bibfnamefont {J.}~\bibnamefont
  {Vijande}}, \bibinfo {author} {\bibfnamefont {E.}~\bibnamefont {Weissman}},
  \bibinfo {author} {\bibfnamefont {A.}~\bibnamefont {Valcarce}}, \ and\
  \bibinfo {author} {\bibfnamefont {N.}~\bibnamefont {Barnea}},\ }\href
  {\doibase 10.1103/PhysRevD.76.094027} {\bibfield  {journal} {\bibinfo
  {journal} {Phys. Rev.}\ }\textbf {\bibinfo {volume} {D76}},\ \bibinfo {pages}
  {094027} (\bibinfo {year} {2007})},\ \Eprint {http://arxiv.org/abs/0710.2516}
  {arXiv:0710.2516 [hep-ph]} \BibitemShut {NoStop}%
\bibitem [{\citenamefont {Zhang}\ \emph {et~al.}(2008)\citenamefont {Zhang},
  \citenamefont {Zhang},\ and\ \citenamefont {Zhang}}]{Zhang:2007mu}%
  \BibitemOpen
  \bibfield  {author} {\bibinfo {author} {\bibfnamefont {M.}~\bibnamefont
  {Zhang}}, \bibinfo {author} {\bibfnamefont {H.~X.}\ \bibnamefont {Zhang}}, \
  and\ \bibinfo {author} {\bibfnamefont {Z.~Y.}\ \bibnamefont {Zhang}},\ }\href
  {\doibase 10.1088/0253-6102/50/2/31} {\bibfield  {journal} {\bibinfo
  {journal} {Commun. Theor. Phys.}\ }\textbf {\bibinfo {volume} {50}},\
  \bibinfo {pages} {437} (\bibinfo {year} {2008})},\ \Eprint
  {http://arxiv.org/abs/0711.1029} {arXiv:0711.1029 [nucl-th]} \BibitemShut
  {NoStop}%
\bibitem [{\citenamefont {Lee}\ and\ \citenamefont {Yasui}(2009)}]{Lee:2009rt}%
  \BibitemOpen
  \bibfield  {author} {\bibinfo {author} {\bibfnamefont {S.~H.}\ \bibnamefont
  {Lee}}\ and\ \bibinfo {author} {\bibfnamefont {S.}~\bibnamefont {Yasui}},\
  }\href {\doibase 10.1140/epjc/s10052-009-1140-x} {\bibfield  {journal}
  {\bibinfo  {journal} {Eur. Phys. J.}\ }\textbf {\bibinfo {volume} {C64}},\
  \bibinfo {pages} {283} (\bibinfo {year} {2009})},\ \Eprint
  {http://arxiv.org/abs/0901.2977} {arXiv:0901.2977 [hep-ph]} \BibitemShut
  {NoStop}%
\bibitem [{\citenamefont {Vijande}\ \emph {et~al.}(2009)\citenamefont
  {Vijande}, \citenamefont {Valcarce},\ and\ \citenamefont
  {Barnea}}]{Vijande:2009kj}%
  \BibitemOpen
  \bibfield  {author} {\bibinfo {author} {\bibfnamefont {J.}~\bibnamefont
  {Vijande}}, \bibinfo {author} {\bibfnamefont {A.}~\bibnamefont {Valcarce}}, \
  and\ \bibinfo {author} {\bibfnamefont {N.}~\bibnamefont {Barnea}},\ }\href
  {\doibase 10.1103/PhysRevD.79.074010} {\bibfield  {journal} {\bibinfo
  {journal} {Phys. Rev.}\ }\textbf {\bibinfo {volume} {D79}},\ \bibinfo {pages}
  {074010} (\bibinfo {year} {2009})},\ \Eprint {http://arxiv.org/abs/0903.2949}
  {arXiv:0903.2949 [hep-ph]} \BibitemShut {NoStop}%
\bibitem [{\citenamefont {Yang}\ \emph {et~al.}(2009)\citenamefont {Yang},
  \citenamefont {Deng}, \citenamefont {Ping},\ and\ \citenamefont
  {Goldman}}]{Yang:2009zzp}%
  \BibitemOpen
  \bibfield  {author} {\bibinfo {author} {\bibfnamefont {Y.}~\bibnamefont
  {Yang}}, \bibinfo {author} {\bibfnamefont {C.}~\bibnamefont {Deng}}, \bibinfo
  {author} {\bibfnamefont {J.}~\bibnamefont {Ping}}, \ and\ \bibinfo {author}
  {\bibfnamefont {T.}~\bibnamefont {Goldman}},\ }\href {\doibase
  10.1103/PhysRevD.80.114023} {\bibfield  {journal} {\bibinfo  {journal} {Phys.
  Rev.}\ }\textbf {\bibinfo {volume} {D80}},\ \bibinfo {pages} {114023}
  (\bibinfo {year} {2009})}\BibitemShut {NoStop}%
\bibitem [{\citenamefont {Valcarce}\ \emph {et~al.}(2011)\citenamefont
  {Valcarce}, \citenamefont {Vijande},\ and\ \citenamefont
  {Carames}}]{Valcarce:2010zs}%
  \BibitemOpen
  \bibfield  {author} {\bibinfo {author} {\bibfnamefont {A.}~\bibnamefont
  {Valcarce}}, \bibinfo {author} {\bibfnamefont {J.}~\bibnamefont {Vijande}}, \
  and\ \bibinfo {author} {\bibfnamefont {T.~F.}\ \bibnamefont {Carames}},\
  }\bibfield  {booktitle} {\emph {\bibinfo {booktitle} {{Proceedings, 4th
  International Workshop on Charm Physics (Charm 2010)}}},\ }\href {\doibase
  10.1142/S2010194511000766} {\bibfield  {journal} {\bibinfo  {journal} {Int.
  J. Mod. Phys. Conf. Ser.}\ }\textbf {\bibinfo {volume} {2}},\ \bibinfo
  {pages} {173} (\bibinfo {year} {2011})},\ \Eprint
  {http://arxiv.org/abs/1012.4627} {arXiv:1012.4627 [hep-ph]} \BibitemShut
  {NoStop}%
\bibitem [{\citenamefont {Carames}\ \emph {et~al.}(2011)\citenamefont
  {Carames}, \citenamefont {Valcarce},\ and\ \citenamefont
  {Vijande}}]{Carames:2011zz}%
  \BibitemOpen
  \bibfield  {author} {\bibinfo {author} {\bibfnamefont {T.~F.}\ \bibnamefont
  {Carames}}, \bibinfo {author} {\bibfnamefont {A.}~\bibnamefont {Valcarce}}, \
  and\ \bibinfo {author} {\bibfnamefont {J.}~\bibnamefont {Vijande}},\ }\href
  {\doibase 10.1016/j.physletb.2011.04.023} {\bibfield  {journal} {\bibinfo
  {journal} {Phys. Lett.}\ }\textbf {\bibinfo {volume} {B699}},\ \bibinfo
  {pages} {291} (\bibinfo {year} {2011})}\BibitemShut {NoStop}%
\bibitem [{\citenamefont {Ohkoda}\ \emph {et~al.}(2012)\citenamefont {Ohkoda},
  \citenamefont {Yamaguchi}, \citenamefont {Yasui}, \citenamefont {Sudoh},\
  and\ \citenamefont {Hosaka}}]{Ohkoda:2012hv}%
  \BibitemOpen
  \bibfield  {author} {\bibinfo {author} {\bibfnamefont {S.}~\bibnamefont
  {Ohkoda}}, \bibinfo {author} {\bibfnamefont {Y.}~\bibnamefont {Yamaguchi}},
  \bibinfo {author} {\bibfnamefont {S.}~\bibnamefont {Yasui}}, \bibinfo
  {author} {\bibfnamefont {K.}~\bibnamefont {Sudoh}}, \ and\ \bibinfo {author}
  {\bibfnamefont {A.}~\bibnamefont {Hosaka}},\ }\href {\doibase
  10.1103/PhysRevD.86.034019} {\bibfield  {journal} {\bibinfo  {journal} {Phys.
  Rev.}\ }\textbf {\bibinfo {volume} {D86}},\ \bibinfo {pages} {034019}
  (\bibinfo {year} {2012})},\ \Eprint {http://arxiv.org/abs/1202.0760}
  {arXiv:1202.0760 [hep-ph]} \BibitemShut {NoStop}%
\bibitem [{\citenamefont {Hyodo}\ \emph {et~al.}(2013)\citenamefont {Hyodo},
  \citenamefont {Liu}, \citenamefont {Oka}, \citenamefont {Sudoh},\ and\
  \citenamefont {Yasui}}]{Hyodo:2012pm}%
  \BibitemOpen
  \bibfield  {author} {\bibinfo {author} {\bibfnamefont {T.}~\bibnamefont
  {Hyodo}}, \bibinfo {author} {\bibfnamefont {Y.-R.}\ \bibnamefont {Liu}},
  \bibinfo {author} {\bibfnamefont {M.}~\bibnamefont {Oka}}, \bibinfo {author}
  {\bibfnamefont {K.}~\bibnamefont {Sudoh}}, \ and\ \bibinfo {author}
  {\bibfnamefont {S.}~\bibnamefont {Yasui}},\ }\href {\doibase
  10.1016/j.physletb.2013.02.045} {\bibfield  {journal} {\bibinfo  {journal}
  {Phys. Lett.}\ }\textbf {\bibinfo {volume} {B721}},\ \bibinfo {pages} {56}
  (\bibinfo {year} {2013})},\ \Eprint {http://arxiv.org/abs/1209.6207}
  {arXiv:1209.6207 [hep-ph]} \BibitemShut {NoStop}%
\bibitem [{\citenamefont {Silbar}\ and\ \citenamefont
  {Goldman}(2014)}]{Silbar:2013dda}%
  \BibitemOpen
  \bibfield  {author} {\bibinfo {author} {\bibfnamefont {R.~R.}\ \bibnamefont
  {Silbar}}\ and\ \bibinfo {author} {\bibfnamefont {T.}~\bibnamefont
  {Goldman}},\ }\href {\doibase 10.1142/S0218301314500918} {\bibfield
  {journal} {\bibinfo  {journal} {Int. J. Mod. Phys.}\ }\textbf {\bibinfo
  {volume} {E23}},\ \bibinfo {pages} {1450091} (\bibinfo {year} {2014})},\
  \Eprint {http://arxiv.org/abs/1304.5480} {arXiv:1304.5480 [nucl-th]}
  \BibitemShut {NoStop}%
\bibitem [{\citenamefont {Karliner}\ and\ \citenamefont
  {Rosner}(2017)}]{Karliner:2017qjm}%
  \BibitemOpen
  \bibfield  {author} {\bibinfo {author} {\bibfnamefont {M.}~\bibnamefont
  {Karliner}}\ and\ \bibinfo {author} {\bibfnamefont {J.~L.}\ \bibnamefont
  {Rosner}},\ }\href {\doibase 10.1103/PhysRevLett.119.202001} {\bibfield
  {journal} {\bibinfo  {journal} {Phys. Rev. Lett.}\ }\textbf {\bibinfo
  {volume} {119}},\ \bibinfo {pages} {202001} (\bibinfo {year} {2017})},\
  \Eprint {http://arxiv.org/abs/1707.07666} {arXiv:1707.07666 [hep-ph]}
  \BibitemShut {NoStop}%
\bibitem [{\citenamefont {Eichten}\ and\ \citenamefont
  {Quigg}(2017)}]{Eichten:2017ffp}%
  \BibitemOpen
  \bibfield  {author} {\bibinfo {author} {\bibfnamefont {E.~J.}\ \bibnamefont
  {Eichten}}\ and\ \bibinfo {author} {\bibfnamefont {C.}~\bibnamefont
  {Quigg}},\ }\href {\doibase 10.1103/PhysRevLett.119.202002} {\bibfield
  {journal} {\bibinfo  {journal} {Phys. Rev. Lett.}\ }\textbf {\bibinfo
  {volume} {119}},\ \bibinfo {pages} {202002} (\bibinfo {year} {2017})},\
  \Eprint {http://arxiv.org/abs/1707.09575} {arXiv:1707.09575 [hep-ph]}
  \BibitemShut {NoStop}%
\bibitem [{\citenamefont {Du}\ \emph {et~al.}(2013)\citenamefont {Du},
  \citenamefont {Chen}, \citenamefont {Chen},\ and\ \citenamefont
  {Zhu}}]{Du:2012wp}%
  \BibitemOpen
  \bibfield  {author} {\bibinfo {author} {\bibfnamefont {M.-L.}\ \bibnamefont
  {Du}}, \bibinfo {author} {\bibfnamefont {W.}~\bibnamefont {Chen}}, \bibinfo
  {author} {\bibfnamefont {X.-L.}\ \bibnamefont {Chen}}, \ and\ \bibinfo
  {author} {\bibfnamefont {S.-L.}\ \bibnamefont {Zhu}},\ }\href {\doibase
  10.1103/PhysRevD.87.014003} {\bibfield  {journal} {\bibinfo  {journal} {Phys.
  Rev.}\ }\textbf {\bibinfo {volume} {D87}},\ \bibinfo {pages} {014003}
  (\bibinfo {year} {2013})},\ \Eprint {http://arxiv.org/abs/1209.5134}
  {arXiv:1209.5134 [hep-ph]} \BibitemShut {NoStop}%
\bibitem [{\citenamefont {Chen}\ \emph {et~al.}(2014)\citenamefont {Chen},
  \citenamefont {Steele},\ and\ \citenamefont {Zhu}}]{Chen:2013aba}%
  \BibitemOpen
  \bibfield  {author} {\bibinfo {author} {\bibfnamefont {W.}~\bibnamefont
  {Chen}}, \bibinfo {author} {\bibfnamefont {T.~G.}\ \bibnamefont {Steele}}, \
  and\ \bibinfo {author} {\bibfnamefont {S.-L.}\ \bibnamefont {Zhu}},\ }\href
  {\doibase 10.1103/PhysRevD.89.054037} {\bibfield  {journal} {\bibinfo
  {journal} {Phys. Rev.}\ }\textbf {\bibinfo {volume} {D89}},\ \bibinfo {pages}
  {054037} (\bibinfo {year} {2014})},\ \Eprint {http://arxiv.org/abs/1310.8337}
  {arXiv:1310.8337 [hep-ph]} \BibitemShut {NoStop}%
\bibitem [{\citenamefont {Bicudo}\ \emph {et~al.}(2015)\citenamefont {Bicudo},
  \citenamefont {Cichy}, \citenamefont {Peters}, \citenamefont {Wagenbach},\
  and\ \citenamefont {Wagner}}]{Bicudo:2015vta}%
  \BibitemOpen
  \bibfield  {author} {\bibinfo {author} {\bibfnamefont {P.}~\bibnamefont
  {Bicudo}}, \bibinfo {author} {\bibfnamefont {K.}~\bibnamefont {Cichy}},
  \bibinfo {author} {\bibfnamefont {A.}~\bibnamefont {Peters}}, \bibinfo
  {author} {\bibfnamefont {B.}~\bibnamefont {Wagenbach}}, \ and\ \bibinfo
  {author} {\bibfnamefont {M.}~\bibnamefont {Wagner}},\ }\href {\doibase
  10.1103/PhysRevD.92.014507} {\bibfield  {journal} {\bibinfo  {journal} {Phys.
  Rev.}\ }\textbf {\bibinfo {volume} {D92}},\ \bibinfo {pages} {014507}
  (\bibinfo {year} {2015})},\ \Eprint {http://arxiv.org/abs/1505.00613}
  {arXiv:1505.00613 [hep-lat]} \BibitemShut {NoStop}%
\bibitem [{\citenamefont {Francis}\ \emph {et~al.}(2017)\citenamefont
  {Francis}, \citenamefont {Hudspith}, \citenamefont {Lewis},\ and\
  \citenamefont {Maltman}}]{Francis:2016hui}%
  \BibitemOpen
  \bibfield  {author} {\bibinfo {author} {\bibfnamefont {A.}~\bibnamefont
  {Francis}}, \bibinfo {author} {\bibfnamefont {R.~J.}\ \bibnamefont
  {Hudspith}}, \bibinfo {author} {\bibfnamefont {R.}~\bibnamefont {Lewis}}, \
  and\ \bibinfo {author} {\bibfnamefont {K.}~\bibnamefont {Maltman}},\ }\href
  {\doibase 10.1103/PhysRevLett.118.142001} {\bibfield  {journal} {\bibinfo
  {journal} {Phys. Rev. Lett.}\ }\textbf {\bibinfo {volume} {118}},\ \bibinfo
  {pages} {142001} (\bibinfo {year} {2017})},\ \Eprint
  {http://arxiv.org/abs/1607.05214} {arXiv:1607.05214 [hep-lat]} \BibitemShut
  {NoStop}%
\bibitem [{\citenamefont {Jaffe}(2005)}]{Jaffe:2004ph}%
  \BibitemOpen
  \bibfield  {author} {\bibinfo {author} {\bibfnamefont {R.~L.}\ \bibnamefont
  {Jaffe}},\ }\bibfield  {booktitle} {\emph {\bibinfo {booktitle}
  {{Proceedings, 6th International Conference on Hyperons, charm and beauty
  hadrons (BEACH 2004)}}},\ }\href {\doibase 10.1016/j.physrep.2004.11.005}
  {\bibfield  {journal} {\bibinfo  {journal} {Phys. Rept.}\ }\textbf {\bibinfo
  {volume} {409}},\ \bibinfo {pages} {1} (\bibinfo {year} {2005})},\ \Eprint
  {http://arxiv.org/abs/hep-ph/0409065} {arXiv:hep-ph/0409065 [hep-ph]}
  \BibitemShut {NoStop}%
\bibitem [{\citenamefont {Richards}\ \emph {et~al.}(1990)\citenamefont
  {Richards}, \citenamefont {Sinclair},\ and\ \citenamefont
  {Sivers}}]{Richards:1990xf}%
  \BibitemOpen
  \bibfield  {author} {\bibinfo {author} {\bibfnamefont {D.~G.}\ \bibnamefont
  {Richards}}, \bibinfo {author} {\bibfnamefont {D.~K.}\ \bibnamefont
  {Sinclair}}, \ and\ \bibinfo {author} {\bibfnamefont {D.~W.}\ \bibnamefont
  {Sivers}},\ }\href {\doibase 10.1103/PhysRevD.42.3191} {\bibfield  {journal}
  {\bibinfo  {journal} {Phys. Rev.}\ }\textbf {\bibinfo {volume} {D42}},\
  \bibinfo {pages} {3191} (\bibinfo {year} {1990})}\BibitemShut {NoStop}%
\bibitem [{\citenamefont {Mihaly}\ \emph {et~al.}(1997)\citenamefont {Mihaly},
  \citenamefont {Fiebig}, \citenamefont {Markum},\ and\ \citenamefont
  {Rabitsch}}]{Mihaly:1996ue}%
  \BibitemOpen
  \bibfield  {author} {\bibinfo {author} {\bibfnamefont {A.}~\bibnamefont
  {Mihaly}}, \bibinfo {author} {\bibfnamefont {H.~R.}\ \bibnamefont {Fiebig}},
  \bibinfo {author} {\bibfnamefont {H.}~\bibnamefont {Markum}}, \ and\ \bibinfo
  {author} {\bibfnamefont {K.}~\bibnamefont {Rabitsch}},\ }\href {\doibase
  10.1103/PhysRevD.55.3077} {\bibfield  {journal} {\bibinfo  {journal} {Phys.
  Rev.}\ }\textbf {\bibinfo {volume} {D55}},\ \bibinfo {pages} {3077} (\bibinfo
  {year} {1997})}\BibitemShut {NoStop}%
\bibitem [{\citenamefont {Stewart}\ and\ \citenamefont
  {Koniuk}(1998)}]{Stewart:1998hk}%
  \BibitemOpen
  \bibfield  {author} {\bibinfo {author} {\bibfnamefont {C.}~\bibnamefont
  {Stewart}}\ and\ \bibinfo {author} {\bibfnamefont {R.}~\bibnamefont
  {Koniuk}},\ }\href {\doibase 10.1103/PhysRevD.57.5581} {\bibfield  {journal}
  {\bibinfo  {journal} {Phys. Rev.}\ }\textbf {\bibinfo {volume} {D57}},\
  \bibinfo {pages} {5581} (\bibinfo {year} {1998})},\ \Eprint
  {http://arxiv.org/abs/hep-lat/9803003} {arXiv:hep-lat/9803003 [hep-lat]}
  \BibitemShut {NoStop}%
\bibitem [{\citenamefont {Michael}\ and\ \citenamefont
  {Pennanen}(1999)}]{Michael:1999nq}%
  \BibitemOpen
  \bibfield  {author} {\bibinfo {author} {\bibfnamefont {C.}~\bibnamefont
  {Michael}}\ and\ \bibinfo {author} {\bibfnamefont {P.}~\bibnamefont
  {Pennanen}} (\bibinfo {collaboration} {UKQCD}),\ }\href {\doibase
  10.1103/PhysRevD.60.054012} {\bibfield  {journal} {\bibinfo  {journal} {Phys.
  Rev.}\ }\textbf {\bibinfo {volume} {D60}},\ \bibinfo {pages} {054012}
  (\bibinfo {year} {1999})},\ \Eprint {http://arxiv.org/abs/hep-lat/9901007}
  {arXiv:hep-lat/9901007 [hep-lat]} \BibitemShut {NoStop}%
\bibitem [{\citenamefont {Pennanen}\ \emph {et~al.}(2000)\citenamefont
  {Pennanen}, \citenamefont {Michael},\ and\ \citenamefont
  {Green}}]{Pennanen:1999xi}%
  \BibitemOpen
  \bibfield  {author} {\bibinfo {author} {\bibfnamefont {P.}~\bibnamefont
  {Pennanen}}, \bibinfo {author} {\bibfnamefont {C.}~\bibnamefont {Michael}}, \
  and\ \bibinfo {author} {\bibfnamefont {A.~M.}\ \bibnamefont {Green}},\
  }\bibfield  {booktitle} {\emph {\bibinfo {booktitle} {{Lattice field theory.
  Proceedings, 17th International Symposium, Lattice'99, Pisa, Italy, June
  29-July 3, 1999}}},\ }\href {\doibase 10.1016/S0920-5632(00)91622-0}
  {\bibfield  {journal} {\bibinfo  {journal} {Nucl. Phys. Proc. Suppl.}\
  }\textbf {\bibinfo {volume} {83}},\ \bibinfo {pages} {200} (\bibinfo {year}
  {2000})},\ \Eprint {http://arxiv.org/abs/hep-lat/9908032}
  {arXiv:hep-lat/9908032 [hep-lat]} \BibitemShut {NoStop}%
\bibitem [{\citenamefont {Cook}\ and\ \citenamefont
  {Fiebig}(2002)}]{Cook:2002am}%
  \BibitemOpen
  \bibfield  {author} {\bibinfo {author} {\bibfnamefont {M.~S.}\ \bibnamefont
  {Cook}}\ and\ \bibinfo {author} {\bibfnamefont {H.~R.}\ \bibnamefont
  {Fiebig}},\ }\href@noop {} {\  (\bibinfo {year} {2002})},\ \Eprint
  {http://arxiv.org/abs/hep-lat/0210054} {arXiv:hep-lat/0210054 [hep-lat]}
  \BibitemShut {NoStop}%
\bibitem [{\citenamefont {Detmold}\ \emph {et~al.}(2007)\citenamefont
  {Detmold}, \citenamefont {Orginos},\ and\ \citenamefont
  {Savage}}]{Detmold:2007wk}%
  \BibitemOpen
  \bibfield  {author} {\bibinfo {author} {\bibfnamefont {W.}~\bibnamefont
  {Detmold}}, \bibinfo {author} {\bibfnamefont {K.}~\bibnamefont {Orginos}}, \
  and\ \bibinfo {author} {\bibfnamefont {M.~J.}\ \bibnamefont {Savage}},\
  }\href {\doibase 10.1103/PhysRevD.76.114503} {\bibfield  {journal} {\bibinfo
  {journal} {Phys. Rev.}\ }\textbf {\bibinfo {volume} {D76}},\ \bibinfo {pages}
  {114503} (\bibinfo {year} {2007})},\ \Eprint
  {http://arxiv.org/abs/hep-lat/0703009} {arXiv:hep-lat/0703009 [HEP-LAT]}
  \BibitemShut {NoStop}%
\bibitem [{\citenamefont {Bali}\ and\ \citenamefont
  {Hetzenegger}(2011)}]{Bali:2011gq}%
  \BibitemOpen
  \bibfield  {author} {\bibinfo {author} {\bibfnamefont {G.}~\bibnamefont
  {Bali}}\ and\ \bibinfo {author} {\bibfnamefont {M.}~\bibnamefont
  {Hetzenegger}} (\bibinfo {collaboration} {QCDSF}),\ }\bibfield  {booktitle}
  {\emph {\bibinfo {booktitle} {{Proceedings, 29th International Symposium on
  Lattice field theory (Lattice 2011)}}},\ }\href@noop {} {\bibfield  {journal}
  {\bibinfo  {journal} {PoS}\ }\textbf {\bibinfo {volume} {LATTICE2011}},\
  \bibinfo {pages} {123} (\bibinfo {year} {2011})},\ \Eprint
  {http://arxiv.org/abs/1111.2222} {arXiv:1111.2222 [hep-lat]} \BibitemShut
  {NoStop}%
\bibitem [{\citenamefont {Wagner}(2011)}]{Wagner:2011ev}%
  \BibitemOpen
  \bibfield  {author} {\bibinfo {author} {\bibfnamefont {M.}~\bibnamefont
  {Wagner}} (\bibinfo {collaboration} {ETM}),\ }\bibfield  {booktitle} {\emph
  {\bibinfo {booktitle} {{Proceedings, 3rd Workshop on Excited QCD 2011}}},\
  }\href {\doibase 10.5506/APhysPolBSupp.4.747} {\bibfield  {journal} {\bibinfo
   {journal} {Acta Phys. Polon. Supp.}\ }\textbf {\bibinfo {volume} {4}},\
  \bibinfo {pages} {747} (\bibinfo {year} {2011})},\ \Eprint
  {http://arxiv.org/abs/1103.5147} {arXiv:1103.5147 [hep-lat]} \BibitemShut
  {NoStop}%
\bibitem [{\citenamefont {Cheung}\ \emph {et~al.}(2017)\citenamefont {Cheung},
  \citenamefont {Thomas}, \citenamefont {Dudek},\ and\ \citenamefont
  {Edwards}}]{Cheung:2017tnt}%
  \BibitemOpen
  \bibfield  {author} {\bibinfo {author} {\bibfnamefont {G.~K.~C.}\
  \bibnamefont {Cheung}}, \bibinfo {author} {\bibfnamefont {C.~E.}\
  \bibnamefont {Thomas}}, \bibinfo {author} {\bibfnamefont {J.~J.}\
  \bibnamefont {Dudek}}, \ and\ \bibinfo {author} {\bibfnamefont {R.~G.}\
  \bibnamefont {Edwards}} (\bibinfo {collaboration} {Hadron Spectrum}),\ }\href
  {\doibase 10.1007/JHEP11(2017)033} {\bibfield  {journal} {\bibinfo  {journal}
  {JHEP}\ }\textbf {\bibinfo {volume} {11}},\ \bibinfo {pages} {033} (\bibinfo
  {year} {2017})},\ \Eprint {http://arxiv.org/abs/1709.01417} {arXiv:1709.01417
  [hep-lat]} \BibitemShut {NoStop}%
\bibitem [{\citenamefont {Brown}\ and\ \citenamefont
  {Orginos}(2012)}]{Brown:2012tm}%
  \BibitemOpen
  \bibfield  {author} {\bibinfo {author} {\bibfnamefont {Z.~S.}\ \bibnamefont
  {Brown}}\ and\ \bibinfo {author} {\bibfnamefont {K.}~\bibnamefont
  {Orginos}},\ }\href {\doibase 10.1103/PhysRevD.86.114506} {\bibfield
  {journal} {\bibinfo  {journal} {Phys. Rev.}\ }\textbf {\bibinfo {volume}
  {D86}},\ \bibinfo {pages} {114506} (\bibinfo {year} {2012})},\ \Eprint
  {http://arxiv.org/abs/1210.1953} {arXiv:1210.1953 [hep-lat]} \BibitemShut
  {NoStop}%
\bibitem [{\citenamefont {Bicudo}\ and\ \citenamefont
  {Wagner}(2013)}]{Bicudo:2012qt}%
  \BibitemOpen
  \bibfield  {author} {\bibinfo {author} {\bibfnamefont {P.}~\bibnamefont
  {Bicudo}}\ and\ \bibinfo {author} {\bibfnamefont {M.}~\bibnamefont {Wagner}}
  (\bibinfo {collaboration} {ETM}),\ }\href {\doibase
  10.1103/PhysRevD.87.114511} {\bibfield  {journal} {\bibinfo  {journal} {Phys.
  Rev.}\ }\textbf {\bibinfo {volume} {D87}},\ \bibinfo {pages} {114511}
  (\bibinfo {year} {2013})},\ \Eprint {http://arxiv.org/abs/1209.6274}
  {arXiv:1209.6274 [hep-ph]} \BibitemShut {NoStop}%
\bibitem [{\citenamefont {Bicudo}\ \emph {et~al.}(2016)\citenamefont {Bicudo},
  \citenamefont {Scheunert},\ and\ \citenamefont {Wagner}}]{Bicudo:2016jwl}%
  \BibitemOpen
  \bibfield  {author} {\bibinfo {author} {\bibfnamefont {P.}~\bibnamefont
  {Bicudo}}, \bibinfo {author} {\bibfnamefont {J.}~\bibnamefont {Scheunert}}, \
  and\ \bibinfo {author} {\bibfnamefont {M.}~\bibnamefont {Wagner}},\
  }\bibfield  {booktitle} {\emph {\bibinfo {booktitle} {{Proceedings, 34th
  International Symposium on Lattice Field Theory (Lattice 2016): Southampton,
  UK, July 24-30, 2016}}},\ }\href@noop {} {\bibfield  {journal} {\bibinfo
  {journal} {PoS}\ }\textbf {\bibinfo {volume} {LATTICE2016}},\ \bibinfo
  {pages} {103} (\bibinfo {year} {2016})},\ \Eprint
  {http://arxiv.org/abs/1609.00548} {arXiv:1609.00548 [hep-lat]} \BibitemShut
  {NoStop}%
\bibitem [{\citenamefont {Aaij}\ \emph {et~al.}(2017)\citenamefont {Aaij} \emph
  {et~al.}}]{Aaij:2017ueg}%
  \BibitemOpen
  \bibfield  {author} {\bibinfo {author} {\bibfnamefont {R.}~\bibnamefont
  {Aaij}} \emph {et~al.} (\bibinfo {collaboration} {LHCb}),\ }\href {\doibase
  10.1103/PhysRevLett.119.112001} {\bibfield  {journal} {\bibinfo  {journal}
  {Phys. Rev. Lett.}\ }\textbf {\bibinfo {volume} {119}},\ \bibinfo {pages}
  {112001} (\bibinfo {year} {2017})},\ \Eprint
  {http://arxiv.org/abs/1707.01621} {arXiv:1707.01621 [hep-ex]} \BibitemShut
  {NoStop}%
\bibitem [{\citenamefont {Czarnecki}\ \emph {et~al.}(2018)\citenamefont
  {Czarnecki}, \citenamefont {Leng},\ and\ \citenamefont
  {Voloshin}}]{Czarnecki:2017vco}%
  \BibitemOpen
  \bibfield  {author} {\bibinfo {author} {\bibfnamefont {A.}~\bibnamefont
  {Czarnecki}}, \bibinfo {author} {\bibfnamefont {B.}~\bibnamefont {Leng}}, \
  and\ \bibinfo {author} {\bibfnamefont {M.~B.}\ \bibnamefont {Voloshin}},\
  }\href {\doibase 10.1016/j.physletb.2018.01.034} {\bibfield  {journal}
  {\bibinfo  {journal} {Phys. Lett.}\ }\textbf {\bibinfo {volume} {B778}},\
  \bibinfo {pages} {233} (\bibinfo {year} {2018})},\ \Eprint
  {http://arxiv.org/abs/1708.04594} {arXiv:1708.04594 [hep-ph]} \BibitemShut
  {NoStop}%
\bibitem [{\citenamefont {Mehen}(2017)}]{Mehen:2017nrh}%
  \BibitemOpen
  \bibfield  {author} {\bibinfo {author} {\bibfnamefont {T.}~\bibnamefont
  {Mehen}},\ }\href {\doibase 10.1103/PhysRevD.96.094028} {\bibfield  {journal}
  {\bibinfo  {journal} {Phys. Rev.}\ }\textbf {\bibinfo {volume} {D96}},\
  \bibinfo {pages} {094028} (\bibinfo {year} {2017})},\ \Eprint
  {http://arxiv.org/abs/1708.05020} {arXiv:1708.05020 [hep-ph]} \BibitemShut
  {NoStop}%
\bibitem [{\citenamefont {Heller}\ and\ \citenamefont
  {Tjon}(1985)}]{Heller:1985cb}%
  \BibitemOpen
  \bibfield  {author} {\bibinfo {author} {\bibfnamefont {L.}~\bibnamefont
  {Heller}}\ and\ \bibinfo {author} {\bibfnamefont {J.~A.}\ \bibnamefont
  {Tjon}},\ }\href {\doibase 10.1103/PhysRevD.32.755} {\bibfield  {journal}
  {\bibinfo  {journal} {Phys. Rev.}\ }\textbf {\bibinfo {volume} {D32}},\
  \bibinfo {pages} {755} (\bibinfo {year} {1985})}\BibitemShut {NoStop}%
\bibitem [{\citenamefont {Carlson}\ \emph {et~al.}(1988)\citenamefont
  {Carlson}, \citenamefont {Heller},\ and\ \citenamefont
  {Tjon}}]{Carlson:1987hh}%
  \BibitemOpen
  \bibfield  {author} {\bibinfo {author} {\bibfnamefont {J.}~\bibnamefont
  {Carlson}}, \bibinfo {author} {\bibfnamefont {L.}~\bibnamefont {Heller}}, \
  and\ \bibinfo {author} {\bibfnamefont {J.~A.}\ \bibnamefont {Tjon}},\ }\href
  {\doibase 10.1103/PhysRevD.37.744} {\bibfield  {journal} {\bibinfo  {journal}
  {Phys. Rev.}\ }\textbf {\bibinfo {volume} {D37}},\ \bibinfo {pages} {744}
  (\bibinfo {year} {1988})}\BibitemShut {NoStop}%
\bibitem [{\citenamefont {Manohar}\ and\ \citenamefont
  {Wise}(1993)}]{Manohar:1992nd}%
  \BibitemOpen
  \bibfield  {author} {\bibinfo {author} {\bibfnamefont {A.~V.}\ \bibnamefont
  {Manohar}}\ and\ \bibinfo {author} {\bibfnamefont {M.~B.}\ \bibnamefont
  {Wise}},\ }\href {\doibase 10.1016/0550-3213(93)90614-U} {\bibfield
  {journal} {\bibinfo  {journal} {Nucl. Phys.}\ }\textbf {\bibinfo {volume}
  {B399}},\ \bibinfo {pages} {17} (\bibinfo {year} {1993})},\ \Eprint
  {http://arxiv.org/abs/hep-ph/9212236} {arXiv:hep-ph/9212236 [hep-ph]}
  \BibitemShut {NoStop}%
\bibitem [{\citenamefont {Iritani}\ \emph {et~al.}(2017)\citenamefont
  {Iritani}, \citenamefont {Aoki}, \citenamefont {Doi}, \citenamefont
  {Hatsuda}, \citenamefont {Ikeda}, \citenamefont {Inoue}, \citenamefont
  {Ishii}, \citenamefont {Nemura},\ and\ \citenamefont
  {Sasaki}}]{Iritani:2017rlk}%
  \BibitemOpen
  \bibfield  {author} {\bibinfo {author} {\bibfnamefont {T.}~\bibnamefont
  {Iritani}}, \bibinfo {author} {\bibfnamefont {S.}~\bibnamefont {Aoki}},
  \bibinfo {author} {\bibfnamefont {T.}~\bibnamefont {Doi}}, \bibinfo {author}
  {\bibfnamefont {T.}~\bibnamefont {Hatsuda}}, \bibinfo {author} {\bibfnamefont
  {Y.}~\bibnamefont {Ikeda}}, \bibinfo {author} {\bibfnamefont
  {T.}~\bibnamefont {Inoue}}, \bibinfo {author} {\bibfnamefont
  {N.}~\bibnamefont {Ishii}}, \bibinfo {author} {\bibfnamefont
  {H.}~\bibnamefont {Nemura}}, \ and\ \bibinfo {author} {\bibfnamefont
  {K.}~\bibnamefont {Sasaki}},\ }\href {\doibase 10.1103/PhysRevD.96.034521}
  {\bibfield  {journal} {\bibinfo  {journal} {Phys. Rev.}\ }\textbf {\bibinfo
  {volume} {D96}},\ \bibinfo {pages} {034521} (\bibinfo {year} {2017})},\
  \Eprint {http://arxiv.org/abs/1703.07210} {arXiv:1703.07210 [hep-lat]}
  \BibitemShut {NoStop}%
\bibitem [{\citenamefont {Francis}\ \emph {et~al.}(2016)\citenamefont
  {Francis}, \citenamefont {Hudspith}, \citenamefont {Lewis},\ and\
  \citenamefont {Maltman}}]{Francis:2016nmj}%
  \BibitemOpen
  \bibfield  {author} {\bibinfo {author} {\bibfnamefont {A.}~\bibnamefont
  {Francis}}, \bibinfo {author} {\bibfnamefont {R.~J.}\ \bibnamefont
  {Hudspith}}, \bibinfo {author} {\bibfnamefont {R.}~\bibnamefont {Lewis}}, \
  and\ \bibinfo {author} {\bibfnamefont {K.}~\bibnamefont {Maltman}},\
  }\bibfield  {booktitle} {\emph {\bibinfo {booktitle} {{Proceedings, 34th
  International Symposium on Lattice Field Theory (Lattice 2016): Southampton,
  UK, July 24-30, 2016}}},\ }\href@noop {} {\bibfield  {journal} {\bibinfo
  {journal} {PoS}\ }\textbf {\bibinfo {volume} {LATTICE2016}},\ \bibinfo
  {pages} {132} (\bibinfo {year} {2016})}\BibitemShut {NoStop}%
\bibitem [{\citenamefont {Namekawa}\ \emph {et~al.}(2013)\citenamefont
  {Namekawa} \emph {et~al.}}]{Namekawa:2013vu}%
  \BibitemOpen
  \bibfield  {author} {\bibinfo {author} {\bibfnamefont {Y.}~\bibnamefont
  {Namekawa}} \emph {et~al.} (\bibinfo {collaboration} {PACS-CS}),\ }\href
  {\doibase 10.1103/PhysRevD.87.094512} {\bibfield  {journal} {\bibinfo
  {journal} {Phys. Rev.}\ }\textbf {\bibinfo {volume} {D87}},\ \bibinfo {pages}
  {094512} (\bibinfo {year} {2013})},\ \Eprint {http://arxiv.org/abs/1301.4743}
  {arXiv:1301.4743 [hep-lat]} \BibitemShut {NoStop}%
\bibitem [{\citenamefont {Hudspith}\ \emph {et~al.}(2017)\citenamefont
  {Hudspith}, \citenamefont {Francis}, \citenamefont {Lewis},\ and\
  \citenamefont {Maltman}}]{Hudspith:2017bbh}%
  \BibitemOpen
  \bibfield  {author} {\bibinfo {author} {\bibfnamefont {R.~J.}\ \bibnamefont
  {Hudspith}}, \bibinfo {author} {\bibfnamefont {A.}~\bibnamefont {Francis}},
  \bibinfo {author} {\bibfnamefont {R.}~\bibnamefont {Lewis}}, \ and\ \bibinfo
  {author} {\bibfnamefont {K.}~\bibnamefont {Maltman}},\ }\bibfield
  {booktitle} {\emph {\bibinfo {booktitle} {{Proceedings, 34th International
  Symposium on Lattice Field Theory (Lattice 2016): Southampton, UK, July
  24-30, 2016}}},\ }\href@noop {} {\bibfield  {journal} {\bibinfo  {journal}
  {PoS}\ }\textbf {\bibinfo {volume} {LATTICE2016}},\ \bibinfo {pages} {133}
  (\bibinfo {year} {2017})}\BibitemShut {NoStop}%
\bibitem [{\citenamefont {Aoki}\ \emph {et~al.}(2009)\citenamefont {Aoki} \emph
  {et~al.}}]{Aoki:2008sm}%
  \BibitemOpen
  \bibfield  {author} {\bibinfo {author} {\bibfnamefont {S.}~\bibnamefont
  {Aoki}} \emph {et~al.} (\bibinfo {collaboration} {PACS-CS}),\ }\href
  {\doibase 10.1103/PhysRevD.79.034503} {\bibfield  {journal} {\bibinfo
  {journal} {Phys. Rev.}\ }\textbf {\bibinfo {volume} {D79}},\ \bibinfo {pages}
  {034503} (\bibinfo {year} {2009})},\ \Eprint {http://arxiv.org/abs/0807.1661}
  {arXiv:0807.1661 [hep-lat]} \BibitemShut {NoStop}%
\bibitem [{\citenamefont {Sheikholeslami}\ and\ \citenamefont
  {Wohlert}(1985)}]{Sheikholeslami:1985ij}%
  \BibitemOpen
  \bibfield  {author} {\bibinfo {author} {\bibfnamefont {B.}~\bibnamefont
  {Sheikholeslami}}\ and\ \bibinfo {author} {\bibfnamefont {R.}~\bibnamefont
  {Wohlert}},\ }\href {\doibase 10.1016/0550-3213(85)90002-1} {\bibfield
  {journal} {\bibinfo  {journal} {Nucl. Phys.}\ }\textbf {\bibinfo {volume}
  {B259}},\ \bibinfo {pages} {572} (\bibinfo {year} {1985})}\BibitemShut
  {NoStop}%
\bibitem [{\citenamefont {Iwasaki}(1985)}]{Iwasaki:1985we}%
  \BibitemOpen
  \bibfield  {author} {\bibinfo {author} {\bibfnamefont {Y.}~\bibnamefont
  {Iwasaki}},\ }\href {\doibase 10.1016/0550-3213(85)90606-6} {\bibfield
  {journal} {\bibinfo  {journal} {Nucl. Phys.}\ }\textbf {\bibinfo {volume}
  {B258}},\ \bibinfo {pages} {141} (\bibinfo {year} {1985})}\BibitemShut
  {NoStop}%
\bibitem [{\citenamefont {Lang}\ \emph {et~al.}(2014)\citenamefont {Lang},
  \citenamefont {Leskovec}, \citenamefont {Mohler}, \citenamefont {Prelovsek},\
  and\ \citenamefont {Woloshyn}}]{Lang:2014yfa}%
  \BibitemOpen
  \bibfield  {author} {\bibinfo {author} {\bibfnamefont {C.~B.}\ \bibnamefont
  {Lang}}, \bibinfo {author} {\bibfnamefont {L.}~\bibnamefont {Leskovec}},
  \bibinfo {author} {\bibfnamefont {D.}~\bibnamefont {Mohler}}, \bibinfo
  {author} {\bibfnamefont {S.}~\bibnamefont {Prelovsek}}, \ and\ \bibinfo
  {author} {\bibfnamefont {R.~M.}\ \bibnamefont {Woloshyn}},\ }\href {\doibase
  10.1103/PhysRevD.90.034510} {\bibfield  {journal} {\bibinfo  {journal} {Phys.
  Rev.}\ }\textbf {\bibinfo {volume} {D90}},\ \bibinfo {pages} {034510}
  (\bibinfo {year} {2014})},\ \Eprint {http://arxiv.org/abs/1403.8103}
  {arXiv:1403.8103 [hep-lat]} \BibitemShut {NoStop}%
\bibitem [{\citenamefont {Hudspith}(2015)}]{Hudspith:2014oja}%
  \BibitemOpen
  \bibfield  {author} {\bibinfo {author} {\bibfnamefont {R.~J.}\ \bibnamefont
  {Hudspith}} (\bibinfo {collaboration} {RBC, UKQCD}),\ }\href {\doibase
  10.1016/j.cpc.2014.10.017} {\bibfield  {journal} {\bibinfo  {journal}
  {Comput. Phys. Commun.}\ }\textbf {\bibinfo {volume} {187}},\ \bibinfo
  {pages} {115} (\bibinfo {year} {2015})},\ \Eprint
  {http://arxiv.org/abs/1405.5812} {arXiv:1405.5812 [hep-lat]} \BibitemShut
  {NoStop}%
\bibitem [{\citenamefont {L{\"u}scher}(2005)}]{Luscher:2005rx}%
  \BibitemOpen
  \bibfield  {author} {\bibinfo {author} {\bibfnamefont {M.}~\bibnamefont
  {L{\"u}scher}},\ }\href {\doibase 10.1016/j.cpc.2004.10.004} {\bibfield
  {journal} {\bibinfo  {journal} {Comput. Phys. Commun.}\ }\textbf {\bibinfo
  {volume} {165}},\ \bibinfo {pages} {199} (\bibinfo {year} {2005})},\ \Eprint
  {http://arxiv.org/abs/hep-lat/0409106} {arXiv:hep-lat/0409106 [hep-lat]}
  \BibitemShut {NoStop}%
\bibitem [{\citenamefont {El-Khadra}\ \emph {et~al.}(1997)\citenamefont
  {El-Khadra}, \citenamefont {Kronfeld},\ and\ \citenamefont
  {Mackenzie}}]{ElKhadra:1996mp}%
  \BibitemOpen
  \bibfield  {author} {\bibinfo {author} {\bibfnamefont {A.~X.}\ \bibnamefont
  {El-Khadra}}, \bibinfo {author} {\bibfnamefont {A.~S.}\ \bibnamefont
  {Kronfeld}}, \ and\ \bibinfo {author} {\bibfnamefont {P.~B.}\ \bibnamefont
  {Mackenzie}},\ }\href {\doibase 10.1103/PhysRevD.55.3933} {\bibfield
  {journal} {\bibinfo  {journal} {Phys. Rev.}\ }\textbf {\bibinfo {volume}
  {D55}},\ \bibinfo {pages} {3933} (\bibinfo {year} {1997})},\ \Eprint
  {http://arxiv.org/abs/hep-lat/9604004} {arXiv:hep-lat/9604004 [hep-lat]}
  \BibitemShut {NoStop}%
\bibitem [{\citenamefont {Aoki}\ \emph {et~al.}(2003)\citenamefont {Aoki},
  \citenamefont {Kuramashi},\ and\ \citenamefont {Tominaga}}]{Aoki:2001ra}%
  \BibitemOpen
  \bibfield  {author} {\bibinfo {author} {\bibfnamefont {S.}~\bibnamefont
  {Aoki}}, \bibinfo {author} {\bibfnamefont {Y.}~\bibnamefont {Kuramashi}}, \
  and\ \bibinfo {author} {\bibfnamefont {S.-I.}\ \bibnamefont {Tominaga}},\
  }\href {\doibase 10.1143/PTP.109.383} {\bibfield  {journal} {\bibinfo
  {journal} {Prog. Theor. Phys.}\ }\textbf {\bibinfo {volume} {109}},\ \bibinfo
  {pages} {383} (\bibinfo {year} {2003})},\ \Eprint
  {http://arxiv.org/abs/hep-lat/0107009} {arXiv:hep-lat/0107009 [hep-lat]}
  \BibitemShut {NoStop}%
\bibitem [{\citenamefont {Christ}\ \emph {et~al.}(2007)\citenamefont {Christ},
  \citenamefont {Li},\ and\ \citenamefont {Lin}}]{Christ:2006us}%
  \BibitemOpen
  \bibfield  {author} {\bibinfo {author} {\bibfnamefont {N.~H.}\ \bibnamefont
  {Christ}}, \bibinfo {author} {\bibfnamefont {M.}~\bibnamefont {Li}}, \ and\
  \bibinfo {author} {\bibfnamefont {H.-W.}\ \bibnamefont {Lin}},\ }\href
  {\doibase 10.1103/PhysRevD.76.074505} {\bibfield  {journal} {\bibinfo
  {journal} {Phys. Rev.}\ }\textbf {\bibinfo {volume} {D76}},\ \bibinfo {pages}
  {074505} (\bibinfo {year} {2007})},\ \Eprint
  {http://arxiv.org/abs/hep-lat/0608006} {arXiv:hep-lat/0608006 [hep-lat]}
  \BibitemShut {NoStop}%
\bibitem [{\citenamefont {Oktay}\ and\ \citenamefont
  {Kronfeld}(2008)}]{Oktay:2008ex}%
  \BibitemOpen
  \bibfield  {author} {\bibinfo {author} {\bibfnamefont {M.~B.}\ \bibnamefont
  {Oktay}}\ and\ \bibinfo {author} {\bibfnamefont {A.~S.}\ \bibnamefont
  {Kronfeld}},\ }\href {\doibase 10.1103/PhysRevD.78.014504} {\bibfield
  {journal} {\bibinfo  {journal} {Phys. Rev.}\ }\textbf {\bibinfo {volume}
  {D78}},\ \bibinfo {pages} {014504} (\bibinfo {year} {2008})},\ \Eprint
  {http://arxiv.org/abs/0803.0523} {arXiv:0803.0523 [hep-lat]} \BibitemShut
  {NoStop}%
\bibitem [{\citenamefont {Namekawa}\ \emph {et~al.}(2011)\citenamefont
  {Namekawa} \emph {et~al.}}]{Namekawa:2011wt}%
  \BibitemOpen
  \bibfield  {author} {\bibinfo {author} {\bibfnamefont {Y.}~\bibnamefont
  {Namekawa}} \emph {et~al.} (\bibinfo {collaboration} {PACS-CS}),\ }\href
  {\doibase 10.1103/PhysRevD.84.074505} {\bibfield  {journal} {\bibinfo
  {journal} {Phys. Rev.}\ }\textbf {\bibinfo {volume} {D84}},\ \bibinfo {pages}
  {074505} (\bibinfo {year} {2011})},\ \Eprint {http://arxiv.org/abs/1104.4600}
  {arXiv:1104.4600 [hep-lat]} \BibitemShut {NoStop}%
\bibitem [{\citenamefont {Lewis}\ and\ \citenamefont
  {Woloshyn}(2009)}]{Lewis:2008fu}%
  \BibitemOpen
  \bibfield  {author} {\bibinfo {author} {\bibfnamefont {R.}~\bibnamefont
  {Lewis}}\ and\ \bibinfo {author} {\bibfnamefont {R.~M.}\ \bibnamefont
  {Woloshyn}},\ }\href {\doibase 10.1103/PhysRevD.79.014502} {\bibfield
  {journal} {\bibinfo  {journal} {Phys. Rev.}\ }\textbf {\bibinfo {volume}
  {D79}},\ \bibinfo {pages} {014502} (\bibinfo {year} {2009})},\ \Eprint
  {http://arxiv.org/abs/0806.4783} {arXiv:0806.4783 [hep-lat]} \BibitemShut
  {NoStop}%
\bibitem [{\citenamefont {Gray}\ \emph {et~al.}(2005)\citenamefont {Gray},
  \citenamefont {Allison}, \citenamefont {Davies}, \citenamefont {Dalgic},
  \citenamefont {Lepage}, \citenamefont {Shigemitsu},\ and\ \citenamefont
  {Wingate}}]{Gray:2005ur}%
  \BibitemOpen
  \bibfield  {author} {\bibinfo {author} {\bibfnamefont {A.}~\bibnamefont
  {Gray}}, \bibinfo {author} {\bibfnamefont {I.}~\bibnamefont {Allison}},
  \bibinfo {author} {\bibfnamefont {C.~T.~H.}\ \bibnamefont {Davies}}, \bibinfo
  {author} {\bibfnamefont {E.}~\bibnamefont {Dalgic}}, \bibinfo {author}
  {\bibfnamefont {G.~P.}\ \bibnamefont {Lepage}}, \bibinfo {author}
  {\bibfnamefont {J.}~\bibnamefont {Shigemitsu}}, \ and\ \bibinfo {author}
  {\bibfnamefont {M.}~\bibnamefont {Wingate}},\ }\href {\doibase
  10.1103/PhysRevD.72.094507} {\bibfield  {journal} {\bibinfo  {journal} {Phys.
  Rev.}\ }\textbf {\bibinfo {volume} {D72}},\ \bibinfo {pages} {094507}
  (\bibinfo {year} {2005})},\ \Eprint {http://arxiv.org/abs/hep-lat/0507013}
  {arXiv:hep-lat/0507013 [hep-lat]} \BibitemShut {NoStop}%
\bibitem [{\citenamefont {Brown}\ \emph {et~al.}(2014)\citenamefont {Brown},
  \citenamefont {Detmold}, \citenamefont {Meinel},\ and\ \citenamefont
  {Orginos}}]{Brown:2014ena}%
  \BibitemOpen
  \bibfield  {author} {\bibinfo {author} {\bibfnamefont {Z.~S.}\ \bibnamefont
  {Brown}}, \bibinfo {author} {\bibfnamefont {W.}~\bibnamefont {Detmold}},
  \bibinfo {author} {\bibfnamefont {S.}~\bibnamefont {Meinel}}, \ and\ \bibinfo
  {author} {\bibfnamefont {K.}~\bibnamefont {Orginos}},\ }\href {\doibase
  10.1103/PhysRevD.90.094507} {\bibfield  {journal} {\bibinfo  {journal} {Phys.
  Rev.}\ }\textbf {\bibinfo {volume} {D90}},\ \bibinfo {pages} {094507}
  (\bibinfo {year} {2014})},\ \Eprint {http://arxiv.org/abs/1409.0497}
  {arXiv:1409.0497 [hep-lat]} \BibitemShut {NoStop}%
\bibitem [{\citenamefont {Patrignani}\ \emph {et~al.}(2016)\citenamefont
  {Patrignani} \emph {et~al.}}]{Patrignani:2016xqp}%
  \BibitemOpen
  \bibfield  {author} {\bibinfo {author} {\bibfnamefont {C.}~\bibnamefont
  {Patrignani}} \emph {et~al.} (\bibinfo {collaboration} {Particle Data
  Group}),\ }\href {\doibase 10.1088/1674-1137/40/10/100001} {\bibfield
  {journal} {\bibinfo  {journal} {Chin. Phys.}\ }\textbf {\bibinfo {volume}
  {C40}},\ \bibinfo {pages} {100001} (\bibinfo {year} {2016})}\BibitemShut
  {NoStop}%
\bibitem [{\citenamefont {Luscher}(1986)}]{Luscher:1986pf}%
  \BibitemOpen
  \bibfield  {author} {\bibinfo {author} {\bibfnamefont {M.}~\bibnamefont
  {Luscher}},\ }\href {\doibase 10.1007/BF01211097} {\bibfield  {journal}
  {\bibinfo  {journal} {Commun. Math. Phys.}\ }\textbf {\bibinfo {volume}
  {105}},\ \bibinfo {pages} {153} (\bibinfo {year} {1986})}\BibitemShut
  {NoStop}%
\bibitem [{\citenamefont {Luscher}(1991)}]{Luscher:1990ux}%
  \BibitemOpen
  \bibfield  {author} {\bibinfo {author} {\bibfnamefont {M.}~\bibnamefont
  {Luscher}},\ }\href {\doibase 10.1016/0550-3213(91)90366-6} {\bibfield
  {journal} {\bibinfo  {journal} {Nucl. Phys.}\ }\textbf {\bibinfo {volume}
  {B354}},\ \bibinfo {pages} {531} (\bibinfo {year} {1991})}\BibitemShut
  {NoStop}%
\bibitem [{\citenamefont {Ali}\ \emph {et~al.}(2018)\citenamefont {Ali},
  \citenamefont {Qin},\ and\ \citenamefont {Wang}}]{Ali:2018xfq}%
  \BibitemOpen
  \bibfield  {author} {\bibinfo {author} {\bibfnamefont {A.}~\bibnamefont
  {Ali}}, \bibinfo {author} {\bibfnamefont {Q.}~\bibnamefont {Qin}}, \ and\
  \bibinfo {author} {\bibfnamefont {W.}~\bibnamefont {Wang}},\ }\href {\doibase
  10.1016/j.physletb.2018.09.018} {\bibfield  {journal} {\bibinfo  {journal}
  {Phys. Lett.}\ }\textbf {\bibinfo {volume} {B785}},\ \bibinfo {pages} {605}
  (\bibinfo {year} {2018})},\ \Eprint {http://arxiv.org/abs/1806.09288}
  {arXiv:1806.09288 [hep-ph]} \BibitemShut {NoStop}%
\bibitem [{\citenamefont {Thacker}\ and\ \citenamefont
  {Lepage}(1991)}]{Thacker:1990bm}%
  \BibitemOpen
  \bibfield  {author} {\bibinfo {author} {\bibfnamefont {B.~A.}\ \bibnamefont
  {Thacker}}\ and\ \bibinfo {author} {\bibfnamefont {G.~P.}\ \bibnamefont
  {Lepage}},\ }\href {\doibase 10.1103/PhysRevD.43.196} {\bibfield  {journal}
  {\bibinfo  {journal} {Phys. Rev.}\ }\textbf {\bibinfo {volume} {D43}},\
  \bibinfo {pages} {196} (\bibinfo {year} {1991})}\BibitemShut {NoStop}%
\bibitem [{\citenamefont {Lepage}\ \emph {et~al.}(1992)\citenamefont {Lepage},
  \citenamefont {Magnea}, \citenamefont {Nakhleh}, \citenamefont {Magnea},\
  and\ \citenamefont {Hornbostel}}]{Lepage:1992tx}%
  \BibitemOpen
  \bibfield  {author} {\bibinfo {author} {\bibfnamefont {G.~P.}\ \bibnamefont
  {Lepage}}, \bibinfo {author} {\bibfnamefont {L.}~\bibnamefont {Magnea}},
  \bibinfo {author} {\bibfnamefont {C.}~\bibnamefont {Nakhleh}}, \bibinfo
  {author} {\bibfnamefont {U.}~\bibnamefont {Magnea}}, \ and\ \bibinfo {author}
  {\bibfnamefont {K.}~\bibnamefont {Hornbostel}},\ }\href {\doibase
  10.1103/PhysRevD.46.4052} {\bibfield  {journal} {\bibinfo  {journal} {Phys.
  Rev.}\ }\textbf {\bibinfo {volume} {D46}},\ \bibinfo {pages} {4052} (\bibinfo
  {year} {1992})},\ \Eprint {http://arxiv.org/abs/hep-lat/9205007}
  {arXiv:hep-lat/9205007 [hep-lat]} \BibitemShut {NoStop}%
\bibitem [{\citenamefont {Davies}\ \emph {et~al.}(1994)\citenamefont {Davies},
  \citenamefont {Hornbostel}, \citenamefont {Langnau}, \citenamefont {Lepage},
  \citenamefont {Lidsey}, \citenamefont {Shigemitsu},\ and\ \citenamefont
  {Sloan}}]{Davies:1994mp}%
  \BibitemOpen
  \bibfield  {author} {\bibinfo {author} {\bibfnamefont {C.~T.~H.}\
  \bibnamefont {Davies}}, \bibinfo {author} {\bibfnamefont {K.}~\bibnamefont
  {Hornbostel}}, \bibinfo {author} {\bibfnamefont {A.}~\bibnamefont {Langnau}},
  \bibinfo {author} {\bibfnamefont {G.~P.}\ \bibnamefont {Lepage}}, \bibinfo
  {author} {\bibfnamefont {A.}~\bibnamefont {Lidsey}}, \bibinfo {author}
  {\bibfnamefont {J.}~\bibnamefont {Shigemitsu}}, \ and\ \bibinfo {author}
  {\bibfnamefont {J.~H.}\ \bibnamefont {Sloan}},\ }\href {\doibase
  10.1103/PhysRevD.50.6963} {\bibfield  {journal} {\bibinfo  {journal} {Phys.
  Rev.}\ }\textbf {\bibinfo {volume} {D50}},\ \bibinfo {pages} {6963} (\bibinfo
  {year} {1994})},\ \Eprint {http://arxiv.org/abs/hep-lat/9406017}
  {arXiv:hep-lat/9406017 [hep-lat]} \BibitemShut {NoStop}%
\end{thebibliography}%

\end{document}